\documentclass[aps,reprint,superscriptaddress,nofootinbib]{revtex4-1}

\usepackage[utf8]{inputenc}
\usepackage[english]{babel}
\usepackage{amssymb}
\usepackage{amsmath}
\usepackage{amsfonts}
\usepackage{mathtools}
\usepackage{graphicx}
\usepackage{float}
\usepackage{dsfont} 
\usepackage{cancel}
\usepackage{enumitem}
\usepackage{appendix}
\usepackage{simplewick}
\usepackage{appendix}
\usepackage{bbm}
\usepackage[dvipsnames]{xcolor}
\usepackage{hyperref}

\newcommand{\ket}[1]{\left| #1 \right>}
\newcommand{\bra}[1]{\left< #1 \right|}

\newcommand{\Tr}{\text{Tr}}

\newcommand{\bea}{\begin{eqnarray}}
\newcommand{\eea}{\end{eqnarray}}

\newcommand{\nn}{\nonumber}

\newcommand{\be}{\begin{equation}}
\newcommand{\ee}{\end{equation}}
\newcommand{\Op}{\mathcal{O}}
\newtheorem{theorem}{Theorem}

\newtheorem{corollary}{Corollary}

\newcommand{\QUICKTIKZ}{
    \usepackage{tikz}
    \usetikzlibrary{positioning}
    \usetikzlibrary{shapes.geometric, arrows}
    \usetikzlibrary{decorations.markings}
    \usetikzlibrary{arrows.meta}
    \usetikzlibrary{intersections}
    
    \newenvironment{arrowdiagram}{
        \begin{tikzpicture}[
            align = center,
            nodecirc/.style={circle, draw, scale=0.9},
            noderect/.style={rectangle, draw, scale=0.9},
            nodebasc/.style={scale=0.9},
            endarrow/.style={thick,->,>=stealth},
            arrow/.style={
                thick,
                postaction={
                    decorate,
                    decoration={
                        markings,
                        mark=at position .5 with {\arrow[xshift=1mm]{stealth[length = 2.5mm, width = 3mm]}}
                    }
                }
            },
            arrow2/.style={
                thick,
                postaction={
                    decorate,
                    decoration={
                        markings,
                        mark=at position .5 with {\arrow[xshift=1mm]{stealth[length = 2mm, width = 3mm]stealth[length = 2mm, width = 3mm]}}
                    }
                }
            },
            arrow3/.style={
                thick,
                postaction={
                    decorate,
                    decoration={
                        markings,
                        mark=at position .5 with {\arrow[xshift=1.5mm]{stealth[length = 1.5mm, width = 3mm]stealth[length = 1.5mm, width = 3mm]stealth[length = 1.5mm, width = 3mm]}}
                    }
                }
            },
            loopup/.style ={in = 60, out = 120, loop},
            loopright/.style ={in = -30, out = 30, loop},
            loopleft/.style ={in = 150, out = 210, loop},
            loopdown/.style ={in = 240, out = 300, loop},
            label/.style = {scale=0.7}
        ]
    }{
        \end{tikzpicture}
    }
}

\newcommand{\QUICKPLUS}{
                    \newcommand{\half}{\frac{1}{2}}
        }

\newcommand{\QUICKCALCULUS}{
        \newcommand{\dx}{\partial}
                }

\newcommand{\QUICKBRACKETS}{
        \newcommand{\gr}[1]{{\left(##1\right)}} 
    \newcommand{\cb}[1]{{\left\{##1\right\}}} \newcommand{\set}[1]{\left\{ ##1 \right\}}         \newcommand{\abs}[1]{{\left|##1\right|}}
        \newcommand{\avg}[1]{\left\langle ##1 \right\rangle} \newcommand{\gen}[1]{\langle ##1 \rangle}     
            
                    \newcommand{\ev}[1]{\left.##1\right\rvert}

    \newcommand{\braket}[2]{\left\langle##1\middle| ##2 \right\rangle}             
        \newcommand{\processor}[1]{
        \def\theeqn{##1}
        \proc
        \theeqn
    }
    \newcommand{\proc}{}    \newcommand{\setrules}[1]{\renewcommand{\proc}{##1}}
    \newcommand{\subrule}[2]{\StrSubstitute{\theeqn}{##1}{##2}[\theeqn]}
    \newcommand{\eq}[1]{
        $\processor{##1}$
    }
    \newcommand{\eqn}[1]{
        \begin{equation}
            \processor{##1}
        \end{equation}
    }
    \newcommand{\eqnanum}[1]{
        \begin{equation*}
            \processor{##1}
        \end{equation*}
    }
    \newcommand{\eqm}[1]{
        \begin{align}
            \processor{##1}
        \end{align}
    }
    \newcommand{\eqases}[1]{
        \begin{cases}
            ##1
        \end{cases}
    }
}

\newcommand{\QUICKLINALG}{
            \DeclareMathOperator{\tr}{tr}
            \newcommand{\matx}[1]{\begin{bmatrix}##1\end{bmatrix}}
                }

\QUICKTIKZ
\QUICKBRACKETS
\QUICKCALCULUS
\QUICKPLUS
\QUICKLINALG

\newcommand{\im}{i}
\newcommand{\primal}[1]{\ket{#1}}
\newcommand{\dual}[1]{\bra{#1}}

\newcommand{\defaultop}{H}
\newcommand{\statesymbol}{K}
\newcommand{\alanczos}{a}
\newcommand{\blanczos}{b}
\newcommand{\moments}{\mu}

\begin{document}

\title{Quantum chaos and the complexity of spread of states}

\author{Vijay Balasubramanian}
\affiliation{David Rittenhouse Laboratory, University of Pennsylvania, 
Philadelphia, PA 19104, USA.}
\affiliation{Theoretische Natuurkunde, Vrije Universiteit Brussel, 
Pleinlaan 2, B-1050 Brussels, Belgium.}
\author{Pawel  Caputa}
\affiliation{Faculty of Physics, University of Warsaw, ul. Pasteura 5, 02-093 Warsaw, Poland.}
\author{Javier M. Magan}
\affiliation{David Rittenhouse Laboratory, University of Pennsylvania, 
Philadelphia, PA 19104, USA.}
\affiliation{Theoretische Natuurkunde, Vrije Universiteit Brussel, 
Pleinlaan 2, B-1050 Brussels, Belgium.}
\author{Qingyue Wu}
\affiliation{David Rittenhouse Laboratory, University of Pennsylvania,
Philadelphia, PA 19104, USA.}

\begin{abstract}
We propose a measure of quantum state complexity defined by minimizing the spread of the wave-function over all choices of basis. Our measure is controlled by the ``survival amplitude'' for a state to remain unchanged, and can be efficiently computed in theories with discrete spectra. For continuous Hamiltonian evolution, it generalizes Krylov operator complexity to quantum states. We apply our methods to the harmonic and inverted oscillators, particles on group manifolds, the Schwarzian theory, the SYK model, and random matrix models. For time-evolved thermofield double states in chaotic systems  our measure shows four regimes: a linear {\it ramp} up to a {\it peak} that is exponential in the entropy, followed by a {\it slope} down to a {\it plateau}. These regimes arise in the same physics producing the slope-dip-ramp-plateau structure of the Spectral Form Factor. Specifically, the complexity slope arises from spectral rigidity, distinguishing  different random matrix ensembles.
\end{abstract}

\maketitle

\section{Introduction} \label{sec:I}

In classic work, Kolmogorov proposed that the complexity of a sequence could be defined as the length of the shortest Turing machine program producing it \cite{Kolmogorov}.  In information theory, Rissanen likewise suggested that the complexity of an ensemble of messages can be defined as their minimal codelength, averaged over the ensemble \cite{Rissanen}.   Similarly, complexity of a problem class can be measured 
by
the size (depth, width, or number of gates) of the smallest circuit, defined in terms of a fixed gate set, that performs the computation.   The fundamental idea here is that the complexity of an object, or class of objects, should be understood as 
the number of simple components required to assemble it.  As such, there is an ambiguity -- the measure of complexity depends on the basis of simple components.  For this reason, most discussions of complexity in computer science deal with scaling laws, and seek to establish that, for reasonable choices of basis, these laws are universal up to  polynomial factors.  

Similar ambiguities arise in attempts to measure the complexity of quantum processes.  For example, consider Nielsen's definition of the complexity of the time evolution operator $U(t) = e^{-i H t}$ as  the minimal distance in the unitary group between  the identity and $U(t)$ \cite{Nielsen}.  This definition requires the choice of a ``complexity metric'' on the group manifold,  penalizing ``hard'' vs. ``easy''  operations, typically defined in terms of their degree of locality. Recent studies \cite{Brown:2016wib,Magan2018,Balasubramanian:2019wgd,Balasubramanian:2021mxo,Bueno:2019ajd,Brandao:2019sgy,Bulchandani2021,brown2021quantum} investigate such choices of complexity metric for describing physical time evolution, with consequences for separating integrability and chaos.

In physical systems it is also interesting to quantify  complexity of individual states. A reasonable definition should regard many-body factorized Gaussian wavefunctions as ``simple'', while widely dispersed, non-locally entangled states are ``complex''.  Dynamically speaking, we expect chaos to produce inherently complex states, perhaps even with complexity increasing over exponential durations,  
e.g., for processes associated to black hole formation \cite{SusskindQC}.  
It would be natural to quantify the complexity of a quantum state in terms of the spread of the wavefunction over some fixed basis.
But  such schemes face an ambiguity -- what basis should we pick?

We propose a measure of quantum state complexity that avoids this ambiguity.  
Suppose we start with an initial state, and allow it to spread through time evolution over some basis.
Following the spirit of Kolmogorov,  we define the complexity of the state by minimizing the spread of the wavefunction  over all possible bases. We show that this minimum is uniquely attained, throughout a finite time interval for continuous  evolution, and for all time for discrete evolution, by an orthonormal basis produced by the Lanczos recursion method \cite{Lanczosbook}. We will call it ``spread complexity''.

The notion of spread complexity is controlled by the ``survival amplitude'' for a state to remain unchanged, a fact we use to develop efficient computational methods. 
We apply our techniques analytically to particle dynamics on group manifolds, and numerically to the Schwarzian theory,  SYK model, and random matrix models.
Applied to chaotic systems including the SYK model and matrix models, we show that  complexity displays four dynamical regimes:  a linearly increasing {\it ramp}  ending in a {\it peak}, followed by a downward {\it slope}  terminating in a {\it plateau}.   The durations of the ramp and slope, and the heights of the peak and plateau, are  exponential in the entropy.  These effects can be compared to the characteristic slope-dip-ramp-plateau structure of the spectral form factor (SFF)  \cite{GUHR1998189,SFF1}. The complexity slope,  like the SFF ramp, arises from spectral rigidity \cite{GUHR1998189,RM1,RM2,RM3,REVIEW1995}, and thus  differs between  Gaussian Unitary, Orthogonal, and Symplectic random matrix ensembles.

Some alternative approaches to  state complexity based on Nielsen's geometric methods appear in \cite{Jefferson:2017sdb,Chapman:2017rqy,Caputa:2018kdj,Balasubramanian:2018hsu,Bueno:2019ajd,Erdmenger:2020sup,Chagnet:2021uvi,Erdmenger:2021wzc,Balasubramanian:2021mxo}, and proposals targeted to CFTs include \cite{Caputa:2017yrh,2018Bartek}. A definition in terms of wave-function spread with respect to the so-called ``computational basis''
was considered in \cite{perm}, and a proposal based on group cohomology is in \cite{czech2022holographic}. An interesting direction based on discrete time evolution and random circuits appears in \cite{haferkamp2021linear}.

\section{Defining Spread Complexity} \label{II.II}

Consider a quantum system with a  time-independent Hamiltonian $H$.  Time evolution of a state $\vert \psi (t)\rangle$ is governed by the Schr\"{o}dinger equation
\be
i\partial_t\vert \psi (t)\rangle=H\vert \psi (t)\rangle \; .
\label{eq:se}
\ee
The solution
$
\vert \psi (t)\rangle= e^{-iHt}\vert \psi (0)\rangle
$
has a formal power series expansion
\be\label{exp}
\vert \psi (t)\rangle=\sum^\infty_{n=0}\frac{(-it)^n}{n!}\vert \psi_n\rangle\;,
\ee
where 
$
\vert \psi_n\rangle=H^n\vert \psi(0) \rangle 
$. 
The Gram–Schmidt procedure applied to $\vert\psi_n\rangle$  generates an ordered, orthonormal basis
$\mathcal{K}=\set{\ket{K_n}: n=0,1,2,\cdots}$
for the  part of the Hilbert space explored by time development of $|\psi(0) \rangle \equiv |K_0\rangle$. The basis $\mathcal{K}$, sometimes called the Krylov basis in the recent literature, may have fewer elements than the dimension of the Hilbert space, depending on  the dynamics and the choice of initial state.

We expect more complex time evolution will spread  $\vert \psi(t) \rangle$ more widely over the Hilbert space relative to the initial state $\vert \psi \rangle$. To quantify this idea,  we define a cost function relative to a complete, orthonormal, ordered basis,  
$\mathcal{B} = 
\set{\vert B_n \rangle : n=0,1,2,\cdots}$ for the Hilbert space
\begin{equation}
C_\mathcal{B}(t) 
=\sum_n c_n\abs{\braket{\psi(t)}{B_n}}^2\equiv \sum_n c_n\,p_{\mathcal{B}}(n,t)  \, ,
\label{eq:cost}
\end{equation}
where the $c_n$ are a positive, increasing sequence of real numbers, and the $p_{\mathcal{B}}(n,t)$ are probabilities of being in each basis vector. Completeness of the basis $\mathcal{B}$, together with the unitarity of time evolution, namely $\sum_n p_{\mathcal{B}}(n,t) = 1$, implies that the cost of a wavefunction increases if it spreads deeper into the basis. We will generally take $c_n = n$ so that the cost measures the average depth in the basis of the support of $\ket{\psi(t)}$.

We could try to define a natural notion of complexity as the minimum of this cost function over all bases $\mathcal{B}$   
\begin{equation}
    C(t) = \min_\mathcal{B}C_\mathcal{B}(t) \, .
    \label{eq:complexitydef}
\end{equation}
At a time $t_0$, any basis with $\ket{B_0} = \ket{\psi(t_0)}$ will minimize (\ref{eq:complexitydef}), achieving $C(t_0) = c_0$. So performing a minimization at each time separately is trivial and it does not provide any information about the spreading dynamics. We seek for a ``functional minimization'' encoding information about the spread of the state over a finite amount of time. 
We will show that, under a reasonable choice of functional minimization, there is an essentially unique basis minimizing (\ref{eq:complexitydef}) across a finite time  domain.

To motivate our functional minimization, let $C^{(m)}_\mathcal{B} \equiv C^{(m)}_\mathcal{B}(0) = d^m C_\mathcal{B}(t)/ dt^m |_{t=0}$, and suppose that the cost functions for bases $\mathcal{B}_1$ and $\mathcal{B}_2$ have convergent Taylor expansions over $0\leq t \leq T$.  Then, if there is a $k$ such that $C^{(m)}_{\mathcal{B}_1} = C^{(m)}_{\mathcal{B}_2}$ for $m<k$ and $C^{(m)}_{\mathcal{B}_1} < C^{(m)}_{\mathcal{B}_2}$ for $m=k$, then  $C_{\mathcal{B}_1}(t) <  C_{\mathcal{B}_2}(t) $ in a domain $0\leq t \leq \tau $ for some  $\tau<T$.
We want to find the basis that minimizes the cost in this sense in the vicinity of $t=0$.  We can formalize this condition in terms of the sequence of derivatives of the cost function at $t=0$:
\begin{equation}
    S_\mathcal{B} = \left( C^{(0)}_\mathcal{B}, C^{(1)}_\mathcal{B},C^{(2)}_\mathcal{B}, \cdots \right) \, .
    \label{eq:derivseq}
\end{equation}
We write $ S_{\mathcal{B}_1} < S_{\mathcal{B}_2}$ if there is some $k$ such that $C^{(m)}_{\mathcal{B}_1} = C^{(m)}_{\mathcal{B}_2}$ for $m<k$ and $C^{(m)}_{\mathcal{B}_1} < C^{(m)}_{\mathcal{B}_2}$ for $m=k$.

In what follows, we say that an ordered basis $\mathcal{B}$ is a complete Krylov basis $\mathcal{K}_c$ if its initial elements are the Krylov basis in the correct order. In more detail, say the Krylov basis has $K$ vectors.  $K$ might be  smaller than the Hilbert space dimension, so in such cases the usual Krylov basis does not span the full Hilbert space. We call $\mathcal{B}$ a complete Krylov basis if $| B_n \rangle=| K_n \rangle$, for $n=0,\cdots, K-1$. The rest of the basis is unspecified for the concerns of this definition. This defines a class of bases for which the number of unspecified elements is the dimension of the Hilbert space minus the dimension of the Krylov basis. We will  prove that any complete Krylov basis ${\cal K}_c$, as defined above, minimizes the derivative sequence $S$ and hence has a lower cost than any other basis, at least in the vicinity of $t=0$.

\begin{theorem}
For any basis $\mathcal{B}$, $S_{\mathcal{K}}\leq S_{\mathcal{B}}$, with equality only for the complete Krylov bases $\mathcal{B}={\cal K}_c$.
\end{theorem}

{\noindent {\bf Proof: }} We will prove the theorem by induction by showing that any orthonormal basis $\mathcal{B}$ whose first $N$ elements coincide with the Krylov basis satisfies $S_{\mathcal{B}} < S_{\mathcal{B}'}$ for all bases $\mathcal{B}'$ whose first $N$ elements do not coincide with $\mathcal{K}$.

The first element of the Krylov basis is $|K_0\rangle = |\psi(0)\rangle$. Suppose now that the first element of $\mathcal{B}$ is $|B_0\rangle = |\psi_0\rangle$. Then the cost is  $C_{\mathcal{B}_1}^{(0)} = C_{\mathcal{B}}(0) = \sum_n c_n |\langle \psi(0) | B_n \rangle |^2 = c_0 $ because  $|B_{i>0}\rangle$ are orthogonal to $|\psi(0)\rangle$.  Any basis $\mathcal{B}'$ which does not include $|\psi_0\rangle$ as its first element will have a higher cost, because, from (\ref{eq:cost}) it will be a weighted average of $c_{n\geq0}$, and hence be larger than $c_0$ since $c_n$ increases with $n$.

To prove the induction step we must evaluate time derivatives of the cost $C^{(m)}_\mathcal{B}(t) = d^m C_\mathcal{B}(t)/ dt^m$. Applying the derivatives to the right  side of (\ref{eq:cost}) and using (\ref{eq:se}) gives $C_\mathcal{B}^{(m)}(0)=\sum_n c_n p_\mathcal{B}^{(m)}(n,0)$, 
where
\bea\label{eq:alphamdef}
&&p^{(m)}_\mathcal{B}(n,t)=\frac{d^m\,p_\mathcal{B}(n,t)}{dt^m} =
\\
& &i^m \sum_{k=0}^{m} (-1)^{k}\binom{m}{k}\bra{\psi(t)}H^{m-k}\ket{B_n}\bra{B_n}H^{k}\ket{\psi(t)}\;.\nonumber
\eea
Now, let us assume that the first $N$ elements of  $\mathcal{B}$ coincide with the first $N$ elements of $\mathcal{K}$, i.e. $|B_i\rangle = |K_i\rangle$ for $i = 0,1,\cdots N-1$.  Following  (\ref{eq:alphamdef}), this means that $p^{(m)}_\mathcal{B}(n,t)=p^{(m)}_\mathcal{K}(n,t)$ for basis elements $n < N$ and all derivatives $m$.  To complete the proof we need two lemmas.

\vspace{0.1in}
{\it \underline{Lemma 1}:}
{\it Suppose the first $N$ elements $\mathcal{B}$ are the first $N$ elements of $\mathcal{K}$, up to a phase factor. Then $p^{(m)}_\mathcal{B}(n,0)=0$ for $n\geq N,m< 2N$.}

\vspace{0.05 in}
{\it \underline{Proof}:}  For $k<N$, $H^k\ket{\psi(0)}$  is a linear combination of $\ket{B_0},\ldots,\ket{B_{N-1}}$, since these vectors equal the first $N$ elements of the Krylov basis.
Orthogonality of the basis $\mathcal{B}$ then implies that $\bra{\psi(0)}H^{k}\ket{B_n}=\bra{B_n}H^{k}\ket{\psi(0)}=0$ for any $n\geq N$ and $k<N$. For $m\leq 2N-1$,  we know that for any integer $k$, either $m-k$ or $k$ is less than $N$.  Since every term in (\ref{eq:alphamdef}) involves either $\bra{\psi(0)}H^{k}\ket{B_n}$ or $\bra{\psi(0)}H^{m-k}\ket{B_n}$ (or their conjugates) we conclude that $p^{(m)}_\mathcal{B}(n,0)=0$ for $n\geq N$ with $m\leq 2N-1$.

\vspace{0.1 in}
{\it \underline{Lemma 2}:}
{\it
Suppose  $|B_i\rangle = |K_i\rangle$ for $i=0,\cdots N-1$, up to phases. Then,
$C_\mathcal{B}^{(2N)}(0)\geq C_\mathcal{K}^{(2N)}(0)$, with equality  when $\mathcal{K}$ contains precisely $N$ vectors, in which case $\mathcal{B}$ is a complete Krylov basis, or when $\ket{B_N}$ also equals $\ket{K_N}$ up to a phase factor.
}

\vspace{0.05 in}
{\it \underline{Proof}:} 
Since the first $N$ basis vectors agree between $\mathcal{B}$ and $\mathcal{K}$, Lemma 1 has already shown that for $n\geq N$, $p^{(m)}_\mathcal{B}(n,0)=0$ when $m\leq 2N-1$.  So we consider $m = 2N$.  Examination of (\ref{eq:alphamdef}) shows that for $n\geq N$ there is a single non-zero term in $p^{(2N)}_\mathcal{B}(n,0)$, namely 
\begin{equation}
p^{(2N)}_\mathcal{B}(n,0)=\binom{2N}{N}\bra{\psi}H^{N}\ket{B_n}\bra{B_n}H^{N}\ket{\psi}.
\end{equation}
Let $\ket{X}$, which is not necessarily normalized, be the component of $H^N \ket{\psi}$ orthogonal to $\ket{B_0},\ldots,\ket{B_{N-1}}$. By the definition of the Krylov basis $\mathcal{K}$, $\ket{X}\propto \ket{K_N}$. Due to orthogonality, for $n\geq N$ we have
\begin{equation}
p^{(2N)}_\mathcal{B}(n,0)=\binom{2N}{N}\braket{X}{B_n}\braket{B_n}{X}.
\label{eq:p2n}
\end{equation}
By completeness of bases, $\sum_n \braket{X}{B_n}\braket{B_n}{X}=\braket{X}{X}$. If $\ket{X}=0$, then $\mathcal{K}$ only contains $N$ vectors, $\mathcal{B}$ is a complete Krylov basis and $C_\mathcal{B}^{(2N)}(0)=C_\mathcal{K}^{(2N)}(0)$. Otherwise, $\braket{X}{X}>0$ and
\eqm{&C_\mathcal{B}^{(2N)}(0)=\sum_n c_n\, p^{(2N)}_\mathcal{B}(n,0)
 \\
&\qquad =\sum_{n=0}^{N-1} c_n\, p^{(2N)}_\mathcal{B}(n,0)+\binom{2N}{N} \sum_{n=N}^{D} c_n\braket{X}{B_n}\braket{B_n}{X} \nonumber \\
&\qquad \geq \sum_{n=0}^{N-1} c_n\, p^{(2N)}_\mathcal{K}(n,0)+\binom{2N}{N} c_N\braket{X}{X}=C_\mathcal{K}^{(2N)}(0)\;,
\nonumber
}
where $D$ is the dimension of the full Hilbert space, which could be infinite. In the last line we used the fact that $c_n$ is increasing, that $\sum_n \braket{X}{B_n}\braket{B_n}{X}=\braket{X}{X}$, and that the first $N$ basis vectors of $\mathcal{B}$ and $\mathcal{K}$ are equal. Equality is achieved only when $\ket{B_N}\propto \ket{X}\propto \ket{K_N}$, up to a phase. Otherwise we have a strict inequality.

Given these lemmas, suppose that a basis $\mathcal{B}$ coincides with the Krylov basis $\mathcal{K}$ up to phases in the first $N$ basis elements, and deviates thereafter. Lemma 1 tells us that the first $2N$ derivatives of the cost function are the same as those of the Krylov basis, because the other basis elements contribute zero. Lemma 2 tells us that if $\ket{B_N}$ is not $\ket{K_N}$ up to a phase, then its $2N$th derivative will be larger. Since the first $2N$ derivatives are equal and the $2N$th derivative of $C_\mathcal{B}(t)$ is larger, $S_\mathcal{B}>S_\mathcal{K}$, completing the proof of the theorem.

\begin{corollary}
Any cost function of the form (\ref{eq:cost})
defined in terms of an increasing, positive sequence $c_n$ and a basis $\mathcal{B}$ is minimized near $t=0$ by a complete Krylov basis $\mathcal{K}_c$.   Thus the associated spread complexity function (\ref{eq:complexitydef}) is $C(t) = C_\mathcal{K}(t)$.
\end{corollary}
 
We have arrived at a basis-independent definition for the complexity, relative to the initial condition, of a quantum state evolving continuously in time.

\subsection{Minimization for discrete time evolution}

The above results can be extended to show that, for discrete time evolution, the Krylov basis minimizes the cost (\ref{eq:cost}) for all times. 
Suppose the discrete time evolution is given by $U_n |\psi(0)\rangle = \vert \psi_n\rangle$, for a sequence of unitaries $U_n$ with $U_0=1$ and $n=0,1,\cdots$. We define the Krylov basis by choosing $|K_0\rangle = |\psi_0\rangle$ and then recursively orthogonalizing each $|\psi_n\rangle$ with all the $|K_j\rangle$ for $j<n$.  As in the continuous time proof, we must choose the initial state as part of the basis that minimizes the cost, i.e., it should be the first state of the Krylov basis $\vert \psi_0\rangle =\vert K_0\rangle$.   Now assume the first $N$ vectors of certain basis $\mathcal{B}$ agree with the Krylov basis, namely $|B_i\rangle = |K_i\rangle$ for $i=0,\cdots N-1$. By assumption
\be 
n\leq N-1\,\,\,\,\,\rightarrow\,\,\,\,\,\vert \psi_n\rangle =\sum\limits_{j=0}^{N-1} \langle K_j\vert \psi_n\rangle \vert K_j\rangle   ,
\ee
and the costs of both bases  are the same until discrete time $n=N-1$. Now we can decompose the next state into a part belonging to the Krylov subspace $|K_i\rangle$, for $i=0,\cdots N-1$, and a part perpendicular to it. Since the bases are defined up to phases, we necessarily have something of the form
\be 
\vert \psi_N\rangle =p_{\perp}  \vert K_N\rangle +p_{\parallel} \vert \chi_{\parallel}\rangle ,
\ee
where $\vert K_N\rangle$ is the next element of the Krylov basis by definition, and $\vert \chi_{\parallel}\rangle$ can be expanded in terms of $|K_i\rangle$, for $i=0,\cdots N-1$. A basis different from the Krylov one would necessarily not include $\vert K_N\rangle$. Therefore, the cost at discrete time $N$ would be larger, since we would have to express $\vert K_N\rangle$ in the new basis, which would require at least two vectors. Since the contribution to the cost from the part $\vert \chi_{\parallel}\rangle$ is the same in both bases, the cost must increase when we divide $\vert K_N\rangle$ into several contributions, since $c_n$ is a strictly increasing function of $n$.

This completes the proof that the Krylov basis minimizes the cost function for all times.  In this argument we have ignored irrelevant phases in the choice of basis elements.

These types of discrete unitary evolutions arise naturally when considering quantum circuits. Given a computational task that produces a certain target state from a certain input state, every protocol performing the task has an assigned quantum state complexity, as defined here. It would be interesting to analyze the complexity of known quantum protocols in this light.

\subsection{Complexity as the exponential of an entropy}
\label{sec:entcomplexity}
It is natural to quantify the spread of the wavefunction as the exponential of the entropy of the probability distribution of weights in an orthonormal basis $\mathcal{B}$.  This provides an alternative definition of complexity 
\be 
C_{H_\mathcal{B}}=e^{H_\mathcal{B}}\;,
\ee
where
\be
H_\mathcal{B}(t)=-\sum\limits_{n}\,p_{\mathcal{B}}(n,t)\,\log \,p_{\mathcal{B}}(n,t)\;
\label{ecom}
\ee
is the Shannon entropy of the basis weight distribution.  Complexity defined in this way measures the minimum Hilbert space dimension required to store the probability distribution of basis weights.

We can again eliminate the basis ambiguity by defining  quantum state complexity as the minimum over all choices of basis. In fact, this entropic definition of complexity is also minimized in the Krylov basis. To show this, suppose that $\mathcal{B}$ does not contain the entire Krylov basis. Then for some $N$, the first $N$ elements of the Krylov basis are in $\mathcal{B}$,  up to a phase factor, and the $(N+1)^{\rm th}$ element is not present. Since the entropy function is independent of the order of the basis, we can let these be the first $N$ elements of the basis.  Therefore, for $n<N$ we have have $p_{\mathcal{B}}(n,t) = p_{\mathcal{K}}(n,t)$ for all  $t$.  So to see the difference between the entropies we just need to analyze $p_{\mathcal{B}}(n,t)$ for $n>N$.  

Now, by Lemma 1, for  $n\geq N$, the first $2N$ derivatives of the probability vanish. More concretely $p^{(m)}_{\mathcal{B}}(n,0) = d^m p_{\mathcal{B}}(n,0)/dt^m = 0$ for  $n\geq N$  and $m<2N$. Expanding $p_{\mathcal{B}}(n,t)$, for $n\geq N$ as a Taylor series in $t$ around $t=0$, the first non-vanishing term is
\be
p_{\mathcal{B}}(n,t)=\frac{p^{(2N)}_{\mathcal{B}}(n,0)\,t^{2N}}{(2N)!}+O(t^{2N+1})\;.
\label{eq:pnt}
\ee
The difference in entropy between two bases that agree in the first $N$ Krylov vectors lies in the sum $-\sum\limits_n p_n\log p_n$, for $n\geq N$. So we now introduce the expansion (\ref{eq:pnt}) in the entropy sum $-\sum\limits_{n\geq N} p_n\log p_n$, and split the logarithm of $p_n$  to obtain two sums, the first involving $\log(t^{2N})$ and the second involving $\log(p_{\mathcal B}^{(2N)}(n,0)/(2N)!)$.

The first sum, after
dropping terms of $O(t^{2N+1}\log t)$ coming from the corrections in (\ref{eq:pnt}), is
\be 
-\frac{t^{2N}\log (t)}{(2N-1)!}\,\sum\limits_{n\geq N }p^{(2N)}_{\mathcal{B}}(n,0)\;.
\label{eq:firstsum}
\ee
From the proof of Lemma 2 above, Eq.~\ref{eq:p2n} shows that $\sum_{n\geq N} p^{(2N)}_\mathcal{B}(n,0)= \sum_{n\geq N} \binom{2N}{N}\braket{X}{B_n}\braket{B_n}{X}$ where $\ket{X} \propto \ket{K_N}$ is the component of $H^{N}\ket{\psi}$ orthogonal to the first $N$ elements of the Krylov basis.  Hence $\ket{X}$ is also orthogonal to $|B_{n<N}\rangle$. Thus we can extend the sum above to get $\sum_{n\geq N} p^{(2N)}_\mathcal{B}(n,0)= \sum_{n\geq 0} \binom{2N}{N}\braket{X}{B_n}\braket{B_n}{X}$.   By completeness of the basis we can then write $\sum_{n\geq N} p^{(2N)}_{\mathcal{B}}(n,0) = \binom{2N}{N}\braket{X}{X}$.
Hence this first term in the sum will not be affected by the remaining elements of the basis beyond the first $N$ elements that were assumed to be the same as those of the Krylov basis.   

The second sum is
\be
-t^{2N}\sum\limits_{n\geq N} \frac{p^{(2N)}_{\mathcal{B}}(n,0)}{(2N)!}\log \gr{\frac{p^{(2N)}_{\mathcal{B}}(n,0)}{(2N)!}}.
\label{eq:secondsum}
\ee
For this sum, note that $-\frac{x}{(2N)!}\log \gr{\frac{x}{(2N)!}}$ is a strictly convex function for $x>0$. Since the probability is always positive, and for $n\geq N$, $p_{\mathcal{B}}(n,0) = 0$, the leading order term in the Taylor expansion in (\ref{eq:pnt}), $p^{(2N)}_{\mathcal{B}}(n,0)$ must be positive. Since the sequence $(\binom{2N}{N}\braket{X}{X},0,0,\ldots)$ majorizes any sequence of positive numbers that sum to $\binom{2N}{N}\braket{X}{X}$, Karamata's inequality implies that the coefficient of $t^{2N}$ in the expansion will always be larger than or equal to the case where $p^{(2N)}_{\mathcal{B}}(n,0)=0$ for all $n$ except one particular $n_*$ where $p^{(2N)}_{\mathcal{B}}(n_*,0)=\binom{2N}{N}\braket{X}{X}$. Due to the strict convexity, this inequality is strict except for the case when the previous two equations are exactly satisfied, which can only happen if some element in the basis were proportional to $\ket{X}\propto \ket{K_N}$.

Given two functions of the form $f_0(t)=\alpha_0 \,t^{2N}+O(t^{2N+1}\log t)$ and $f_1(t)=\alpha_1 \,t^{2N}+O(t^{2N+1}\log t)$ with $\alpha_0<\alpha_1$, there is some $t_0$ such that for $t<t_0$, $f_0(t)<f_1(t)$. Since the first sum (\ref{eq:firstsum}) is the same  for both the Krylov basis and $\mathcal{B}$, and the second sum (\ref{eq:secondsum}) has the form
$\alpha\, t^{2N}+O(t^{2N+1}\log t)$ there exists some $t_0$ such that $H_\mathcal{K}(t)<H_\mathcal{B}(t)$ for $t<t_0$.

We conclude that the Krylov basis also minimizes complexity when defined in terms of the entropy of the spread of the initial state over a basis. Following the same line of reasoning, we can also prove that the participation ratio associated with a given basis $\mathcal{B}$, defined as
\be 
P_{\mathcal{B}}=\frac{1}{\sum\limits_{n} p_n^2}\;,
\ee
is also minimized by the Krylov basis for small times.

\section{Computing Spread Complexity} \label{sec:comC}

Following Corollary 1, to calculate the spread complexity we must derive the Krylov basis ${\cal K}$. We can do this via the Lanczos algorithm \cite{Lanczosbook}, which recursively applies the Gram–Schmidt procedure to $\vert \psi_n\rangle = H^n |\psi(0)\rangle$ to generate an orthonormal basis
$\mathcal{K}=\set{\ket{K_n}: n=0,1,2,\cdots}$: 
\be
|A_{n+1}\rangle=(H-a_{n})|K_n\rangle-b_n|K_{n-1}\rangle,\quad |K_n\rangle=b^{-1}_n|A_n\rangle\;.
\label{eq:Lrecursion}
\ee
The Lanczos coefficients  $a_n$ and $b_n$ are defined as
\be
a_n=\langle K_n|H|K_n\rangle,\qquad b_n=\langle A_n|A_n\rangle^{1/2}\;,
\ee
with $b_0 \equiv 0$ and $|K_0\rangle=|\psi(0)\rangle$ being the initial state. Observe that the Lanczos algorithm (\ref{eq:Lrecursion}) implies that
\be\label{Hact}
H|K_n\rangle=a_n|K_n\rangle+b_{n+1}|K_{n+1}\rangle+b_n|K_{n-1}\rangle \; .
\ee
This means that the Hamiltonian becomes a tri-diagonal matrix in the Krylov basis.
For finite-dimensional systems, this is known as the ``Hessenberg form'' of the Hamiltonian.

\subsection{Krylov basis from the Hessenberg form}
Numerically stable algorithms for computing the Hessenberg form of a matrix, using Householder reflections instead of the Gram-Schmidt procedure, are commonly implemented in libraries like SciPy \cite{python-SciPy,LAPACK-Hessenberg} and Mathematica. There are two caveats. First, the ``initial state'' used in these implementations is typically fixed at $(1,0,0,\ldots)^T$. To start with an arbitrary initial state, we must first perform a change of basis so that the desired initial vector has those special coordinates. Second, the off-diagonal values $b_n$ are sometimes negative in these implementations.  This amounts to a choice of phase in the definition of the Krylov basis. Taking the absolute value of all the off-diagonal elements solves this issue, and is equivalent to multiplying some of the vectors of the new basis by $-1$, which does not change the physics. From the Hessenberg form of the Hamiltonian we can directly read off the Lanczos coefficients: the $a_n$ are the diagonal elements, and the $b_n$ are the entries above the diagonal.   The wavefunction in the Krylov basis can be obtained by exponentiating the Hessenberg form and applying to the initial state.   This procedure has the advantage of being numerically stable.

\subsection{Krylov basis from the survival amplitude}
We can also devise a more general method for computing the Lanczos coefficients which remains valid  for infinite dimensional systems and the large $N$ limit of finite dimensional systems.
We  start by showing how to compute the Lanczos coefficients from the ``survival amplitude'', i.e., the amplitude that the state at time $t$ is the same as the state at time zero. Defining the expansion of the evolving state in the Krylov basis as
\be
|\psi(t)\rangle=\sum_{n}\psi_n (t)|K_n\rangle \;,
\label{eq:krylovexap1}
\ee
the survival amplitude is just 
\begin{equation}
    S(t) = \langle \psi(t) | \psi(0) \rangle = \langle \psi(0) | e^{i H t} | \psi(0) \rangle = \psi_0(t)^* \, , 
    \label{eq:survival}
\end{equation}
where we recall that $|K_0\rangle = |\psi(0)\rangle$.  The survival amplitude is also the moment-generating function for the Hamiltonian in the initial state:
\begin{eqnarray}
\mu_n &=& 
\ev{\frac{d^n}{dt^n}S(t)}_{t=0}
=
\ev{
\langle \psi(0) | \frac{d^n}{dt^n} e^{i H t} |\psi(0) \rangle
}_{t=0 }\nonumber \\
&=&
\langle K_0 | (iH)^n | K_0 \rangle \, .
\label{eq:momgenfn}
\end{eqnarray}

\begin{figure}[t]
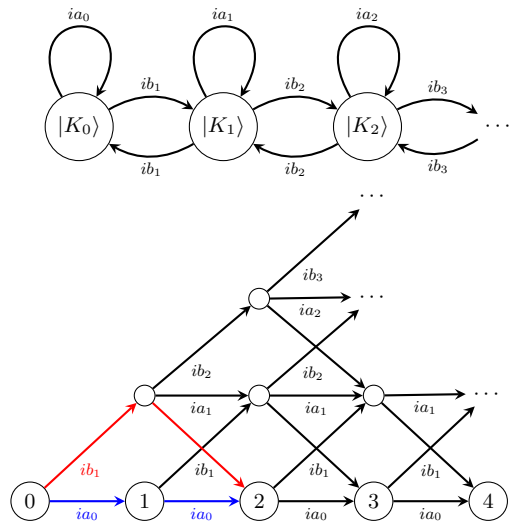

    \centering
        \begin{arrowdiagram}
                \node[nodecirc]    (A)                  {$\primal{K_0}$};
                \node[nodecirc]    (B)[right = of A]    {$\primal{K_1}$};
                \node[nodecirc]    (C)[right = of B]    {$\primal{K_2}$};
                \node[nodebasc]    (D)[right = of C]    {$\cdots$};
                
                \draw[endarrow] (A) to[bend left] node[above, label] {$\im \blanczos_1$} (B);
                \draw[endarrow] (B) to[bend left] node[below, label] {$\im \blanczos_1$} (A);
                \draw (B) edge[endarrow, bend left] node[above, label] {$\im \blanczos_2$} (C);
                \draw (C) edge[endarrow, bend left] node[below, label] {$\im \blanczos_2$} (B);
                \draw (C) edge[endarrow, bend left] node[above, label] {$\im \blanczos_3$} (D);
                \draw (D) edge[endarrow, bend left] node[below, label] {$\im \blanczos_3$} (C);
                
                \draw[endarrow] (A) to[loopup] node[above, label] {$\im \alanczos_0$} (A);
                \draw[endarrow] (B) to[loopup] node[above, label] {$\im \alanczos_1$} (B);
                \draw[endarrow] (C) to[loopup] node[above, label] {$\im \alanczos_2$} (C);
        \end{arrowdiagram}
        \begin{arrowdiagram}
                \node[nodecirc]    (A0)                   {$0$};
                \node[nodecirc]    (B0)[right = of A0]    {$1$};
                \node[nodecirc]    (B1)[above = of B0]    {};
                \node[nodecirc]    (C0)[right = of B0]    {$2$};
                \node[nodecirc]    (C1)[above = of C0]    {};
                \node[nodecirc]    (C2)[above = of C1]    {};
                \node[nodecirc]    (D0)[right = of C0]    {$3$};
                \node[nodecirc]    (D1)[above = of D0]    {};
                \node[nodebasc]    (D2)[above = of D1]    {$\cdots$};
                \node[nodebasc]    (D3)[above = of D2]    {$\cdots$};
                \node[nodecirc]    (E0)[right = of D0]    {$4$};
                \node[nodebasc]    (E1)[above = of E0]    {$\cdots$};
                
                                \draw[endarrow, blue] (A0) to[] node[below, label] {$\im \alanczos_0$} (B0);
                \draw[endarrow, blue] (B0) to[] node[below, label] {$\im \alanczos_0$} (C0);
                \draw[endarrow] (C0) to[] node[below, label] {$\im \alanczos_0$} (D0);
                \draw[endarrow] (D0) to[] node[below, label] {$\im \alanczos_0$} (E0);
                
                \draw[endarrow] (B1) to[] node[below, label] {$\im \alanczos_1$} (C1);
                \draw[endarrow] (C1) to[] node[below, label] {$\im \alanczos_1$} (D1);
                \draw[endarrow] (D1) to[] node[below, label] {$\im \alanczos_1$} (E1);
                
                \draw[endarrow] (C2) to[] node[below, label] {$\im \alanczos_2$} (D2);
                
                                \draw[endarrow, red] (A0) to[] node[below=1.5mm, label] {$\im \blanczos_1$} (B1);
                \draw[endarrow] (B0) to[] node[below=1.5mm, label] {$\im \blanczos_1$} (C1);
                \draw[endarrow] (C0) to[] node[below=1.5mm, label] {$\im \blanczos_1$} (D1);
                \draw[endarrow] (D0) to[] node[below=1.5mm, label] {$\im \blanczos_1$} (E1);
                
                \draw[endarrow] (B1) to[] node[below=1.5mm, label] {$\im \blanczos_2$} (C2);
                \draw[endarrow] (C1) to[] node[below=1.5mm, label] {$\im \blanczos_2$} (D2);
                
                \draw[endarrow] (C2) to[] node[below=1.5mm, label] {$\im \blanczos_3$} (D3);
                
                                \draw[endarrow, red] (B1) to[] (C0);
                \draw[endarrow] (C1) to[] (D0);
                \draw[endarrow] (D1) to[] (E0);
                
                \draw[endarrow] (C2) to[] (D1);
        \end{arrowdiagram}
    \caption{{\bf Top:} ``Markov chain'' representation of $i H$. {\bf Bottom:} ``Unwrapping'' of the Markov chain so that ``time'' goes from left to right. In every vertical column of nodes, the bottom node corresponds to $\primal{\statesymbol_0}$, the first node above corresponds to $\primal{\statesymbol_1}$ and so on. The sum of the weights of the blue and red paths gives $\dual{\statesymbol_0}\gr{\im\defaultop}^2\primal{\statesymbol_0}$.
    }
    \label{LiouvMarkov}
\end{figure}

The particular form of the action of the Hamiltonian $H$ in the Krylov basis~(\ref{Hact}) can be conveniently represented by an un-normalized ``Markov chain''  with transition weights given by the Lanczos coefficients, as shown in the upper panel of Fig. \ref{LiouvMarkov}.  The action of $iH$ on  $\sum_n d_n |K_n\rangle$ is then equivalent to the action of the chain transition matrix on a chain state vector $(d_0, d_1, \cdots )$.  If we start with the vector $(1,0,0,\cdots)$
and iterate the chain $n$ times, the weight of the $i^{\rm th}$ node will be the weight of $|K_i\rangle$
in the state  $(i H)^n |K_0\rangle$.  Thus, after $n$ iterations of the chain, the weight of $|K_0\rangle$ will be the moment $\mu_n = \langle K_0 | (i H)^n | K_0 \rangle$.

If we start with a state localized on $|K_0\rangle$,  it is convenient to ``unwrap'' the Markov chain as shown in the bottom panel of Fig.~\ref{LiouvMarkov}.  In this representation, the nodes of the $j^{\mathrm{th}}$ vertical column represent the  chain after $j$ iterations of the transition matrix.  In each column, labeled by $j$, the bottom node (in row $0$) corresponds to $|K_0\rangle$, the first node above 
(in row $1$) corresponds to $|K_1 \rangle$, and so on.  The transition weights $w(e)$ of edges $e$ between columns  represent the action of $iH$ defined in (\ref{Hact}).  We define the weight of a path of concatenated edges $P = \{e_1, e_2,\cdots \}$ as the product of the included edge weights: $w(P) = \prod_{e\in P} w(e)$.  
Finally, we define the weight of the node $0$ to be $1$, and the weight of any other node as a sum of the weights of all paths from $0$ to that node.  By construction, the weights in the $n^{\rm th}$ column are the amplitudes $\langle K_j | (iH)^n |K_0\rangle$, and, specifically,  the weight of the bottom node (labeled $n$) computes the moments $\mu_n = \langle K_0 | (i H)^n | K_0 \rangle$.

For example we have
    \be 
    \dual{K_0}\gr{\im\defaultop}\primal{K_0}=i\alanczos_0\;,
    \ee
since there is only one path from $0$ to node $1$ with weight $\im \alanczos_0$, and
    \be 
    \dual{K_0}\gr{\im\defaultop}^2\primal{K_0}=-\alanczos_0^2-\blanczos_1^2\;,
    \ee
because there are two paths from node $0$ to node $2$, one with weights $\im \alanczos_0,\im \alanczos_0$ and one with weights $\im \blanczos_1,\im \blanczos_1$.

Computing the values of $\moments_0,\ldots,\moments_n$ from $\alanczos_n,\blanczos_n$ using this path sum takes $O(n^2)$ operations. The weighted path sum from node $0$ to some node $X$ in the graph is the sum of the weighted path sums of all nodes with a transition to $X$, multiplied by the weight of the transition edge. Initializing the weighted path sum of node $0$ to 1 and performing this operation layer by layer gives the values we need on the bottom nodes $|0\rangle_n$.

\begin{figure}[t]
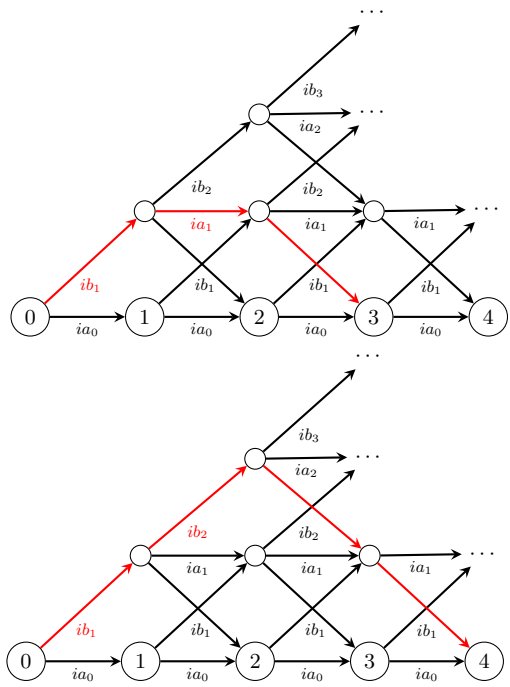

        \centering
            \begin{arrowdiagram}
                \node[nodecirc]    (A0)                   {$0$};
                \node[nodecirc]    (B0)[right = of A0]    {$1$};
                \node[nodecirc]    (B1)[above = of B0]    {};
                \node[nodecirc]    (C0)[right = of B0]    {$2$};
                \node[nodecirc]    (C1)[above = of C0]    {};
                \node[nodecirc]    (C2)[above = of C1]    {};
                \node[nodecirc]    (D0)[right = of C0]    {$3$};
                \node[nodecirc]    (D1)[above = of D0]    {};
                \node[nodebasc]    (D2)[above = of D1]    {$\cdots$};
                \node[nodebasc]    (D3)[above = of D2]    {$\cdots$};
                \node[nodecirc]    (E0)[right = of D0]    {$4$};
                \node[nodebasc]    (E1)[above = of E0]    {$\cdots$};
                
                                \draw[endarrow] (A0) to[] node[below, label] {$\im \alanczos_0$} (B0);
                \draw[endarrow] (B0) to[] node[below, label] {$\im \alanczos_0$} (C0);
                \draw[endarrow] (C0) to[] node[below, label] {$\im \alanczos_0$} (D0);
                \draw[endarrow] (D0) to[] node[below, label] {$\im \alanczos_0$} (E0);
                
                \draw[endarrow,red] (B1) to[] node[below, label] {$\im \alanczos_1$} (C1);
                \draw[endarrow] (C1) to[] node[below, label] {$\im \alanczos_1$} (D1);
                \draw[endarrow] (D1) to[] node[below, label] {$\im \alanczos_1$} (E1);
                
                \draw[endarrow] (C2) to[] node[below, label] {$\im \alanczos_2$} (D2);
                
                                \draw[endarrow,red] (A0) to[] node[below=1.5mm, label] {$\im \blanczos_1$} (B1);
                \draw[endarrow] (B0) to[] node[below=1.5mm, label] {$\im \blanczos_1$} (C1);
                \draw[endarrow] (C0) to[] node[below=1.5mm, label] {$\im \blanczos_1$} (D1);
                \draw[endarrow] (D0) to[] node[below=1.5mm, label] {$\im \blanczos_1$} (E1);
                
                \draw[endarrow] (B1) to[] node[below=1.5mm, label] {$\im \blanczos_2$} (C2);
                \draw[endarrow] (C1) to[] node[below=1.5mm, label] {$\im \blanczos_2$} (D2);
                
                \draw[endarrow] (C2) to[] node[below=1.5mm, label] {$\im \blanczos_3$} (D3);
                
                                \draw[endarrow] (B1) to[] (C0);
                \draw[endarrow,red] (C1) to[] (D0);
                \draw[endarrow] (D1) to[] (E0);
                
                \draw[endarrow] (C2) to[] (D1);
            \end{arrowdiagram}
            \begin{arrowdiagram}
                \node[nodecirc]    (A0)                   {$0$};
                \node[nodecirc]    (B0)[right = of A0]    {$1$};
                \node[nodecirc]    (B1)[above = of B0]    {};
                \node[nodecirc]    (C0)[right = of B0]    {$2$};
                \node[nodecirc]    (C1)[above = of C0]    {};
                \node[nodecirc]    (C2)[above = of C1]    {};
                \node[nodecirc]    (D0)[right = of C0]    {$3$};
                \node[nodecirc]    (D1)[above = of D0]    {};
                \node[nodebasc]    (D2)[above = of D1]    {$\cdots$};
                \node[nodebasc]    (D3)[above = of D2]    {$\cdots$};
                \node[nodecirc]    (E0)[right = of D0]    {$4$};
                \node[nodebasc]    (E1)[above = of E0]    {$\cdots$};
                
                                \draw[endarrow] (A0) to[] node[below, label] {$\im \alanczos_0$} (B0);
                \draw[endarrow] (B0) to[] node[below, label] {$\im \alanczos_0$} (C0);
                \draw[endarrow] (C0) to[] node[below, label] {$\im \alanczos_0$} (D0);
                \draw[endarrow] (D0) to[] node[below, label] {$\im \alanczos_0$} (E0);
                
                \draw[endarrow] (B1) to[] node[below, label] {$\im \alanczos_1$} (C1);
                \draw[endarrow] (C1) to[] node[below, label] {$\im \alanczos_1$} (D1);
                \draw[endarrow] (D1) to[] node[below, label] {$\im \alanczos_1$} (E1);
                
                \draw[endarrow] (C2) to[] node[below, label] {$\im \alanczos_2$} (D2);
                
                                \draw[endarrow,red] (A0) to[] node[below=1.5mm, label] {$\im \blanczos_1$} (B1);
                \draw[endarrow] (B0) to[] node[below=1.5mm, label] {$\im \blanczos_1$} (C1);
                \draw[endarrow] (C0) to[] node[below=1.5mm, label] {$\im \blanczos_1$} (D1);
                \draw[endarrow] (D0) to[] node[below=1.5mm, label] {$\im \blanczos_1$} (E1);
                
                \draw[endarrow,red] (B1) to[] node[below=1.5mm, label] {$\im \blanczos_2$} (C2);
                \draw[endarrow] (C1) to[] node[below=1.5mm, label] {$\im \blanczos_2$} (D2);
                
                \draw[endarrow] (C2) to[] node[below=1.5mm, label] {$\im \blanczos_3$} (D3);
                
                                \draw[endarrow] (B1) to[] (C0);
                \draw[endarrow] (C1) to[] (D0);
                \draw[endarrow,red] (D1) to[] (E0);
                
                \draw[endarrow,red] (C2) to[] (D1);
            \end{arrowdiagram}
        
        \caption{{\bf Top:} Except for the path in red, the weights of every path from node $0$ to node $3$ can be computed with  knowledge just of $\alanczos_0$ and $\blanczos_1$. The weight of the red path can be computed by subtracting the weights of every other path from $\moments_{3}$, and can then be used to compute $\alanczos_1$. {\bf Bottom:} Except for the path in red, the weights of every path from node $0$ to node $4$ can be computed with knowledge just of $\alanczos_0$, $\alanczos_1$, and $\blanczos_1$. The weight of the red path  can be computed by subtracting the weights of every other path from $\moments_{4}$, and can then be used to compute $\blanczos_2$.}
        \label{ExcludedPath}
    \end{figure}

Suppose now that we are given the survival amplitude $S(t)$, or can compute it through other means. By taking derivatives  we can compute the moments $\moments_0,\ldots,\moments_n$. From this data we can calculate the Lanczos coefficients by using the Markov chain described above.
Specifically, suppose we have already calculated $\alanczos_0,\ldots,\alanczos_{k-1}$ and $\blanczos_1,\ldots,\blanczos_k$ and the odd moment $\moments_{2k+1}$.   There is a unique path in the unwrapped Markov chain from node $0$ to node $2k+1$ that passes through an edge with weight $i a_k$ (example in Fig.~\ref{ExcludedPath}).  This follows because any path from $0$ to $2k+1$ must follow precisely $2k+1$ edges since every step necessarily progresses one column to the right.  This means that no path can rise to a row higher than $k$ because the need to descend back to row $0$ would make the path too long.   For the same reason, a path that reaches row $k$ must have precisely $k$ upward diagonal and $k$ downward diagonal edges, allowing a single horizontal edge in the path.  The only way to have this edge in the $k^{\rm th}$ row is to start with $k$ diagonal upward edges, then go one step horizontally in the $k^{\rm th}$ row and then descend $k$ steps diagonally.

By similar reasoning, the remaining paths between nodes $0$ and  $2k+1$ lie  below the $k^{\rm th}$ row and hence only include edges with weights $\alanczos_0,\ldots,\alanczos_{k-1}$, $\blanczos_1,\ldots,\blanczos_k$.  Thus we can compute the path sum for trajectories from node $0$ to node $2k+1$ that do not go through edge with weight $i \alanczos_{k}$, and subtract this sum from $\moments_{2k+1}$. The remainder is the weight of the excluded path, namely $\im^{2k+1}\blanczos_1^2\ldots \blanczos_k^2\alanczos_{k}$. Since we know the $\blanczos_k$'s by assumption, we may divide them out, leaving us with  $\alanczos_{k}$.

Likewise, the even moments $\moments_{2k}$ allow us to extract values of $\blanczos_{k}$. The only path from node $0$ to node $2k$ that goes through an edge of weight $\im \blanczos_k$ has path weight $\blanczos_1^2\ldots\blanczos_k^2$ (example in Fig.~\ref{ExcludedPath}). The weights of all the other paths can be computed using only $\alanczos_0,\ldots,\alanczos_k$ and $\blanczos_1,\ldots,\blanczos_{k-1}$.

To summarize, we can compute the Krylov basis and Lanczos coefficients efficiently through the following algorithm: (1) compute the survival amplitude, and use it to extract the moments of the Hamiltonian in the initial state, (2) apply the recursive algorithm above to systematically compute the Lanczos coefficients to the desired order.   This procedure is potentially sensitive to the accumulation of rounding error, due to the repeated divisions needed to compute $\alanczos_n$ and $\blanczos_n$ from their products. In our numerical analyses we avoided this instability by using the {\tt mpmath} \cite{python-mpmath} library to perform computations to arbitrary precision.  
    
\subsection{Wave function and Complexity}
Above, we described an algorithm for computing the Krylov basis ${\cal K}$ and the associated Lanczos coefficients from the survival amplitude.  To apply our definition of spread complexity to a time-evolving state we must expand it in  ${\cal K}$ as
\be
|\psi(t)\rangle=\sum_{n}\psi_n (t)|K_n\rangle \;,
\label{eq:krylovexap}
\ee
where unitarity requires $\sum_{n}|\psi_n(t)|^2\equiv\sum_{n}p_n(t)=1$.  Applying the Schr\"{o}dinger equation (\ref{eq:se}) to this expression, and then the Lanczos recursion in the form (\ref{Hact}) gives
\be\label{SchrodingerEq}
i\partial_t\psi_n(t)=a_n\psi_n(t)+b_{n+1}\psi_{n+1}(t) + b_n\psi_{n-1}(t) \;.
\ee
The survival amplitude is simply the complex conjugate of $\psi_0(t)$, see~(\ref{eq:survival}). Thus, given $\psi_0(t) = S(t)^*$ and the Lanczos coefficients, (\ref{SchrodingerEq}) defines an algebraic procedure for computing all the  $\psi_n(t)$.  We start by noting that $b_0=0$ and use $\psi_0(t)$ and its time derivative in (\ref{SchrodingerEq}) to compute $\psi_1(t)$.  Then, given $\psi_0(t)$ and $\psi_1(t)$ we can compute $\psi_2(t)$ and so on.

Finally, given $\psi_n(t)$ we  apply our definition of complexity in (\ref{eq:cost}, \ref{eq:complexitydef}):
\be 
C(t) = C_\mathcal{K}(t) =
\sum_{n} n \,p_n(t)=\sum_{n} n \,\vert \psi_n(t)\vert^2\;,
\ee
where we took the complexity coefficients in the cost function (\ref{eq:cost}) to be $c_n=n$.  With this definition, spread complexity measures the average depth of support of a time evolving state in the Krylov basis. Formally, this quantity is the expectation value in the evolving state $|\psi(t)\rangle$ of a ``complexity operator''
\be 
\hat{K}_{\psi}=\sum_{n} n |K_n\rangle\langle K_n\vert\;,
\label{KOper}
\ee
such that the spread complexity reads
\be \label{kex}
C(t)=\langle\psi(t)|\hat{K}_{\psi}|\psi(t)\rangle\;.
\ee
Below we will also consider the entropic definition~(\ref{ecom})
\be 
C_H= e^H=e^{-\sum\limits_n p_n\log p_n}\;,
\label{ecom2}
\ee
which can also be calculated from the $p_n=\vert \psi_n\vert^2$\;. This can also be understood as the exponential of the entropy of the algebra generated by the complexity operator. See \cite{ohya2004quantum} for the definition of the entropy of an operator algebra.

\subsection{Survival, TFD and the partition sum}

We will find it illuminating to study the growth of spread complexity for Thermo-Field Double (TFD) states. These are defined as follows.  Consider a Hamiltonian $H$ acting on a Hilbert space $\mathcal{H}$, with eigenstates $\vert n \rangle$ and eigenvalues $E_n$.  
To purify the thermal ensemble we construct the maximally entangled TFD state 
\be 
\vert\psi_{\beta}\rangle \equiv\frac{1}{\sqrt{Z_{\beta}}}\sum_n e^{-\frac{\beta E_n}{2}}\vert n,n\rangle\;,
\label{TFDdef}
\ee
in the tensor product of the original Hilbert space with itself. This state is invariant under evolution with Hamiltonian $H_L-H_R$, where $H_{L,R} = H$ act independently on the left and right copies of ${\cal H}$. However, the state is not invariant under evolution by the action of a single Hamiltonian, say $H_L\equiv H$. Equivalently, we could evolve by $(H_L + H_R)/2$ but these evolutions are equal because the TFD state is invariant under the action of $(H_L-H_R)/2$.  Unitary evolution with a single Hamiltonian gives
\be 
\vert\psi_{\beta} (t)\rangle=e^{-iHt}\vert\psi_{\beta}\rangle=\vert\psi_{\beta+2it}\rangle\;.
\ee
Notice that the TFD and its time evolution are contained within the subspace spanned by $\set{\ket{n,n}}$. As a result, the finite dimension algorithm for computing the Lanczos coefficients need only work within this small subspace, simplifying numerical evaluations. The maximum dimension of the explored Hilbert space in this time evolution is therefore the dimension of the original Hilbert space $\mathcal{H}$.

In the AdS/CFT correspondence, such TFD states are dual to the eternal black hole \cite{Maldacena:2001kr}. 
The spectrum of the theory is conveniently packaged in the
analytically continued  partition function
\be 
Z_{\beta-it}=\sum_n e^{-(\beta-it) E_n}\;,
\ee
and the related spectral form factor
\be
SFF_{\beta-it}\equiv\frac{\vert Z_{\beta-it}\vert^2}{\vert Z_{\beta}\vert^2} .
\ee
These time-dependent quantities  have been extensively studied in random matrix theory and quantum gravity \cite{GUHR1998189,SFF1}, for example to explore chaotic behavior.

The interesting feature for us is that the survival amplitude for the time evolved TFD state has a simple expression in terms of the partition function
\be 
S(t) =\langle\psi_{\beta+2it}\vert\psi_{\beta}\rangle=\frac{Z_{\beta-it}}{Z_{\beta}}\;.
\label{eq:TFDsurvival}
\ee
The spectral form factor (SFF) is then the survival probability of a dynamical process, corresponding to the evolution of the TFD. We can use this fact to extract the probabilities of the Krylov basis states. The fact that the survival probability of the TFD is the SFF has been used in \cite{delCampo:2017bzr} in relation to quantum speed limits. It would be interesting to understand these speed limits from the present spread complexity perspective.

Given this  survival amplitude, the moments in (\ref{eq:momgenfn})
\eqm{\moments_n \equiv \ev{\frac{d^n}{dt^n}S(t)}_{t=0}
=\frac{1}{Z_\beta}\,\textrm{Tr}\left( e^{-\beta H}\left( i H\right) ^n\right) \;,
}
are thermal expectation values of the Hamiltonian. In holographic theories,  the partition function and the energy moments have simple geometric duals, and, at least in  2d gravity \cite{Saad:2019lba,Stanford:2019vob,Johnson:2019eik,Maxfield:2020ale,Mertens:2020hbs,Witten:2020wvy}, there are non-perturbative definitions of these quantities. Since spread complexity is a functional of the survival amplitude, the relation of the latter  to the partition function provides a path towards understanding the relation between quantum complexity, geometry and quantum gravity, and perhaps the conjectures relating complexity in quantum field theory to spatial volumes and actions in a dual theory of gravity \cite{SStanford,Brown1,Brown2}. Likewise, the relation between the spectrum of the Hamiltonian and the dynamics of  complexity in  TFD states provides a  bridge from the classification of phases of quantum matter via the associated partition functions, to a novel characterization in terms of the dynamics of quantum complexity.

Finally,  although we have shown that complexity dynamics in the TFD state depends only on the spectrum, if we start with a general quantum state $\vert \psi(0)\rangle$, complexity growth will depend both on the spectrum and the structure of  energy eigenstates.  Indeed,  for a general initial state the survival amplitude is
\be
\langle \psi(t)\vert \psi\rangle =\sum\limits_n e^{i\,E_n\, t} \langle \psi(t)\vert n\rangle\langle n\vert \psi(0) \rangle \, ,
\ee
and depends  on overlaps between the evolving state and the  eigenstates.

\section{Relation to Krylov complexity}

Our approach to state complexity is related to the notion of Krylov operator complexity, which has been put forward in  \cite{Parker:2018yvk}, and developed in \cite{
Barbon:2019wsy,Avdoshkin:2019trj,Dymarsky:2019elm,Magan:2020iac,Jian:2020qpp,Rabinovici:2020ryf,Dymarsky:2021bjq,Kar:2021nbm,Caputa:2021sib,Kim:2021okd,Caputa:2021ori,Patramanis:2021lkx,Trigueros:2021rwj,Rabinovici:2021qqt,hornedal2022ultimate,Adhikari:2022whf}, based on the Lanczos approach \cite{Lanczosbook}  to operator dynamics in many-body systems. This approach starts with a Hamiltonian $H$ and a time-dependent operator $\Op(t)$, determined by
\be
\Op(t)=e^{iHt}\,\Op(0)\, e^{-iHt}\;,
\ee
in terms on the initial operator $\Op(0)\equiv\Op$. Taylor expanding in time we obtain
\be\label{expo}
\Op(t)=\sum^\infty_{n=0}\frac{(it)^n}{n!}\tilde{\Op}_n\;,
\ee
where we have defined
\be
\tilde{\Op}_0=\Op,\quad \tilde{\Op}_1=[H,\Op],\quad \tilde{\Op}_2=[H,[H,\Op]],...\,\,\,\;.
\ee
To define a notion of support of the operator
we need to introduce an inner product. At a technical level we need to endow the operator algebra with the structure of a Hilbert space. The choice of such an inner product is one of the potential ambiguities of this approach.

Concretely, suppose we say $|\mathcal{O})$ is a vector in an auxiliary Hilbert space  corresponding to operator $\mathcal{O}$ by considering the following family of inner products \cite{Lanczosbook} 
\be
(A|B)^g_\beta=\int^\beta_0 g(\lambda)\,\langle e^{\lambda H}A^\dagger e^{-\lambda H}B \rangle_\beta\, d\lambda\;, \label{IPGen}
\ee
where $\langle\rangle_\beta$ denotes the thermal expectation value 
\be
\langle A\rangle_\beta=\frac{1}{Z}\Tr\left(e^{-\beta H}A\right),\qquad Z=\Tr\left(e^{-\beta H}\right)\;,
\ee
and we require
\be
g(\lambda)\ge 0,\quad g(\beta-\lambda)=g(\lambda),\quad \frac{1}{\beta}\int^\beta_0d\lambda  \, g(\lambda)=1\;.
\ee
Once we have chosen an inner product, the Lanczos approach  starts from $\vert\tilde{\Op}_n)$ and derives an orthonormal basis $\vert\Op_n)$, the Krylov basis, by using Gram–Schmidt orthogonalization. The Krylov basis is defined recursively as
\be
|A_{n+1})=\mathcal{L}|\Op_n)-b_n|\Op_{n-1}),\quad |\Op_n)=b^{-1}_n|A_n)\;.
\label{eq:Lrecursion2}
\ee
Here $\mathcal{L}$ is the Liouvillian superoperator that produces the commutator with the Hamiltonian
\be 
\mathcal{L}|\mathcal{O})\equiv \vert [H,\mathcal{O}])\;, 
\ee
and we have defined the Lanczos coefficients  $b_n$ as
\be
b_n=(A_n|A_n)^{1/2}\;.
\ee
This iterative process has the initial conditions $b_0 \equiv 0$ and $|K_0)=|\mathcal{O} )$ is the initial state.  The Lanczos coefficients produced in this way are in general different to those produced by the Hamiltonian of theory acting on states in the original Hilbert space (\ref{eq:Lrecursion}).

We can now expand the time-dependent operator in the Krylov basis as
\be
|\Op(t))=\sum_n i^n\varphi_n(t)|\Op_n)\;.\label{OTKrB}
\ee
The amplitudes $\varphi_n(t)$ can be shown to satisfy the following Schrodinger equation
\be\label{SchrodingerEq2}
\partial_t\varphi_n(t)=b_n\varphi_{n-1}(t)-b_{n+1}\varphi_{n+1}(t)\;.
\ee
With this equation and the Lanczos coefficients $b_n$ we can solve for the amplitudes $\varphi_n(t)$ with initial condition $\varphi_n(0)=\delta_{n0}$.

The Lanczos approach suggests a natural measure of operator complexity, dubbed Krylov complexity in  \cite{Parker:2018yvk}. It is defined as the average spread of the operator in the Krylov basis
\be 
K_{\mathcal{O}}\equiv \sum_{n} n \,p_n(t)=\sum_{n} n \,\vert \varphi_n(t)\vert^2\;.
\ee
The relation of Krylov complexity to the present spread complexity is transparent once we have chosen the inner product. Such a choice maps Heisenberg evolution of the operator to Schrodinger evolution in the  auxiliary Hilbert space with the Liouvillian acting as the Hamiltonian.  Applying our approach to such a Hilbert space and the pertinent initial state $\vert\mathcal{O})$, our notion of spread complexity becomes Krylov complexity. In other words, any Krylov complexity can be understood as quantum state complexity in a certain auxiliary Hilbert space.

However, state complexity, as we defined it, is more general, and gives more fine-grained information about the quantum dynamics.   Notice that operator growth, at least as it is conventionally defined, only sees one set of Lanczos coefficients, namely the $b_n$, while for quantum states we typically have both sets of Lanczos coefficients $a_n$ and $b_n$. Equivalently, not all quantum state complexities can be understood as Krylov complexities (at least not in a conventional manner). In this sense, our approach generalizes Krylov complexity.

As developed in \cite{Parker:2018yvk,Magan:2020iac}, many different notions of operator complexity that have recently appeared in the literature can be simply understood as notions of the spread of the operator wavefunction, but with respect to different choices of orthonormal basis. 
We have shown that our measure of state complexity is distinguished in that it minimizes the spread of the operator wavefunction over all choices of basis, eliminating any basis ambiguity. This idea can be applied to operator complexity as well.

In fact, however, Krylov operator complexity has a further ambiguity that is not present in our approach, arising from the choice of inner product. Starting with \cite{Parker:2018yvk}, recent literature \cite{
Barbon:2019wsy,Avdoshkin:2019trj,Dymarsky:2019elm,Magan:2020iac,Jian:2020qpp,Rabinovici:2020ryf,
Dymarsky:2021bjq,Kar:2021nbm,Caputa:2021sib,Kim:2021okd,Caputa:2021ori,Patramanis:2021lkx,Trigueros:2021rwj,Rabinovici:2021qqt,hornedal2022ultimate,Adhikari:2022whf} has mainly focused on the Wightman inner product
\be
(A|B)=\langle e^{ H\beta/2}A^\dagger e^{-H\beta/2}B \rangle_\beta\;,
\label{wightman}
\ee
which corresponds to setting $g(\lambda)=\delta(\lambda-\beta/2)$ in~(\ref{IPGen}). With this  choice, Krylov complexity in chaotic systems grows exponentially fast \cite{Parker:2018yvk}, displaying a Lyapunov exponent that turns out to coincide with the maximal allowed value as defined by out-of-time-ordered correlators \cite{Maldacena:2015waa}. From the present perspective, this choice of inner product measures the quantum state complexity of $\vert\psi(t)\rangle=\rho_\beta^{1/4}\mathcal{O}_L (t)\rho_\beta^{-1/4}\vert \psi_\beta\rangle$, where we remind that $\vert\psi_\beta\rangle$ represents the TFD state.

But this Wightman inner product choice is  arbitrary and calls for a deeper understanding. In fact, \cite{Magan:2020iac} showed that different choices of inner product can be related to each other in a simple manner, but that Krylov complexity behaves differently with respect to each choice. Following the philosophy of the present paper, we could further minimize Krylov complexity over the choice of inner product~(\ref{IPGen}). Although we have not performed this minimization,
Ref.~\cite{Magan:2020iac} actually shows that the Wightman inner product~(\ref{wightman}) corresponds to the slowest growth of Krylov complexity.

Summarizing, the correct Lyapunov exponent arises precisely after minimizing over all possible choices of the ambiguous inner product, and over all choices of basis. This gives further support to our guiding principle that complexity should be defined via a minimization over the possible ambiguous choices.

Note that  Ref. \cite{Parker:2018yvk} argued that Krylov complexity also bounds from above (instead of from below) some other notions of operator complexity. The bound in \cite{Parker:2018yvk} is in fact consistent with our  results. Ref. \cite{Parker:2018yvk} demonstrates their bound for a certain set of highly constrained  operator complexity definitions, that not meet the criteria that we have set out, such as monotonicity of the $c_n$ coefficients.  An interesting example that demonstrates how the notion of spread complexity that we defined is a lower bound appears in the results of  Ref. \cite{perm}, where complexity was defined in terms of spread of the wave-function  with respect to the so-called ``computational basis'', e.g, in a spin model the basis that diagonalize all the spin operators in a chosen direction. This notion was found to saturate to its maximum value at time of $\mathcal{O}(N)$, where $N$ is the number of spins. By contrast, as we will see below, in the Krylov basis the spread  saturates at a time of $\mathcal{O}(e^{N})$.  In other words the spread of the wavefunction in the Krylov basis is much slower.

\section{Analytical Models}
\label{Analytic}

We will consider a class of models in which the spread complexity can be computed analytically by exploiting techniques  developed recently in the context of operator complexity \cite{Caputa:2021sib}. 
Suppose that the Hamiltonian  belongs to the Lie algebra of a symmetry group:
\be \label{liug2}
H=\alpha\,(L_+ +L_-)+\gamma L_0 +\delta \,1 \;,
\ee
where  $L_+$ and $L_-$ are raising  and lowering ladder operators, and $L_0$ belongs to the Cartan subalgebra of the Lie algebra (see  \cite{RevModPhys.62.867} for examples of such theories).  The identity term contributes to a phase to the time evolution and so does not affect the associated quantum complexity. But it can be used to set the ground state energy.
The coefficients $\alpha$ and $\gamma$ are model-dependent;
their meaning will become clearer in the specific examples.

Comparing with the action of the  Hamiltonian in the Krylov basis \eqref{Hact}, namely 
\be
H|K_n\rangle=a_n|K_n\rangle+b_{n+1}|K_{n+1}\rangle+b_n|K_{n-1}\rangle \; ,
\label{eq:recemth}
\ee
we see that, if the initial state is a highest weight state, the Krylov basis states furnish a representation of the  symmetry group. In other words, Eq.~(\ref{eq:recemth}), which provides a solution to the Lanczos recursion method by putting the Hamilton in tridiagonal form, also guarantees that the Krylov basis states form a representation of the symmetry. Moreover, since the action of the ladder operators and the elements of the Cartan subalgebra are fixed by symmetry, we can read off the Lanczos coefficients immediately
\bea 
\alpha L_+|K_n\rangle &=& b_{n+1}|K_{n+1}\rangle,\nonumber\\
\alpha L_-|K_n\rangle &=& b_{n}|K_{n-1}\rangle,\nonumber\\
\gamma L_0|K_{n}\rangle &=& a_n |K_{n}\rangle \;.
\eea
Unitary evolution with the Hamiltonian \eqref{liug2} acting on a highest weight state is determined, up to the irrelevant phase $\delta$, by a generalized Lie group displacement operator $D(\xi,\xi_0)$ 
\be 
D(\xi)\equiv e^{\xi L_+-\bar{\xi} L_- + \, \xi_0 L_0} \;,
\label{eq:gencohstate}
\ee
for $\xi = -i\alpha t$, its conjugate $\bar{\xi}$, and $\xi_0 = -i\gamma t$.   When $\xi_0=0$ this is a conventional displacement operator \cite{coherent1,coherent2}. Thus, we can understand the action of the Hamiltonian as producing generalized coherent states. The amplitudes $\psi_n(t)$ of the time-evolved state in the Krylov basis $|K_n\rangle$ are obtained by expanding these states in an orthonormal basis. The link with coherent states allows us to geometrize the notion of spread complexity following \cite{Caputa:2021sib,Caputa:2021ori}.

Below we study motion on SL(2,R), SU(2) and the Heisenberg-Weyl group. We will see that the $a_n$ coefficients can dramatically change state complexity growth.  For example, suppose the  $b_n$ grow linearly with $n$. Then systems with different $a_n$ can have have very different complexity growth patterns such as quadratic or periodic.  In fact, systems with $b_n\sim n$ and $a_n=0$ will have exponentially growing complexity, as we see by analogy with the operator growth analysis in \cite{Parker:2018yvk}.

\subsection{A particle moving in SL(2,R)}
\label{sec:SL2Rsection}
We start with SL(2,R), a group previously studied in the context of operator growth in the SYK model
\cite{Caputa:2021sib}.  
Here we will realize it as the symmetry controlling time evolution of the TFD state of the harmonic oscillator.

Consider a family of  Hamiltonians
\be\label{HSl2R}
H=\alpha(L_{-1}+L_1)+\gamma L_0+\delta \,1\;,
\ee
where the generators satisfy the SL(2,R) algebra
\be 
[L_0,L_{\pm 1}]=\mp L_{\pm 1},\qquad [L_1,L_{-1}]=2L_0\;.\label{SL2R}
\ee
In the discrete series representation associated with scaling dimension $h$, the generators act as
\bea
L_0\ket{h,n}&=&(h+n)\ket{h,n},\nn\\
L_{-1}\ket{h,n}&=&\sqrt{(n+1)(2h+n)}\ket{h,n+1},\nn\\
L_1\ket{h,n}&=&\sqrt{n(2h+n-1)}\ket{h,n-1}\;.\label{L1On}
\eea
As discussed above, for time evolution starting from a highest weight state,  $\ket{h,n}$ can be interpreted as Krylov basis elements~(\ref{HSl2R}), i.e.,
\be 
\ket{h,n}\rightarrow \vert K_n\rangle\;.
\ee
Acting with the Hamiltonian on the Krylov basis yields Lanczos coefficients
\be
a_n=\gamma(h+n)+\delta,\qquad b_n=\alpha\sqrt{n(2h+n-1)}\;.
\label{eq:sl2rLanczos}
\ee

For some choices of the coefficients in the Hamiltonian (\ref{HSl2R}) this is the time evolution of an oscillator.  To see this, consider the harmonic oscillator with frequency $\omega$, and neglect the vacuum energy since it has no effect on the dynamics. The spectrum and partition function are
\be 
E_n=\omega n\,\,\,\,\Longrightarrow\,\,\,\, Z(\beta)=\frac{e^{\beta\hslash\omega}}{e^{\beta\hslash\omega}-1}\;,
\ee
leading to the TFD survival amplitude (\ref{eq:TFDsurvival})
\be
S(t)^{HO}=\frac{(1-e^{-\beta\omega})}{1-e^{-(\beta-it)\omega}}\;.
\ee
This expression is generally complex but its norm is periodic in $t$. For the inverted harmonic oscillator, $\omega\to -i\omega$, which is strictly speaking not stable, but which we will think of in terms of analytical continuation from the stable oscillator, the norm decays to zero as time increases.

The moments can be easily computed as 
\be
\mu_n^{HO}=(\im\omega)^n\frac{\sum_{k=0}^{n-1} A(n-1,k) e^{k\beta\omega}}{(e^{\beta\omega}-1)^n} = (\im\omega)^n\frac{A_{n-1} \left[ e^{\beta\omega}\right] }{(e^{\beta\omega}-1)^n}\;,
\ee
where $A(n,k)$ and $A_n\left[ t\right] $ are the Euler numbers and polynomials respectively,  with the convention $\mu_0=1$. Starting with $A_0(t)=A_1(t)=1$, the next non-trivial Euler polynomials are 
\be
A_2(t)=1+t,\quad A_3(t)=1+4t+t^2,\cdots\;.
\ee
Using the Lanczos algorithm described above, we can now compute the two sets of Lanczos coefficients
\bea
a_n^{HO} &=& n\frac{\omega}{\tanh\left(\beta\omega/2\right)}+\frac{\omega}{e^{\beta\omega}-1}\;, \nonumber \\ 
b_n^{HO} &=&\frac{\omega}{2\sinh(\beta\omega/2)}
\,n \; .
\label{eq:HOLanczos}
\eea
These results match the SL(2,R) Lanczos coefficients (\ref{eq:sl2rLanczos}) 
if we pick the representation $h=1/2$ with  Hamiltonian coefficients
\be\label{reloh}
\gamma=\frac{\omega}{\tanh(\beta\omega/2)},\quad \delta=-\frac{\omega}{2}. \quad
\alpha=\frac{\omega}{2\sinh(\beta\omega/2)} \; .
\ee
This identification maps the initial TFD state for the oscillator to
\be 
\vert \psi_\beta\rangle =\vert K_0\rangle=\ket{h=1/2}.
\ee

The time evolution of the TFD state is now understood as a time-dependent genereralized coherent state on the group manifold in the sense of (\ref{eq:gencohstate}):
\be
\ket{\Psi(t)}=e^{-iHt}\ket{h=1/2}.
\ee
Using the Baker-Campbell-Hausdorff (BCH) relation, the unitary evolution operator can be rewritten as
\be
e^{-iHt}=e^{i\frac{\omega}{2} t}e^{A L_{-1}}e^{BL_0}e^{CL_1},\label{BCHLs}
\ee
where we have defined
\bea
A&=&C=\frac{2\alpha}{i\sqrt{4\alpha^2-\gamma^2}\coth\left(t\sqrt{\alpha^2-\gamma^2/4}\right)-\gamma},\nn\\
B&=&-2\log \left[\,\cosh\left(t\sqrt{\alpha^2-\gamma^2/4}\right)\right.+\nn\\ &&\left.\frac{i\gamma\sinh\left(t\sqrt{\alpha^2-\gamma^2/4}\right)}{2\sqrt{\alpha^2-\gamma^2/4}}\,\right].
\eea
The key quantity is 
\be
\alpha^2-\gamma^2/4=-\frac{\omega^2}{4}.
\ee
By using (\ref{reloh}) to choose $\alpha$ and $\gamma$, which quantify the growth rate of the $b_n$'s and $a_n$'s respectively, we can discuss the standard or inverted harmonic oscillator. The transition between these scenarios occurs when $\gamma=2\alpha$.

Using the BCH relation \eqref{BCHLs}, a general time-dependent coherent state is given by
\bea
&&\ket{\Psi(t)}=e^{i\frac{\omega}{2} t}e^{B h}\sum^{\infty}_{n=0}\frac{A^n}{n!} L^n_{-1}\ket{h}=\nn\\
&&e^{i\frac{\omega}{2} t}e^{B h}\sum^{\infty}_{n=0}A^n\sqrt{\frac{\Gamma(2h+n)}{n!\Gamma(2h)}}\ket{h,n}\equiv\sum^{\infty}_{n=0}\psi_n\ket{h,n}.\nonumber\\
\eea
Using $h=1/2$ for the harmonic oscillator gives
\bea
\ket{\Psi(t)}&=&e^{i\frac{\omega}{2} t}e^{\frac{B}{2}}\sum^{\infty}_{n=0}A^n\ket{1/2,n}\equiv\sum^{\infty}_{n=0}\psi_n(t)|K_n\rangle\;,\nonumber
\eea
where
\be
\psi_n(t)=\frac{1-e^{-\beta\omega}}{1-e^{-\omega(\beta+it)}}\left(\frac{2i\sin(\omega t/2)\sinh\left(\frac{\omega}{2}(\beta-it)\right)}{\cos(t\omega)-\cosh(\beta\omega)}\right)^n,
\ee
and the Krylov basis is defined as
\be
|K_n\rangle=L^{n}_{-1}\ket{1/2}\;.
\ee
The amplitudes satisfy the Schrodinger equation~(\ref{SchrodingerEq}) and it is simple to verify that
\be
\sum^\infty_{n=0}|\psi_n(t)|^2=1\;.
\ee
The complexity for the standard and inverted ($\omega\to -i\omega_i$) oscillators become respectively
\be
C(t)=\frac{\sin^2(\omega t/2)}{\sinh^2(\beta\omega/2)} ~~;~~ C(t)=\frac{\sinh^2(\omega_i t/2)}{\sin^2(\beta\omega_i/2)}\;.
\ee
We see that complexity is periodic in time for a standard oscillator, but grows exponentially for the unstable oscillator -- results that make intuitive sense.
At large times we can approximate the inverted harmonic oscillator by
\be
C(t) \simeq \frac{1}{4\sin^2(\beta\omega_i/2)}e^{\omega_i t}\equiv e^{\lambda (t-t_*)}\;,
\ee
where
$
\lambda=\omega_i$ and
$t_*=\frac{2}{\omega_i}\log\left(2\sin(\beta\omega_i/2)\right)$.

For general representations $h$ we can compute the probability $p_n$ to be:
\bea
|\psi_n(t)|^2=\frac{\Gamma(2h+n)}{n!\Gamma(2h)}\frac{\left(\frac{\sinh^2\left(\alpha t\sqrt{1-\frac{\gamma^2}{4\alpha^2}}\right)}{\cosh^2\left(\alpha t\sqrt{1-\frac{\gamma^2}{4\alpha^2}}\right)-\frac{\gamma^2}{4\alpha^2}}\right)^{n}}{\left(\frac{\cosh^2\left(\alpha t\sqrt{1-\frac{\gamma^2}{4\alpha^2}}\right)-\frac{\gamma^2}{4\alpha^2}}{1-\frac{\gamma^2}{4\alpha^2}}\right)^{2h}}\;,\nn\\\label{pnSL2R}
\eea
with complexity
\be
C(t)=\sum^\infty_{n=0}n p_n=\frac{2h}{1-\frac{\gamma^2}{4\alpha^2}}\sinh^2\left(\alpha t\sqrt{1-\frac{\gamma^2}{4\alpha^2}}\right).
\label{eq:sl2rgeneralC}
\ee
When $\gamma > 2 \alpha$ the square root is imaginary and complexity is periodic in analogy to the standard oscillator, while when $\gamma < 2 \alpha$ the system has exponentially growing complexity like the unstable oscillator.  At the transition point $\omega=0$ between the standard and inverted oscillators,  $\gamma=2\alpha$ and we have
\be  
a_n=\gamma n= 2\alpha n=2 b_n\;,
\label{freea}
\ee
for large $n$. Taking the limit $\omega \to 0$ limit in (\ref{eq:sl2rgeneralC}) from either above or below, we find that the complexity grows quadratically in time
\be \label{critc}
C_{\gamma=2\alpha}(t)=2h\alpha^2t^2\;.
\ee
Strictly speaking when $\omega=0$ the theory is free and the partition function is not well defined, but we will think about this setting as an analytical continuation of the stable oscillator.

We can give a second, more general argument, for this behavior. We place the sites of the 1d chain in (\ref{SchrodingerEq}) on the real line with spacing $\Delta x_n = \frac{1}{\sqrt{b_n}}$; the evolution can be rewritten as
\eqm{i\partial_t \psi &= b_{n+1}\psi_{n+1}+a_n\psi_n+b_n\psi_{n-1}\\
&= \frac{\frac{b_{n+1}}{b_n}\psi_{n+1}+2\psi_n+\psi_{n-1}}{\Delta x_n^2}+\gr{a_n-2b_n}\psi_n.}
In the large $n$ limit when $a_n-2b_n$ is a constant $V_0$ and $\frac{b_{n+1}}{b_n}\approx 1$, this equation  simplifies to
\eqm{i\partial_t \psi &= \frac{\psi_{n+1}+2\psi_n+\psi_{n-1}}{\Delta x_n^2}+V_0\psi_n.\label{eq:continuumlimit} }
We notice that this bears some similarity to the discretization of the second derivative of a function $\partial_x^2 f(x) = \frac{f(x+\epsilon)-2f(x)+f(x-\epsilon)}{\epsilon^2}$. For large $n$, the spacing $\Delta x_n\to 0$, and their ratios $\Delta x_n /\Delta x_{n+1}\to 1$.   We instead choose a Krylov basis with phases of $-1$ at odd sides, then in the new bases we can smoothly interpolate the values of $\psi_n$: $\psi\gr{\sum_n \Delta x_n}=(-1)^n\psi_n$.  We can then approximate (\ref{eq:continuumlimit}) in the new basis by the Schrodinger equation of a free particle on a line
\eqm{i\partial_t \psi(x) = -\partial_x^2 \psi(x)+V_0\psi(x).}
A free particle travels at a constant velocity, so we expect $\avg{x}\sim t$ in the large $n$ limit. Since in the large $n$ limit, $b_n\sim n$, the position corresponding to site $n$ is $x_n=\sum_{i=1}^n \Delta x_i \sim \sum_{i=0}^n \frac{1}{\sqrt{i}} \sim \sqrt{n}$, we expect $\avg{\sqrt{n}}\sim t$, and therefore $\avg{n}\sim t^2$.

\subsubsection{Entropic complexity and variance: }To further characterize the spread of the wavefunction across the Krylov basis \eqref{pnSL2R}, we can compute the entropic notion of the complexity in Sec.~\ref{sec:entcomplexity} and the variance of the distribution. In the context of operator growth, these were studied in \cite{Barbon:2019wsy,Caputa:2021ori} respectively, where they were dubbed K-entropy and K-variance.

For $h=1/2$ we can compute the entropy in the Krylov basis analytically 
\be
H_{\mathcal{K}}=-\frac{x\ln x+(1-x)\ln(1-x)}{x}\;,
\ee
where  
\be
x=\frac{1-\frac{\gamma^2}{4\alpha^2}}{\cosh^2\left(\alpha t\sqrt{1-\frac{\gamma^2}{4\alpha^2}}\right)-\frac{\gamma^2}{4\alpha^2}}\;,
\ee
Thus for general $\gamma$ and $\alpha$ the entropy grows linearly at late times. When  $\gamma=2\alpha$, which describes the free limit of the the oscillator, entropy at late times shows slower, logarithmic growth
\be
H_{\mathcal{K}}\sim 2\log(\alpha t)+1+O((\alpha t)^{-2})\;.
\ee
This implies a quadratic growth of the entropic definition of complexity
\be 
C_{H_{\mathcal{K}}}\sim \alpha^2 t^2\;,
\ee
similar to our original definition in~(\ref{critc}), but with parametrically smaller growth rate for large scaling dimension $h$.

Similarly, the normalized variance is 
\bea
\delta_n^2\equiv\frac{\sigma_n^2}{C^2}&=&\frac{\sum_n n^2p_n-(\sum_n np_n)^2}{(\sum_n np_n)^2}\nn\\
&=&\frac{1}{2h}\left[1+\frac{1-\frac{\gamma^2}{4\alpha^2}}{\sinh^2\left(\alpha t\sqrt{1-\frac{\gamma^2}{4\alpha^2}}\right)}\right].\label{dK2SL2R}
\eea
The variance is also sensitive to representation $h$ and shows that the distribution is sharply localized around the mean for ``heavy" states. Around $\gamma=2\alpha$, the variance approaches its large time limit more slowly, i.e., as $t^{-2}$ instead of generic exponential of \eqref{dK2SL2R}.

\subsection{A particle moving in SU(2)}
Similarly, consider the $SU(2)$ algebra
\be
[J_0,J_\pm]=\pm J_{\pm},\qquad [J_+,J_-]=2J_0\;,
\ee
and the associated Hamiltonian
\be
H=\alpha(J_{+}+J_{-})+\gamma J_0+\delta \,1\;.
\ee
Then, in the spin-j 
representation we have the action
\bea
J_0\ket{j,-j+n}&=&(-j+n)\ket{j,-j+n},\nn\\
J_+\ket{j,-j+n}&=&\sqrt{(n+1)(2j-n)}\ket{j,-j+n+1},\nn\\
J_-\ket{j,-j+n}&=&\sqrt{n(2j-n+1)}\ket{j,-j+n-1}\;.\label{Jmrepn}
\eea
Thus we identify the $2j+1$ Krylov basis vectors $\ket{K_n}=\ket{j,-j+n}$ and Lanczos coefficients
\be
a_n=\gamma(-j+n)+\delta,\qquad b_n=\alpha\sqrt{n(2j-n+1)}\;.
\ee
The BCH formula gives the time evolution operator
\be
e^{-iHt}=e^{-i\delta t}e^{A J_{+}}e^{B J_{0}}e^{C J_{-}}\;,
\ee
where
\bea
A&=&C=\frac{2\alpha}{i\sqrt{4\alpha^2+\gamma^2}\cot\left(t\sqrt{\alpha^2+\gamma^2/4}\right)-\gamma},\nn\\
B&=&-2\log\left[\cos\left(\frac{t}{2}\sqrt{4\alpha^2+\gamma^2}\right)\right. \nn\\
&+&\left.\frac{i\gamma\sin\left(\frac{t}{2}\sqrt{4\alpha^2+\gamma^2}\right)}{\sqrt{4\alpha^2+\gamma^2}}\right].
\eea
The time evolving state is then
\bea
\ket{\psi(t)}&=&e^{-iHt}\ket{j,-j}\nn\\
&=&e^{-i\delta t}e^{-jB}\sum^{2j}_{n=0}A^n\sqrt{\frac{\Gamma(2j+1)}{n!\Gamma(2j-n+1)}}\ket{j,-j+n}\;.\nn
\eea
This gives the Krylov basis coefficients (\ref{eq:krylovexap})
\be
 \psi_n(t)=e^{-i\delta t}e^{-jB}A^n\sqrt{\frac{\Gamma(2j+1)}{n!\Gamma(2j-n+1)}}\;,
\ee
and the complexity
\be
C(t)=\frac{2j}{1+\frac{\gamma^2}{4\alpha^2}}\sin^2\left(\alpha t\sqrt{1+\frac{\gamma^2}{4\alpha^2}}\right) \;,
\ee
which oscillates in time. 

These results makes sense because all the representations of SU(2) are finite dimensional.   Thus,  complexity growth is upper bounded by the dimension of the representation.  For the time evolution of highest weight states we have showed that the support of the state in the Krylov basis oscillates with time, causing the complexity to be periodic.  This evolution does not show the behavior expected in chaotic theories where the state should remain spread over the complete Hilbert for a long period of time, leading to a plateau in the complexity.

\subsection{A particle moving in the Heisenberg-Weyl group}
Finally, consider the Hamiltonian
\be
H=\lambda(a^\dagger+a)+\omega N+\delta 1\;,
\label{eq:HeisenbergH}
\ee
built from the operators of the Heisenberg-Weyl algebra 
\be
[a,a^\dagger]=1,\qquad [N,a^\dagger]=a^\dagger,\quad [N,a]=-a\;.
\ee
The action of these operators on a representation is given by
\be
a^\dagger\ket{n}=\sqrt{n+1}\ket{n+1},\qquad a\ket{n}=\sqrt{n}\ket{n-1}\;.
\ee
Following the procedure described above, we identify  the Lanczos coefficients
\be
a_n=\omega n+\delta,\qquad b_n=\lambda \sqrt{n}\;.\label{abHW}
\ee
Moreover, the unitary evolution operator can be decomposed as
\be
U(t)=e^{-iHt}=e^{\alpha_4(t)a^\dagger}e^{-\bar{\alpha}_4(t)a}e^{-i\omega t N}e^{\alpha_1(t)}\;,
\ee
where
\bea
\alpha_4(t)&=&-\frac{\lambda}{\omega}\left(1-e^{-i\omega t}\right),\\
\alpha_1(t)&=&-i\delta t+\frac{\lambda^2}{\omega^2}\left(i\omega t-1+e^{-i\omega t}\right)\;.
\eea
Our time evolving state is then
\be
\ket{\psi(t)}=e^{-iHt}\ket{0}=e^{\alpha_1(t)}\sum^\infty_{n=0}\frac{\alpha_4^n}{\sqrt{n!}}\ket{n}\;.
\ee
This leads to Krylov basis coefficients (\ref{eq:krylovexap})
\be
\psi_n(t)=e^{\alpha_1(t)}\frac{\alpha_4^n}{\sqrt{n!}}\;,
\ee
that satisfy the Schrodinger equation (\ref{SchrodingerEq}) with $a_n$ and $b_n$ given by \eqref{abHW}. The survival amplitude is
\be
\psi_0(t)=\exp\left[-i\delta t+\frac{\lambda^2}{\omega^2}\left(i\omega t-1+e^{-i\omega t}\right)\right]\;.
\ee
The complexity of the TFD state governed by this symmetry is
\be
C(t)=\frac{4\lambda^2}{\omega^2}\sin^2\left(\frac{\omega t}{2}\right)\;.
\ee
For $\omega\neq0$ the complexity oscillates even though there is an infinite dimensional Hilbert space.  This is because the inclusion of the $\omega N$ term in the Hamiltonian (\ref{eq:HeisenbergH}) produces a potential energy for excitations, effectively bounding the system. This fact  is realized in the $a_n$ coefficients that grow with $n$ when $\omega \neq 0$, which results ultimately in the oscillating complexity.   When $\omega \to 0$, so that the Hamiltonian does not contain a bounding potential for excitations, $a_n \to {\rm constant}$, and then the complexity grows quadratically in time without bound.

\section{Computational Models}
We now seek to apply our methods to three $0+1$ dimensional systems with chaotic dynamics. These are the Schwarzian theory, random matrix models and the SYK model.  The Schwarzian theory appears in the low energy approximation of the SYK model and particular matrix models, and also in the boundary description of Jackiw-Teitelboim (JT) gravity \cite{kitaev,sachdev,sachdev2,Polchinski:2016xgd,Maldacena:2016hyu,Kristan2016,Maldacena:2016upp,SFF2,Stanford:2017thb,Mertens:2017mtv,Kitaev:2017awl,GABORREVIEW}. Matrix models and their relation to two dimensional gravity have been known for a long time \cite{GUHR1998189,RM1,RM2,RM3,REVIEW1995}, but new holographic gravity applications have appeared recently \cite{Saad:2019lba,Stanford:2019vob,Johnson:2019eik,Maxfield:2020ale,Mertens:2020hbs,Witten:2020wvy}. As such, all of these theories are known to be related via dualities to gravity in 1+1 dimensions, in particular to JT gravity.   In such dualities, eternal black holes in the gravity theory are related to the TFD state of the dual $0+1$ system. 

As we described above, the survival amplitude, and hence the growth of spread  complexity, of TFD states is related to the analytically continued partition function (\ref{eq:TFDsurvival}).   Previous work \cite{GUHR1998189,SFF1} has shown the spectral  form factor (and hence the partition sum) in the thermal theory of the  models considered here initially slopes from a normalized value of 1 to an exponentially small magnitude called the dip. The dip value and its time of occurrence are not universal in chaotic systems. For Gaussian Unitary Ensembles the dip time is of $\mathcal{O}(N^{1/2})$. After the  dip, the spectral form factor grows linearly, a regime called the ramp, until it plateaus after a time exponential  in the entropy of the system  at a value that can be computed from the partition function. The ramp is a feature that can be traced back to the universal statistics of random matrices and chaotic models. The plateau persists until the appearance of Poincar\'{e} recurrences in a finite size system.

Below we will use the relation between the survival amplitude in the TFD state and the partition function (\ref{eq:TFDsurvival}) to  demonstrate related effects in the state complexity. More concretely, we find the spread complexity shows an initial linear ramp that persists for a time exponential in the system size until a peak, that is followed by a slope down to a plateau.   These four regimes arise from the same physics that drives the slope, dip, ramp and plateau of the spectral form factor.
The linear growth and plateau demonstrate the behavior expected for complexity in chaotic systems \cite{SusskindQC}.
However, peak and slope are new features  uncovered by our analysis.  The complexity peak has a height $a\,e^S$, and occurs when the system reaches the ``farthest'' states in the Hilbert space.   The subsequent downward slope bring the complexity to a lower plateau value $b\,e^S$.  The peak and slope are controlled by spectral rigidity, a universal feature seen in the energy levels of matrix models, and presumably chaotic systems in general.

\subsection{The Schwarzian theory}

The Schwarzian theory is $0+1$ dimensional \cite{kitaev,Maldacena:2016hyu,Stanford:2017thb,Mertens:2017mtv,Kitaev:2017awl,GABORREVIEW}, and the thermal Euclidean theory is defined by the action
\be 
S=-c\int_0^{\beta}\,d\tau \left\lbrace F,\tau\right\rbrace \;,
\label{eq:schwarzianaction}
\ee
where $\tau \sim \tau + \beta$, $c$ is the inverse coupling constant (with inverse energy dimensions), $F(\tau)$ is the degree of freedom and the brackets stand for the Schwarzian derivative
\be
\left\lbrace F,\tau\right\rbrace =\frac{F'''(\tau)}{F'(\tau)}-\frac{3}{2}\left( \frac{F''(\tau)}{F'(\tau)}\right) ^2\;.
\ee
The partition function of the Schwarzian theory is given by \cite{Maldacena:2016hyu,Stanford:2017thb,Mertens:2017mtv,Kitaev:2017awl}
\be 
Z_{\beta}=a\,\frac{1}{\beta^{\frac{3}{2}}} e^{\frac{2\pi^2 c}{\beta}}\propto \int_0^{\infty}dE\,\rho_E\,e^{-\beta E}\;,
\ee
where $a$ is a non-universal constant that depends on the regularization scheme. However, it disappears in appropriately normalized physical observables. The density of states is
\be 
\rho_E=\sinh (2\pi\sqrt{2cE})\;.
\ee
For this system, the moments of the Hamiltonian in the TFD state  are equal to the moments in the thermal ensemble, and can be calculated as
\be \label{momS}
\mu_n^S=i^n\frac{\textrm{Tr}\rho_{\beta}\,H^n}{Z_\beta}=\frac{2 i^n}{\sqrt{\pi}}\beta^{-n}\,\Gamma (\frac{3}{2}+n)\,M(-n, \frac{3}{2}, -\frac{2c\pi^2}{\beta})\;,
\ee
where $M(a,b,c)$ is the Kummer's (confluent) hypergeometric function.  In the semiclassical $\frac{c}{\beta}\gg 1$ limit,   
the moments of fixed $n$ become
\be 
\mu_n^S\xrightarrow[\frac{c}{\beta}\gg 1]{} i^n E^n\,\left(1+\frac{n(n+\frac{1}{2})}{\beta\,E}\right)\;,
\label{eq:HmomentsS}
\ee
where $E$ is the semiclassical one-point function
\be\label{avesc}
E=\frac{2c\pi^2}{\beta^2}\;.
\ee
We can combine this with the second moment in (\ref{eq:HmomentsS}) to show that the variance is much smaller than the mean squared,  $\sigma_E^2=2E/\beta\ll E^{2}$, i.e. this is indeed a semiclassical limit.

We can  compute the survival probability from the analytically continued partition sum.   For small times $t\ll\beta$ we have
\be\label{smt}
\sqrt{S(t) S(t)^*}\sim e^{-\frac{E\,t^2}{\beta}}=e^{-\frac{\sigma_E^2\,t^2}{2}} \;.
\ee
This type of survival probability was studied in \cite{Caputa:2021sib}, where it was established that 
\begin{equation}
a_n = 0 ~~~~;~~~b_n = \sigma_E \, \sqrt{n}\; .
\label{leadingshorttime}
\end{equation}
In our case (\ref{smt}) is an approximation, and thus we expect some non-zero $a_n$ and corrections to these values of $b_n$.  We will evaluate these numerically below.  However, if we simply keep the leading dependence (\ref{leadingshorttime}), then we know from \cite{Caputa:2021sib} that the  corresponding state complexity growth is
\be 
C(t)=\sigma_E^2\,t^2\;.
\label{eq:shortime}
\ee
Namely, complexity grows quadratically in time.  Our arguments have showed this in the semiclassical limit for sufficiently small $n$ which is a good aproximation for small times.

We can now consider the opposite limit of the moments, namely we let $n$  grow to infinity while keeping $c/\beta$ fixed. This limit becomes relevant at sufficiently late times where the power series expansion of the unitary time evolution operator $e^{-iHt}$ requires many terms to provide a good approximation.  In this limit,  Kummer's hypergeometric function scales as
\be 
M\left(-n, \frac{3}{2}, -\frac{2c\pi^2}{\beta}\right) \sim \left(\frac{2nc\pi^2}{\beta}\right)^{-\frac{1}{4}}\;.
\ee
This means that the scaling of the moments with $n$ is just controlled by the Gamma function factor in~(\ref{momS}):
\be 
\mu_n^S\sim \beta^{-n}\, n!\sim \left( \frac{n}{\beta}\right) ^n\;.
\ee
In our algorithm in  Sec.~\ref{sec:comC}, the $\mu_n$ moments were related to polynomials of $O(n)$ involving $a_k$ and $b_k$ with $k<n$. Thus, to achieve the $n^n$ growth of the moments, the $a_n$, $b_n$, or both  must scale linearly with $n$.

\begin{figure}
\centering
\begin{tabular}{c}
  \includegraphics[width=0.98\linewidth]{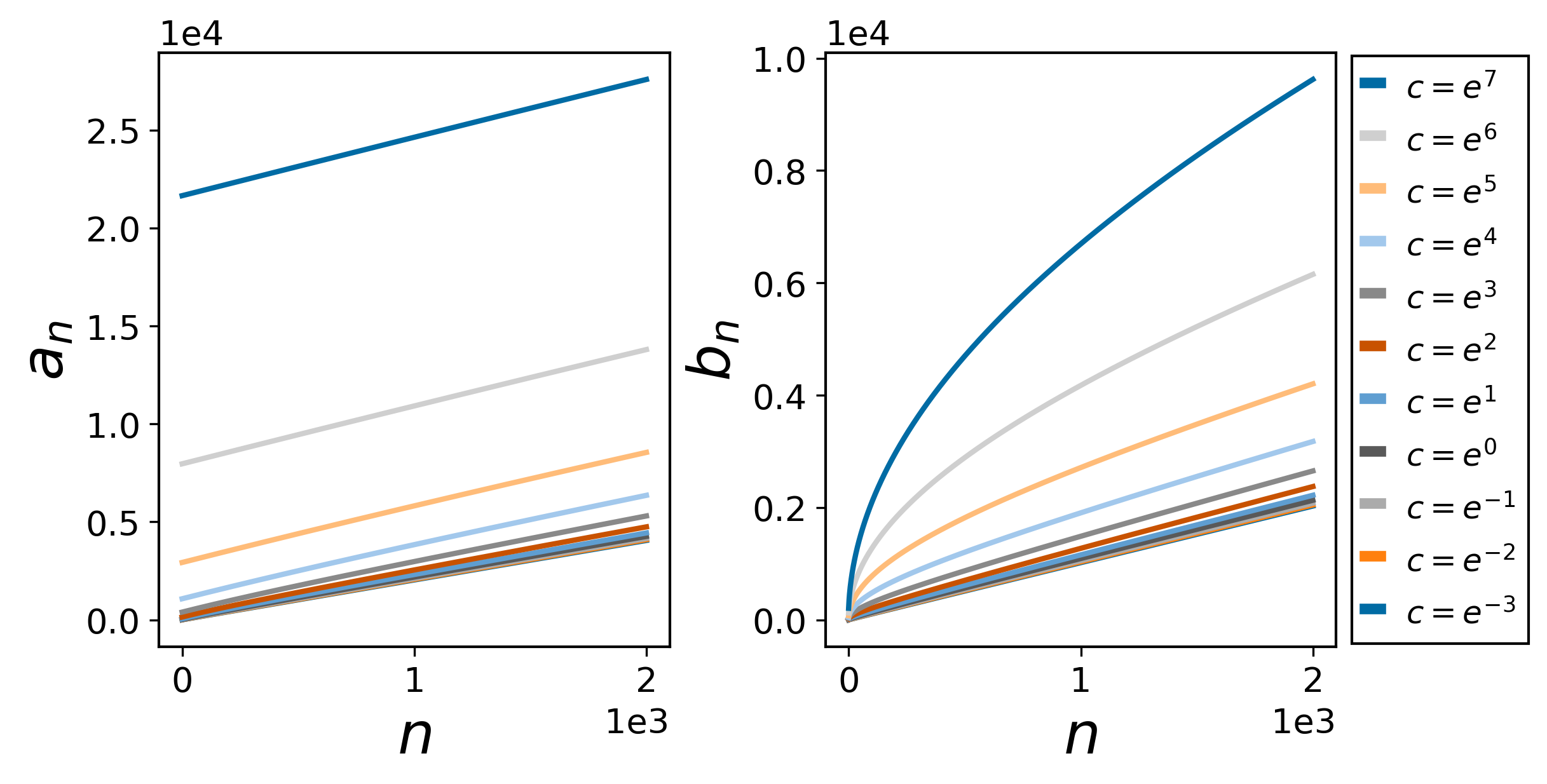}
\end{tabular}
\caption{ {\bf Left:} $a_n$ as a function of $n$ for the TFD state of the Schwarzian theory at $\beta=1$ and $c/\beta=e^{-3},\ldots, e^{7}$. {\bf Right:} $b_n$ as a function of $n$ for the same parameters. At large $c$ and $n$, both the $a_n$ and $b_n$ can be closely approximated by fitting the $a_n,b_n$ of a harmonic oscillator.
}
\label{figabns}
\end{figure}

\begin{figure}
\centering
\begin{tabular}{c}
  \includegraphics[width=0.98\linewidth]{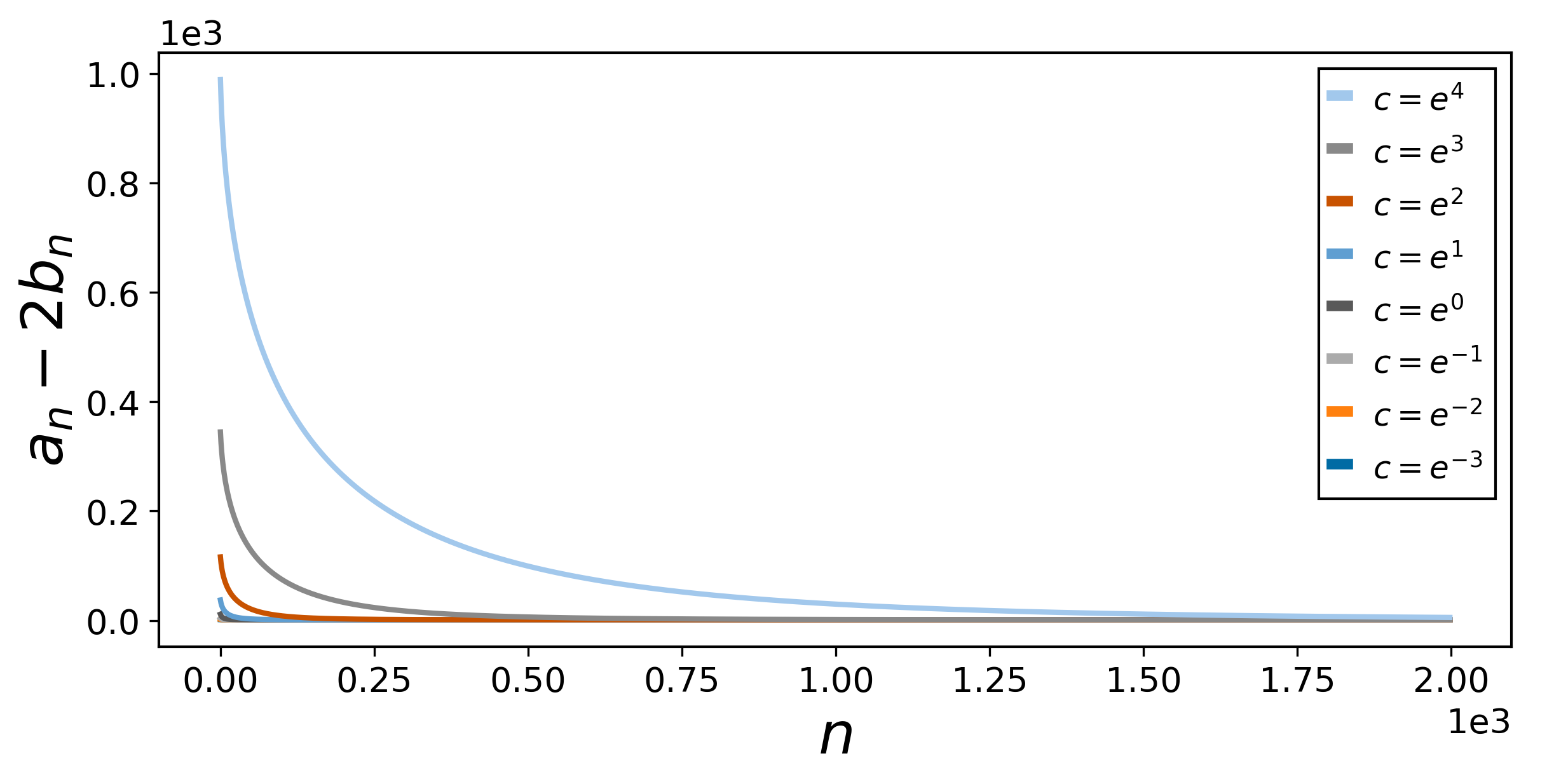}
\end{tabular}
\caption{$a_n-2b_n$ as a function of $n$ for $\beta=1$ and $c/\beta=e^{-3},\ldots, e^{4}$. $a_n-2b_n$ approaches a constant for large $n$, corresponding to the free potential limit of the harmonic oscillator.
}
\label{figratab}
\end{figure}

The previous observations tell us that we should expect two regimes for both sets of Lanczos coefficients and for the spread complexity. In the first regime we should expect a square root law for the $b_n$'s and quadratically growing behavior of complexity. In the second regime we should expect some linear behavior of the Lanczos coefficients.   As we have seen in  Sec.~\ref{Analytic}, the growth of complexity is then going to depend sensitively on how $a_n$ and $b_n$ both grow, including their coefficients.

Fig.~\ref{figabns} uses (\ref{momS}) to numerically evaluate  both Lanczos coefficients for different values of $c/\beta$, and displays the two regimes mentioned earlier. At large $n$ both of them grow linearly. The relative rate of growth is depicted in Fig.~\ref{figratab}, where it is shown that at large $n$ 
\be 
a_n=2b_n\;.
\label{eq:anbn}
\ee
This was precisely the relation between the Lanczos coefficients obtained for the TFD of the harmonic oscillator in the free limit~(\ref{freea}). Recall that this was a particular limit of motion in the SL(2,R) group.
Thus, at large times, when the wavefunction also has substantial support on Krylov basis elements $|K_n\rangle$ for large $n$, the time evolution of the TFD in the Schwarzian theory can be approximated by motion in the $SL(2,R)$ group.

In fact, for any $n$ the $SL(2,R)$ description becomes increasingly better in the semiclassical limit as we let $c/\beta$ grow.
Thus, we see that in the infinite $c/\beta$ limit, where the Schwarzian theory is well described by $AdS_2$ \cite{kitaev,Maldacena:2016hyu,Kristan2016,Maldacena:2016upp,Kitaev:2017awl,GABORREVIEW}, the evolution of the TFD is well described by motion in the $SL(2,R)$ group at all times, and the Hamiltonian and  complexity operator~(\ref{KOper}) closes an $SL(2,R)$ ``complexity algebra'' acting in the physical Hilbert space, which in turn furnishes a representation of the group. This notion of complexity algebra has been recently studied for operator growth \cite{Caputa:2021sib}.

\begin{figure}
\centering
\begin{tabular}{cc}
  \includegraphics[width=0.98\linewidth]{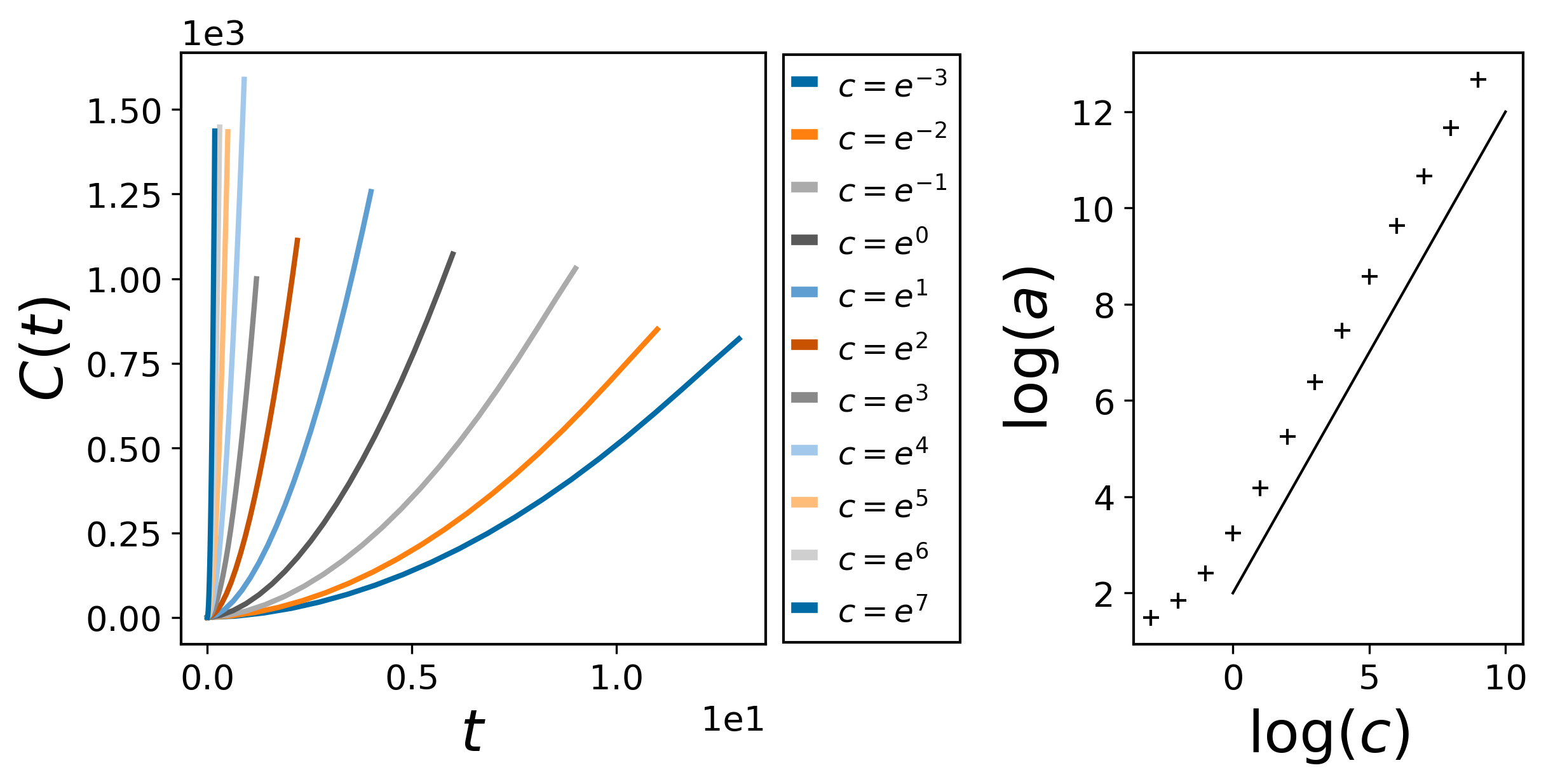}
\end{tabular}
\caption{({\bf Left}) Spread complexity as a function of time for the TFD state of the Schwarzian theory with $\beta=1$ and varying $c/\beta=e^{-3},\ldots, e^{4}$. The complexity grows approximately as $C(t)=at^2$ for some constant $a$. ({\bf Right }) $a$ (obtained from the first coefficient of a quadratic fit) vs $c$ on a (natural) log-log scale. Line with slope 1 for comparison.  We see that $a \propto c$, and from (\ref{avesc}) $\sigma_E^2 \propto c$, so that $a \propto \sigma_E^2$}
\label{kcsth}
\end{figure}

In Sec.~\ref{sec:SL2Rsection} we found that when $a_n = 2b_n \sim n$ complexity grows quadratically in time (\ref{critc}).  Thus the large $n$ result in (\ref{eq:anbn}) implies that at large times the Schwarzian theory has quadratically growing complexity.
In (\ref{eq:shortime}) we found this quadratic growth at small times also. Thus we expect quadratic growth at all times. In Fig.~(\ref{kcsth}) we confirm these expectations numerically. Fig.~(\ref{kcsth})b shows that the rate of growth is controlled by the variance of the energy, namely,
\be 
C(t)\propto \sigma_E^2 t^2\;.
\ee
As we discussed, in the semiclassical limit these relations become exact because the $SL(2,R)$ description becomes accurate at all times, and we can use the analytical results of the previous section.

\subsection{Random Matrices}

\begin{figure}
\centering
\begin{tabular}{c}
  \includegraphics[width=0.98\linewidth]{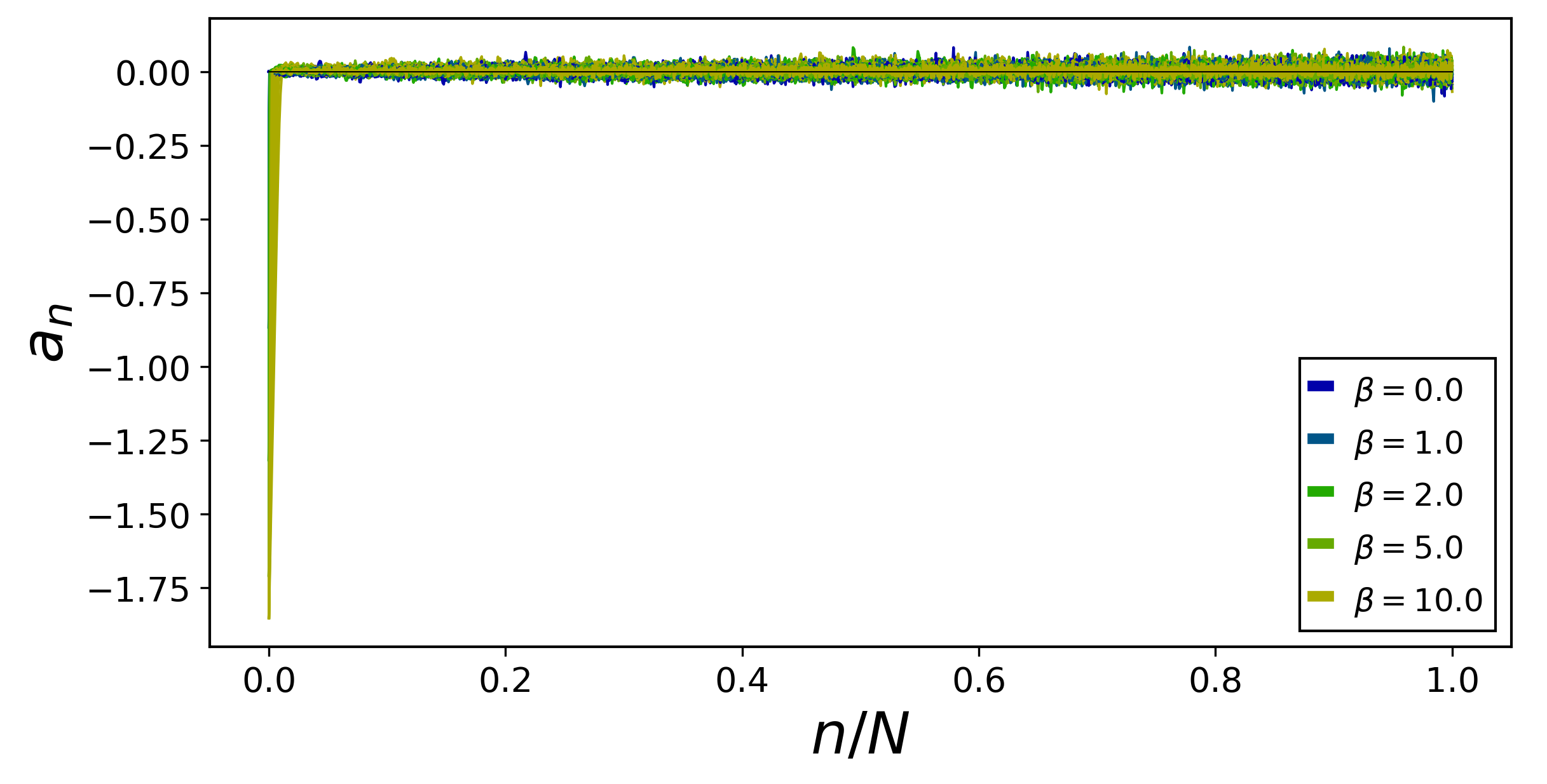}   
 \end{tabular}
\caption{Full set of Lanczos coefficients $a_n$ as a function of $n/N$ for the time evolution of the TFD state in the GUE ensemble, for different values of $\beta=\{0, 1, 2,5,10\}$ and $N=\{1024, 1280, 1536, 1792, 2048, 2560, 3072, 3584, 4096\}$. Transition between regimes analyzed in Fig.~\ref{fig_LZansmall}.
} 
\label{fig_anfull}
\end{figure}

\begin{figure}
\centering
\begin{tabular}{c}
    \includegraphics[width=0.98\linewidth]{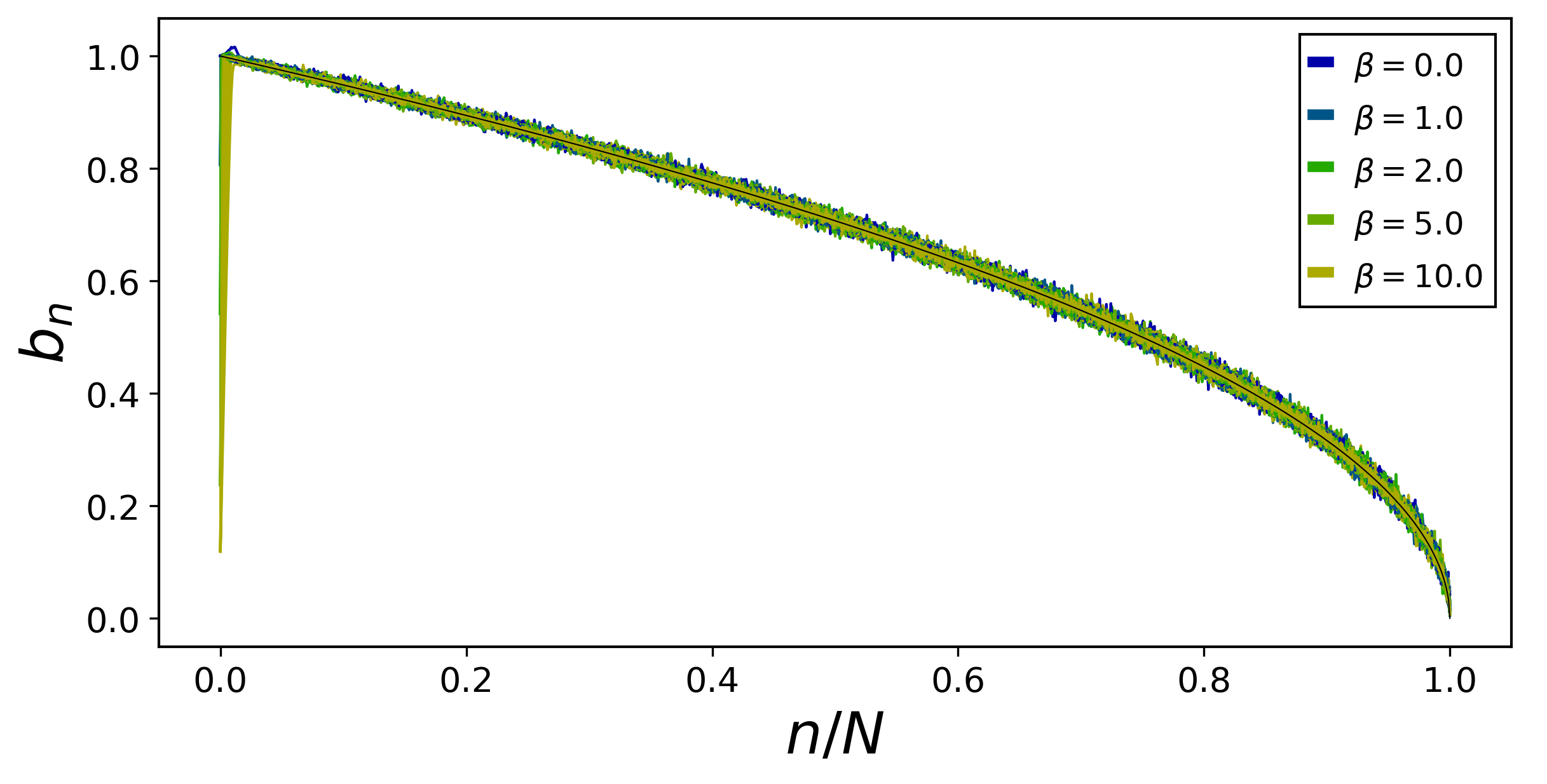}
\end{tabular}
\caption{Full set of Lanczos coefficients $b_n$ as a function of $n/N$ for time evolution of the TFD state in the GUE ensemble, for different values of $\beta=\{0, 1, 2,5,10\}$ and $N=\{1024, 1280, 1536, 1792, 2048, 2560, 3072, 3584, 4096\}$. Transition between regimes analyzed in Fig.~\ref{fig_LZbnsmall}.
}
\label{fig_bnfull}
\end{figure}

A basic conjecture states  that the fine grained structure of the spectrum of a quantum chaotic Hamiltonian is well approximated by the statistics of random matrices \cite{Dyson,Bohigas} (see the reviews \cite{GUHR1998189,RM1,RM2,RM3}). Then, given the energy-time uncertainty principle, we expect that aspects of the long time dynamics in chaotic systems will be well described by statistics of nearby eigenvalues of the Hamiltonian in the random matrix approximation.  For example, this was shown to be the case for the spectral form factor of the SYK model \cite{SFF1}. Therefore, since we seek to understand universal aspects of black holes and more general chaotic systems, it is natural to start by considering random Hamiltonians.  We will study all three universality classes: the Gaussian Unitary Ensemble (GUE), the Gaussian Orthogonal Ensemble (GOE), and the Gaussian symplectic ensemble (GSE).

A random matrix theory is defined by the specifying the probability  for finding a particular instance of a matrix in a given ensemble. The GUE is an ensemble of Hermitian $N\times N$ matrices $H_{ij}$ with gaussian measure
\be 
\frac{1}{Z_{\textrm{GUE}}}e^{-\frac{N}{2E^2_0} \textrm{Tr}(H^2)}
\;,\ee
For  numerical computations, we chose units so that $E_0=1$. Then  $Z_{\textrm{GUE}}=2^{N/2}\pi ^{N^2/2}$ is the partition function of the matrix model and normalizes the probability distribution.

We now perform the following procedure. We take an instance of a random Hamiltonian  from the GUE ensemble, compute its eigenvalues, and construct the TFD state~(\ref{TFDdef}). Next we study the unitary evolution on one side of the TFD by applying the recursion method described earlier. 
We repeat this computation for different instances of the random Hamiltonian, different values of $N$, and different values of $\beta$.

\begin{figure}
\centering
\begin{tabular}{c}
  \includegraphics[width=0.98\linewidth]{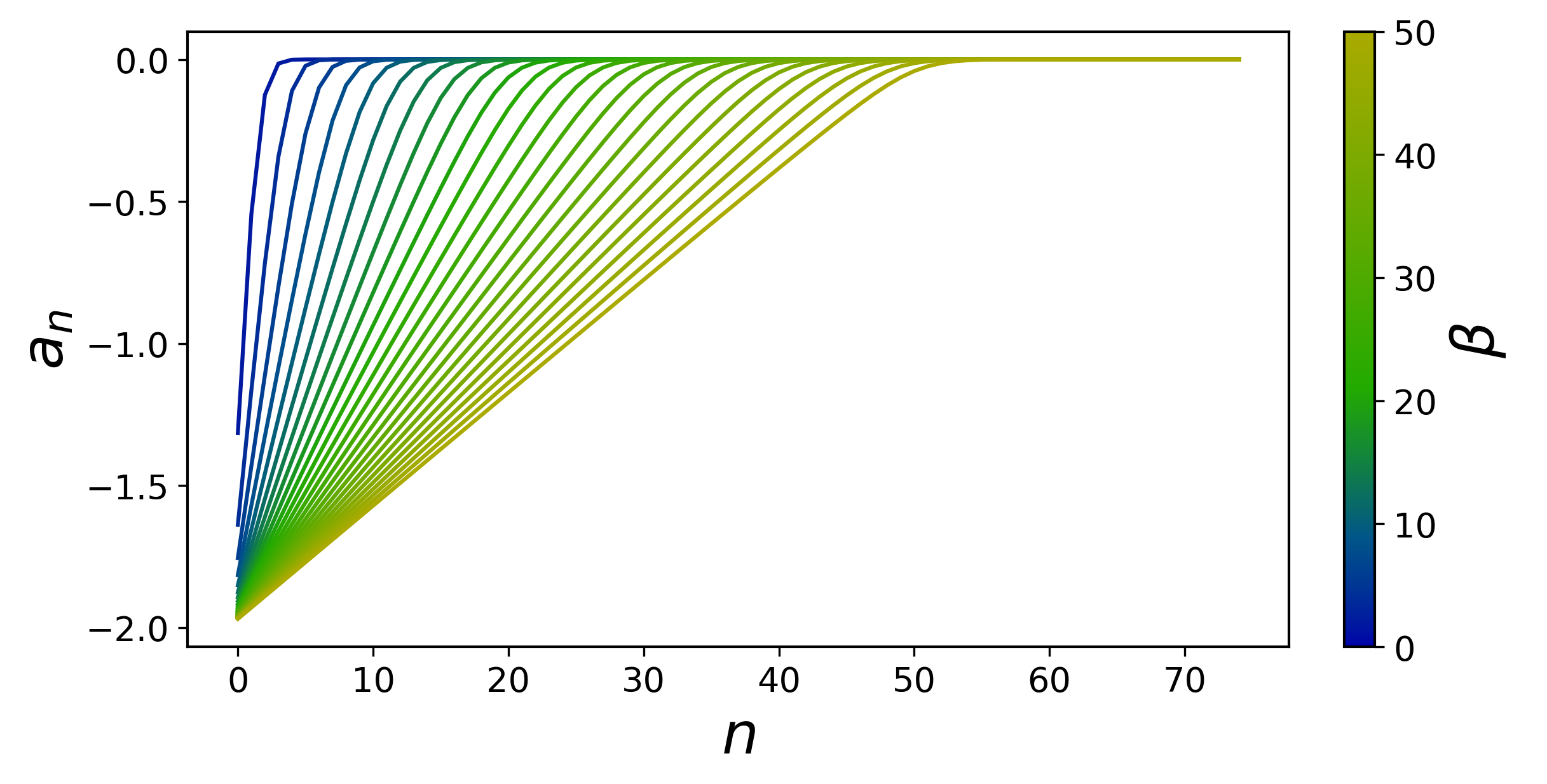}
\end{tabular}
\caption{The Lanczos coefficients $a_n$ as a function of $n$ in the large $N$ limit and for several values of $\beta$ between $2$ and $50$, for the GUE ensemble associated with the time evolution of TFD. The transition to the plateau occurs at $n\sim\mathcal{O}(1)$. The color bar indicates the value of $\beta$ for each curve.
}
\label{fig_LZansmall}
\end{figure}

\begin{figure}
\centering
\begin{tabular}{c}
  \includegraphics[width=0.98\linewidth]{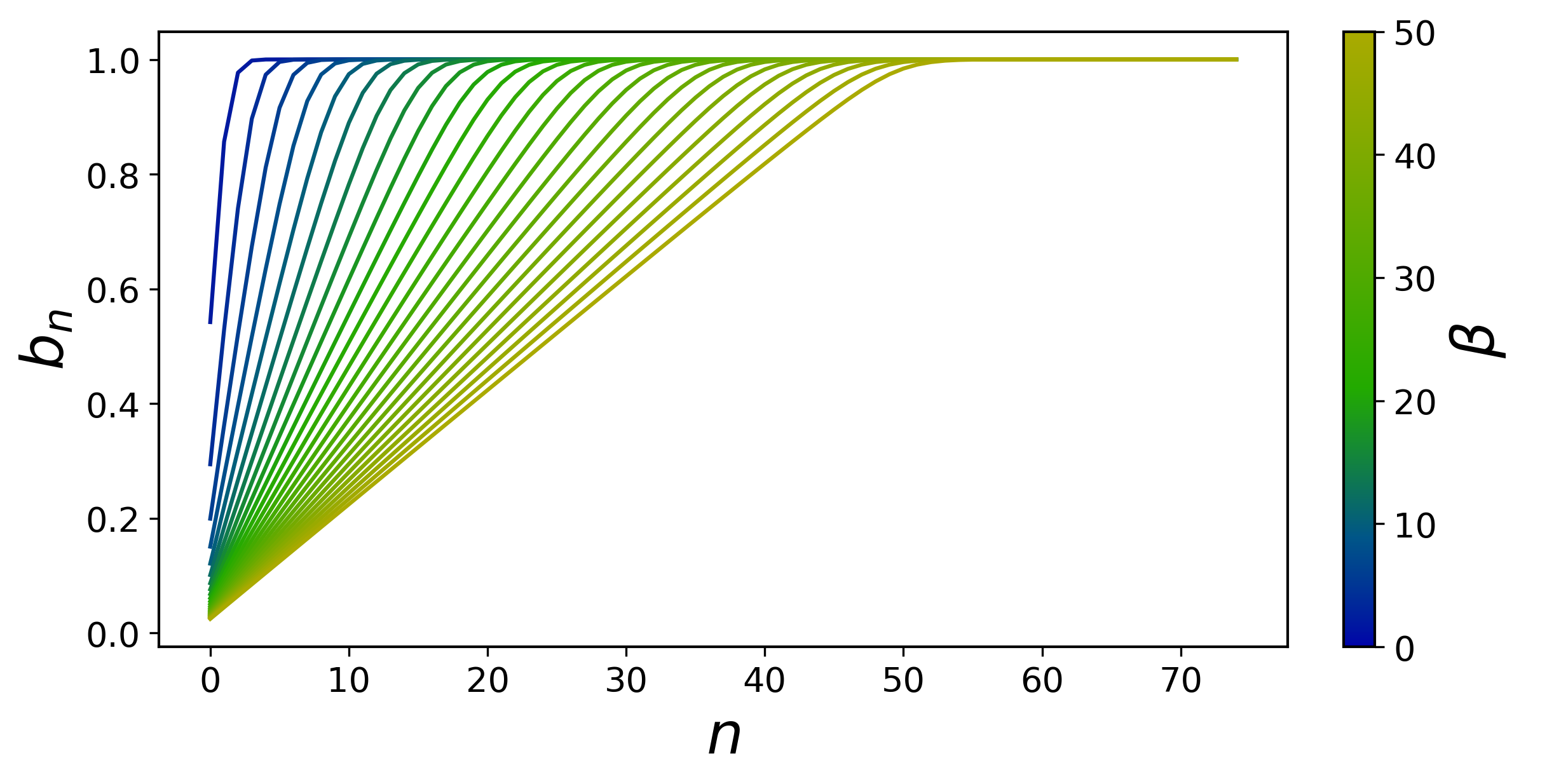}
\end{tabular}
\caption{The Lanczos coefficients $b_n$ as a function of $n$ in the large $N$ limit and for several values of $\beta$ between $2$ and $50$, for the GUE ensemble associated with the time evolution of TFD. The transition to the plateau occurs at $n\sim\mathcal{O}(1)$. The color bar indicates the value of $\beta$ for each curve.
}
\label{fig_LZbnsmall}
\end{figure}

To solve the recursion method described earlier we use
known numerically stable algorithms for computing the Hessenberg form of any given $N\times N$ Hamiltonian \cite{python-SciPy,LAPACK-Hessenberg}. As discussed in Sec.~\ref{sec:comC}, these algorithms use Householder reflections instead of the Gram-Schmidt procedure. From the Hessenberg form we can read off the Lanczos coefficients. To compute the wavefunction in the Krylov basis, we exponentiate the Hessenberg form to obtain the matrix representing unitary evolution, and apply this matrix to the initial state. This procedure directly provides the wavefunction in the Krylov basis. From the wavefunction, we can compute the probability of being in any given Krylov basis state, and hence the complexity.

In Figs.~\ref{fig_anfull} and~\ref{fig_bnfull} we plot the full set of Lanczos coefficients as a function of $n/N$, for  $N=\{1024, 1280, 1536, 1792, 2048, 2560, 3072, 3584, 4096\}$ and temperatures $\beta=\{0, 1, 2,5,10\}$. The figures show the global structure of the Lanczos coefficients as a function of $n$, which indexes the Krylov basis elements from $0$ to $N-1$, where $N$ is the dimension of the Hilbert space. As expected for a chaotic model, the Krylov basis expands the full Hilbert space, and unitary evolution explores the full Hilbert space.  Interestingly, the GUE ensemble includes all possible Hermitian matrices as Hamiltonians. Hence, our results show that most unitary theories are chaotic in this way.

Both sets of Lanczos coefficients show two behaviors as a function $n$. For the $a_n$ there is a linearly growing regime, followed by a near-plateau (Fig.~\ref{fig_anfull}). The plateau for $a_n$ is approximately at zero, up to fluctuations. The transition from the ramp in $n$ to the plateau in $n$ seems to occur at $n\ll N$ (Fig.~\ref{fig_anfull}). We numerically confirmed that the transition indeed occurs at $n$ of $\mathcal{O}(1)$ in the large $N$ limit (Fig~\ref{fig_LZansmall}).
The fact that, most of the $a_n \approx 0$ is a significant simplification for analytical methods.

For the $b_n$ we again see a sharp ramp at small values of $n$ followed by a gradual decay with a slope of $O(1/N)$ to zero as $n \to N$ (Fig.~\ref{fig_bnfull},\ref{fig_LZbnsmall}). The transition occurs at $n\sim\mathcal{O}(1)$ (Fig.~\ref{fig_LZbnsmall}), and in fact in the large $N$ limit the decay to zero is so slow that  for any fixed interval of $n$, $b_n$ is approximately constant.  The decay to zero at $n \to N$ occurs because the $b_n$ are hopping coefficients in the one-dimensional Krylov-basis chain (\ref{SchrodingerEq}).  Thus, for any finite system size, $b_n$ must vanish when we reach the edge of the chain. Similar  behavior for the $b_n$ has been found in the context of operator growth \cite{Barbon:2019wsy,Rabinovici:2020ryf,Kar:2021nbm}, except that the transition between the ramp and approximate plateau happens here at $n\sim \mathcal{O}(1)$ while for operator growth it occurs at $n\sim\mathcal{O}(\log N)$ namely at the order of the entropy.

\begin{figure}
\centering
\begin{tabular}{c}
  \includegraphics[width=0.98\linewidth]{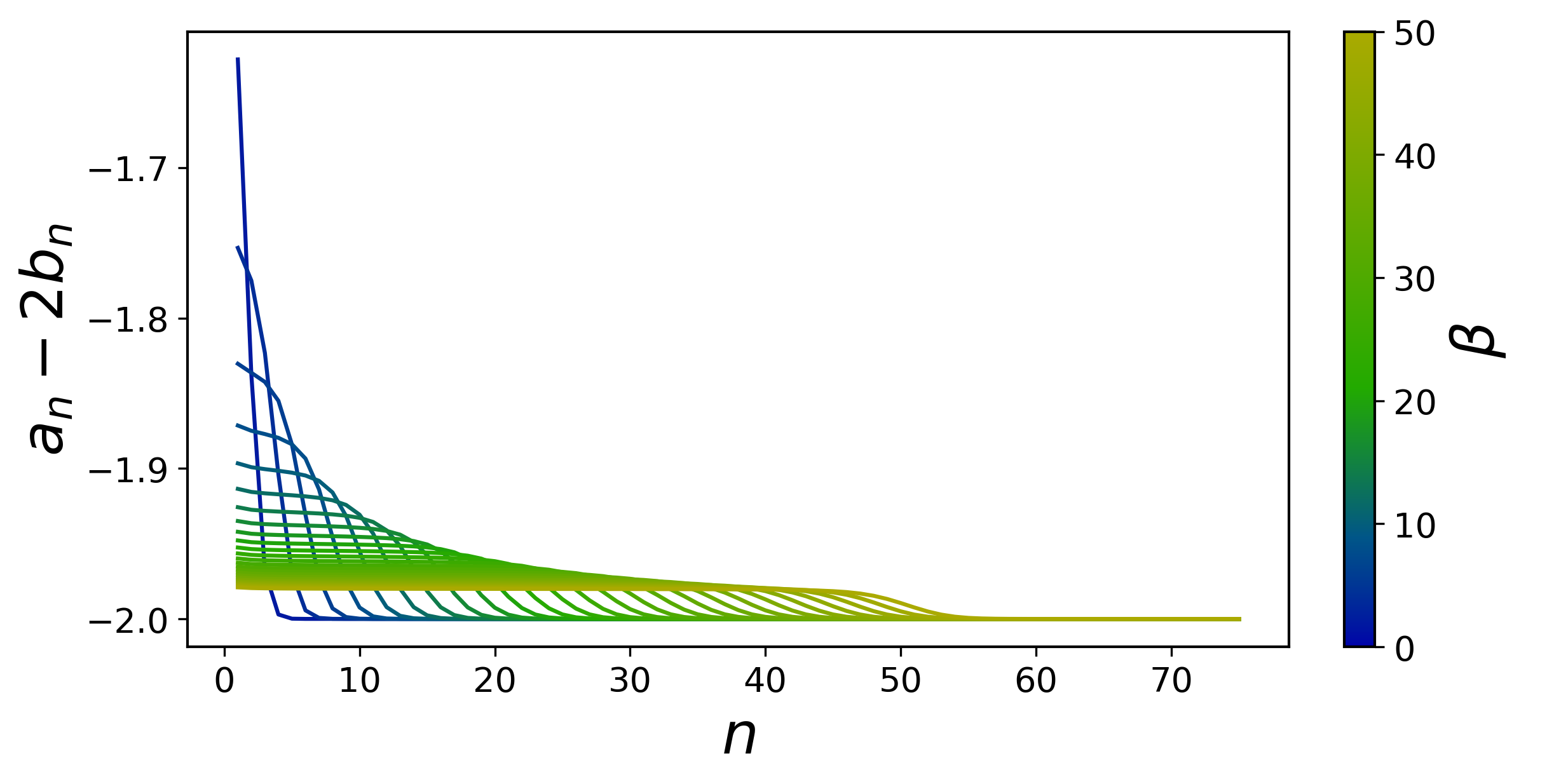}
\end{tabular}
\caption{At small $n$ ---between zero and the transition to plateau behavior--- and large enough $\beta$, $a_n-2b_n$ is approximately constant. As shown in fig \ref{fig_LZbnsmall} and \ref{fig_LZansmall}, $a_n,b_n$ grow linearly, so we find ourselves in the free limit of the harmonic oscillator~(\ref{freea}). The color bar indicates the value of $\beta$ for each curve.
}
\label{fig_LZa-2b_GUE_Ninf_smalln}
\end{figure}

To understand the behavior of spread complexity, as described in Sec.~\ref{Analytic}, we need to find the precise relation between the rate of growth of the $a_n$ and the rate of growth of the $b_n$. Fig~(\ref{fig_LZa-2b_GUE_Ninf_smalln}) shows that, in the range of small $n$ where both Lanczos coefficients grow linearly, $a_n+d=2 b_n$ with $d$ a constant,
like in the  Schwarzian theory. This relation between $a_n$ and $b_n$ manifests as the first plateau in the plot of $a_n-2b_n$ in Fig.~(\ref{fig_LZa-2b_GUE_Ninf_smalln}).  The second plateau in this figure at larger $n$ occurs because $a_n$ and $b_n$ are both changing very slowly in this regime.

Recall from (\ref{exp}) and (\ref{SchrodingerEq}) that at short times, the time-evolving state has most of its support on Krylov basis elements $|K_n\rangle$ with small $n$.  As discussed above $a_n+d=2b_n\sim n$ in this range, just as in the  free limit of the particle moving in the SL(2,R) group~(\ref{freea}). In analogy we expect that complexity grows quadratically at early times.  At later times the time evolution will acquire support on Krylov basis elements with larger $n$.  As we discussed above, in the large $N$ limit $a_n=0$ beyond some $n$ of $O(1)$, and $b_n$ is roughly constant for any fixed interval of $n$.  Using these conditions, the Schrodinger equation in the Krylov basis~(\ref{SchrodingerEq}) becomes a free wave equation in one dimension, whose solutions are plane waves moving at constant speed.  This implies that the mean position in the Krylov basis, and hence the complexity grows linearly with time. This is the same regime as the one found in \cite{Barbon:2019wsy,Rabinovici:2020ryf,Kar:2021nbm} for operator growth at large times. This regime was also found in the context of Nielsen's complexity in \cite{Magan2018}. Using random quantum circuits it has been found recently in  \cite{haferkamp2021linear}.

\begin{figure}
\centering
\begin{tabular}{c}
  \includegraphics[width=0.98\linewidth]{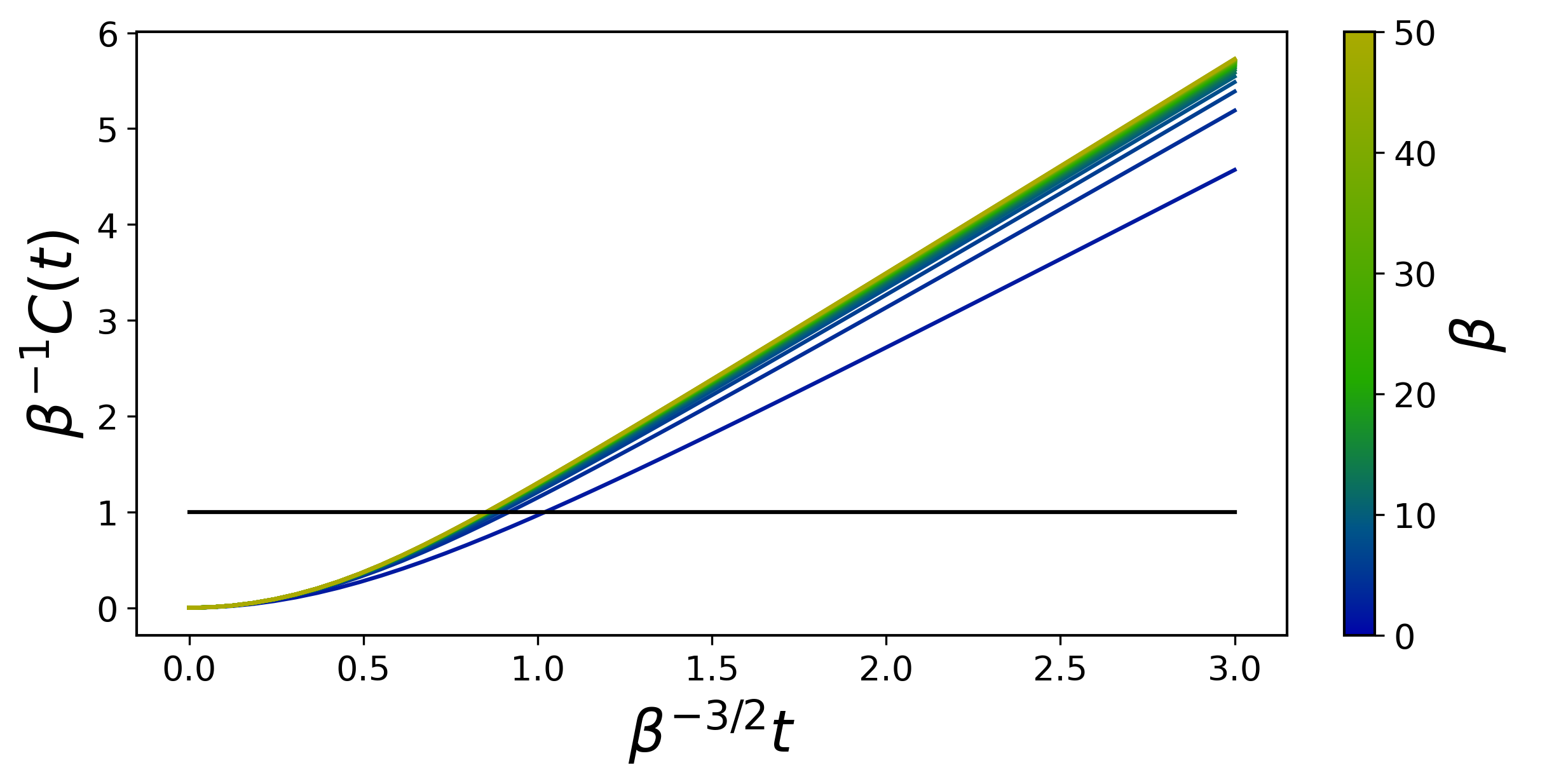}
\end{tabular}
\caption{Spread complexity of the time evolved TFD for small times, and at infinite $N$, for several values of $\beta$ between $2$ and $50$, corresponding to the GUE ensemble of random matrices. Complexity starts growing quadratically and transitions to linear growth at time of order $\beta$. The color bar indicates the value of $\beta$ for each curve.
}
\label{fig_Ninf_shortT}
\end{figure}

These regimes of complexity growth are confirmed in Fig.~\ref{fig_Ninf_shortT}, where we see a transition from initial quadratic growth to linear growth at a time of order  $\beta^{3/2}$ (recall that we are working in units where $E_0=1$). In the quadratic growth regime, we checked numerically that, just like in the Schwarzian theory, the growth rate is controlled by the variance in energy which is of order $1/\beta^2$. 

As we discussed above, although the $b_n$ are approximately constant over any finite  interval in $n$ in the large $N$ limit, over  intervals of $\mathcal{O}(N)$ they do gradually decay to zero. This is because the Lanczos algorithm must halt when we reach the dimension of the Hilbert space. This means the support of the state in the Krylov basis cannot keep growing, but it is possible for the support to narrow back again.   This means that, at large times, spread complexity should reach a maximum and then may decay or plateau.   For chaotic systems we indeed expect the maximum in the complexity to be of $\mathcal{O}(N)$ and a plateau at this order as well.

\begin{figure}
\centering
\begin{tabular}{c}
  \includegraphics[width=0.98\linewidth]{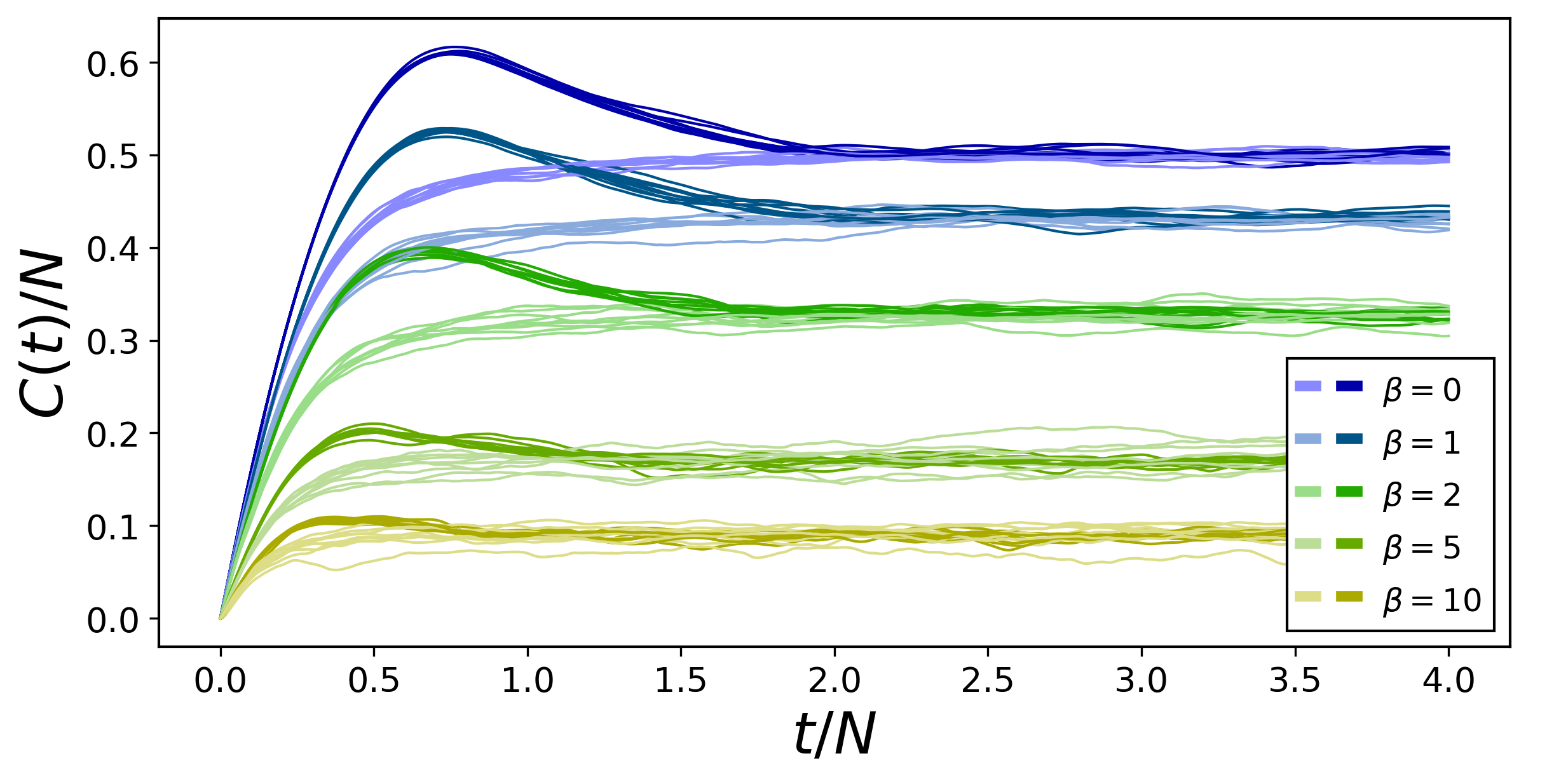}  
 \end{tabular}
\caption{Spread complexity of the time evolved TFD over an exponentially large period of time for different values of $N$ and $\beta$, as described in the main text. {\bf Dark Hues: } GUE ensemble.  Going from highest (blue) to lowest (yellow)  curves  we have $\beta=\{0, 1, 2,5,10\}$.  In each case we have plotted ensembles with $N=\{1024, 1280, 1536, 1792, 2048, 2560, 3072, 3584, 4096\}$.
Complexity grows linearly to a peak, followed by a downward slope to a plateau. 
{\bf Light Hues: } Ensemble with the same density of states as GUE, but without correlations between eigenvalues. In this case, the curves  plateau without reaching a peak followed by a downward slope.
}
\label{fig_KC_GUE_and_Spectral_plateau}
\end{figure}

\begin{figure}
\centering
\begin{tabular}{c}
  \includegraphics[width=0.98\linewidth]{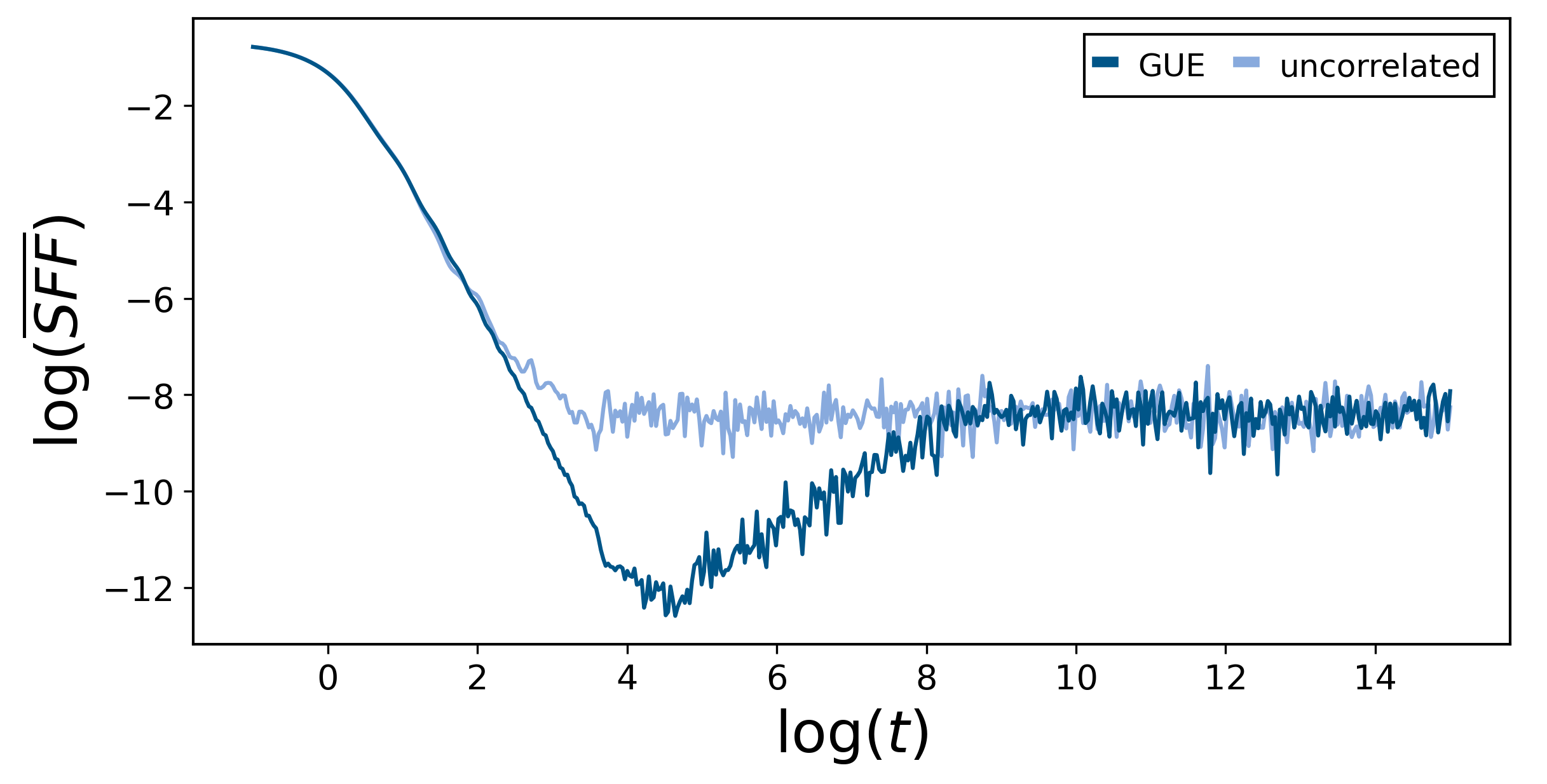}
\end{tabular}
\caption{Spectral Form Factor (survival probability of the time evolved TFD) over an exponentially large period of time for $N=4096$ and $\beta=1$, averaged over $10$ samples of the GUE ensemble. {\bf Dark blue}: The GUE esemble of random matrices displays a ramp followed by a plateau. {\bf Light blue}: 
For an ensemble with the same density of states as the GUE but with no correlations between eigenvalues, the spectral form factor displays a plateau without a ramp.
}
\label{SFFGUE}
\end{figure}

The dark hued curves in Fig.~\ref{fig_KC_GUE_and_Spectral_plateau}  show how state complexity changes in a variety of GUE ensembles until times of  order  the size of the Hilbert space ($t/N \sim O(1)$).  It is immediately clear that the complexity dynamics displays a characteristic overall structure: a linear {\it ramp} for times that are exponentially  large in the entropy, followed by a {\it peak}, and then a downward {\it slope} to saturation at an exponentially large {\it plateau}.  The onset times and heights of the peak  and plateau in the complexity are $\mathcal{O}(N)$, i.e., exponentially large in the entropy of the system.  The initial linear ramp and plateau were conjectured for chaotic systems by \cite{SusskindQC}.  We propose that the subsequent peak overshooting the plateau, followed by a downward slope are also universal characteristics of the complexity dynamics.

The dynamics of state complexity that we have displayed is related to the behavior of the spectral form factor in chaotic theories \cite{GUHR1998189,SFF1}. Recall that we showed that the spectral form factor computes the survival probability of the TFD state, namely
\be 
\textrm{SFF}=\vert \frac{Z_{\beta-it}}{Z_{\beta}}\vert^2=\vert \langle \psi_{\beta+2it}\vert \psi\rangle\vert^2\;.
\ee
This is also the time-dependent probability of the first state in the Krylov basis, i.e., the initial TFD state.  Fig.~\ref{SFFGUE} shows the spectral form factor as a function of time. The dark blue line, corresponding to the GUE ensemble, shows a downward {\it slope}, that lasts until the {\it dip} (i.e., the minimum), followed by an upward {\it ramp} and then a {\it plateau}. The plateau occurs because the system has a finite, discrete spectrum.   From our state evolution perspective, the time development of the TFD  explores the state space broadly and at late times the distribution over the Krylov basis is uniform after some coarse-graining over time. See \cite{VijayGabor} for a tractable model of why such coarse-graining is needed. The shape of the ramp is determined by the universal statistics of random matrices \cite{GUHR1998189}. In particular, it depends on the universal correlations between eigenvalues, a structure  known as spectral rigidity.

\begin{figure}
\centering
\begin{tabular}{c}
  \includegraphics[width=0.98\linewidth]{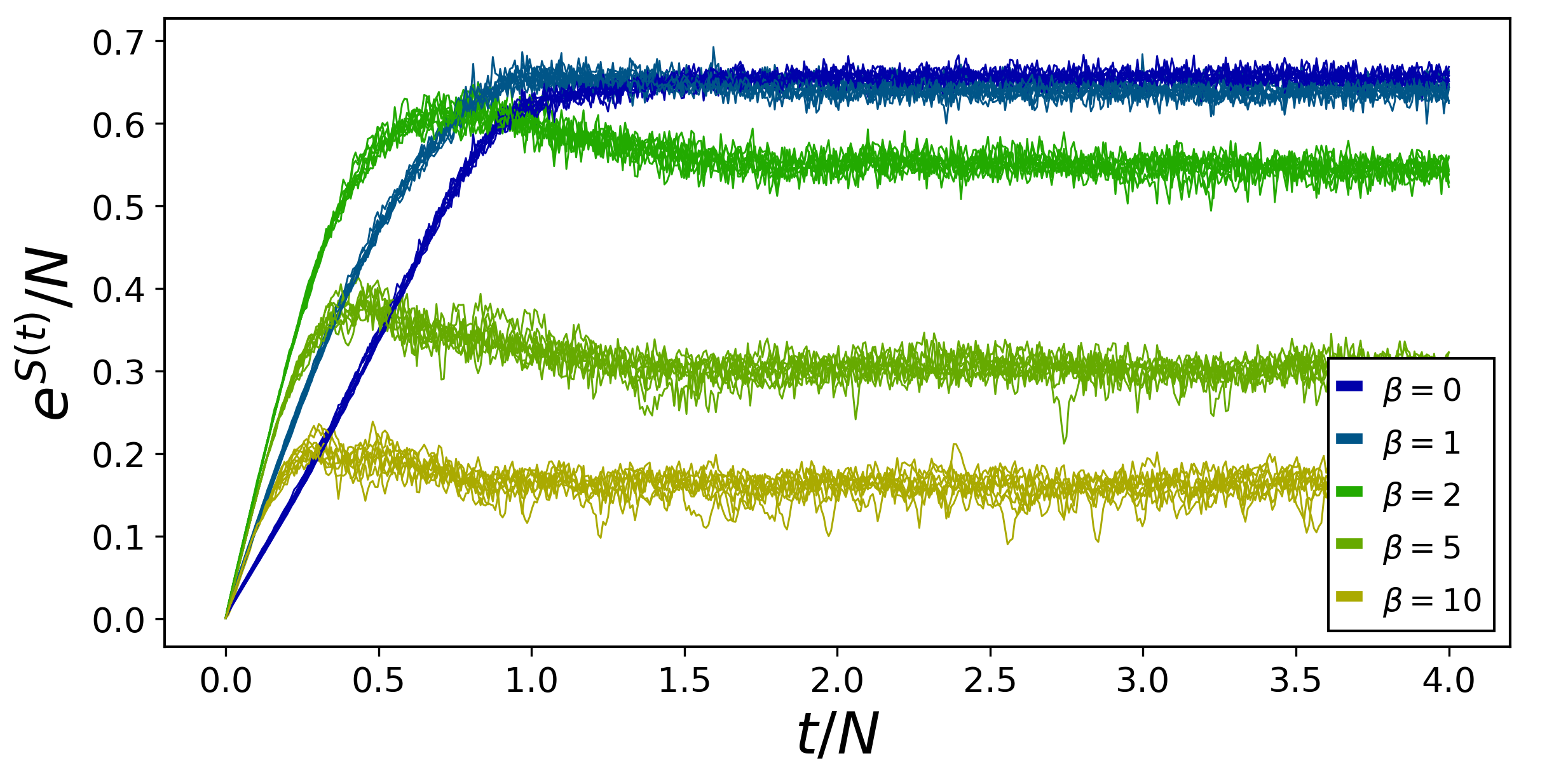}  
\end{tabular}
\caption{Entropic complexity~(\ref{ecom}) for the GUE ensemble of the time evolved TFD state over an exponentially large period of time for different $\beta=\{0, 1, 2,5,10\}$.  In each case we have plotted ensembles with $N=\{1024, 1280, 1536, 1792, 2048, 2560, 3072, 3584, 4096\}$.
}
\label{KEnt_GUE_plateau}
\end{figure}

\begin{figure}
\centering
\begin{tabular}{cc}
  \includegraphics[width=0.9\linewidth]{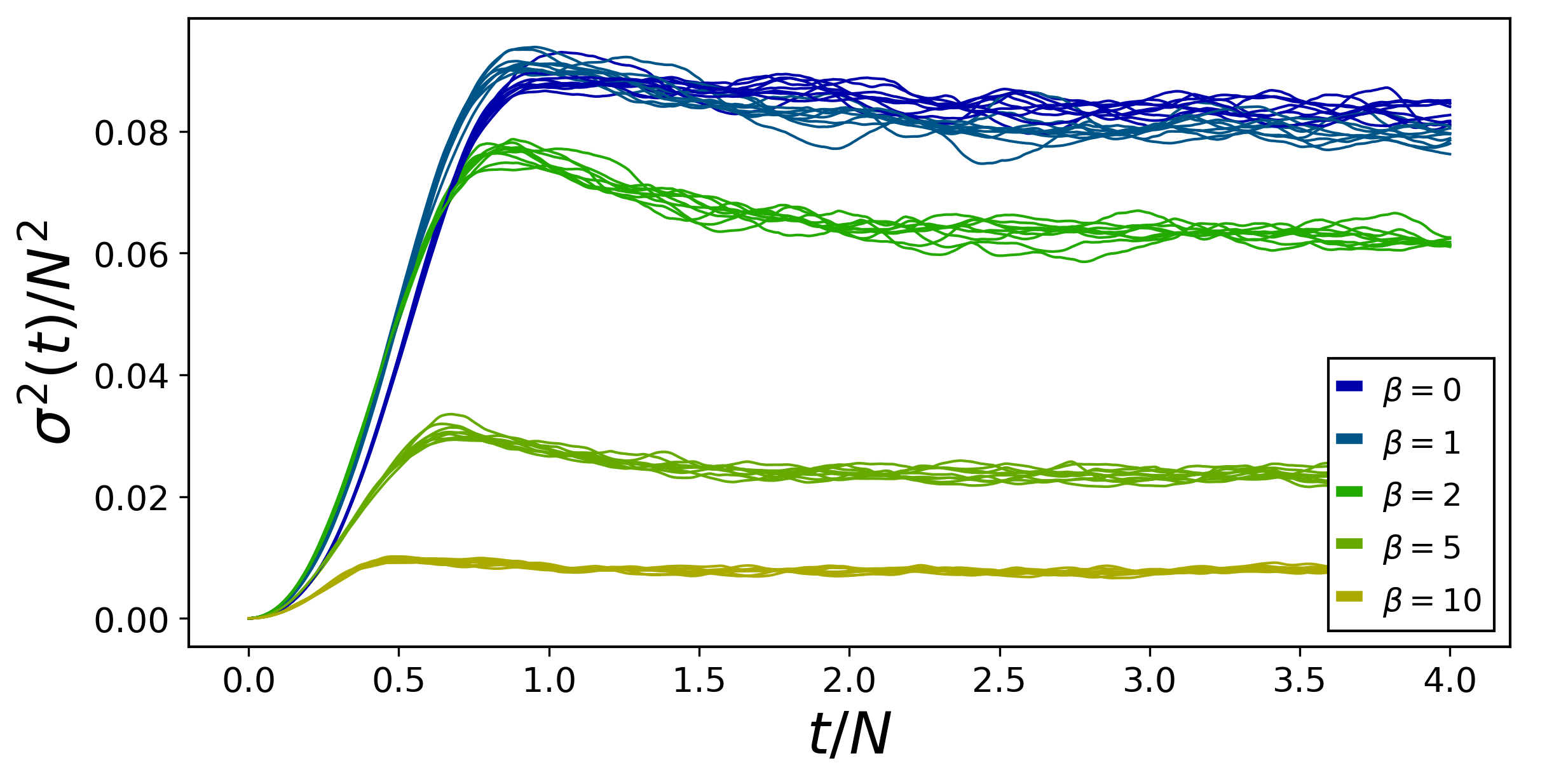}
\end{tabular}
\caption{Variance of the position in the Krylov basis for the GUE ensemble of the TFD state evolved over an exponentially large  time, for different values of $\beta=\{0, 1, 2,5,10\}$.  In each case we have plotted ensembles with $N=\{1024, 1280, 1536, 1792, 2048, 2560, 3072, 3584, 4096\}$.}
\label{KVar_GUE_plateau}
\end{figure}

We can argue that the  ramp, peak, slope and plateau of complexity are in analogy to the slope, dip, ramp and plateau of the spectral form factor. With regard to the plateau this analogy is transparent. Complexity achieves the plateau when the Krylov basis probabilities reach  approximate stationarity after some coarse-graining in time.  More precisely, as we will see below, the probabilities actually continue to fluctuate in the plateau region, but they are stationary after time averaging, or after averaging over small neighbourhood of $n$, the Krylov basis index. The spectral form factor is one of these basis state probabilities -- it is the probability of staying in the initial TFD state.  Thus it must saturate as well. With regard to the peak and downward slope in the complexity, the analogy is more subtle. As shown in \cite{SFF1},  the dip (i.e., minimum) in the spectral form factor occurs at a time of $\mathcal{O}(\sqrt{N})$, while the complexity peak occurs at a time of  $\mathcal{O}(N)$ (see  Fig.~\ref{fig_KC_GUE_and_Spectral_plateau}). Indeed, since complexity grows linearly, it can only achieve a size of $\mathcal{O}(N)$, as required for the state to acquire support everywhere on the Hilbert space, at a time of $\mathcal{O}(N)$. Thus, the time of the dip in the spectral form factor and the time of the peak in the complexity do not scale in the same way with $N$.  Of course, since the complexity is a sum over all the Krylov basis probabilities, only one of which is directly related to the spectral form factor, the time scales in these quantities need not be same.

Nevertheless, we can show that the downward complexity slope after the peak likely arises from eigenvalue correlations, just like the ramp in spectral form factor after the dip.
To show this, we 
construct an ensemble with the same density of states as the GUE, but without spectral correlations. To do so we can draw eigenvalues at random from the Wigner semicircle distribution, rather than from a specific random Hamiltonian. 
Alternatively, we can take sufficiently many instances of random Hamiltonians with their associated spectra, and then construct another spectrum by randomly sampling eigenvalues from the different Hamiltonians \cite{astreicher}. For the present discussion it is sufficient, and simpler, to draw eigenvalues from the Wigner semicircle distribution. Complexity growth of the TFD state assuming such a spectrum is plotted in light hues in Fig.~\ref{fig_KC_GUE_and_Spectral_plateau}. We see the peak and downward slope disappear. A related effect can be seen in the spectral form factor, where the dip and ramp disappear (light blue line in Fig.~\ref{SFFGUE}). This is a strong hint that the  downward complexity slope after the peak is controlled by spectral correlations. Further evidence arises from the GOE and GSE ensembles considered below.

\begin{figure}
\centering
\begin{tabular}{c}
  \includegraphics[width=0.98\linewidth]{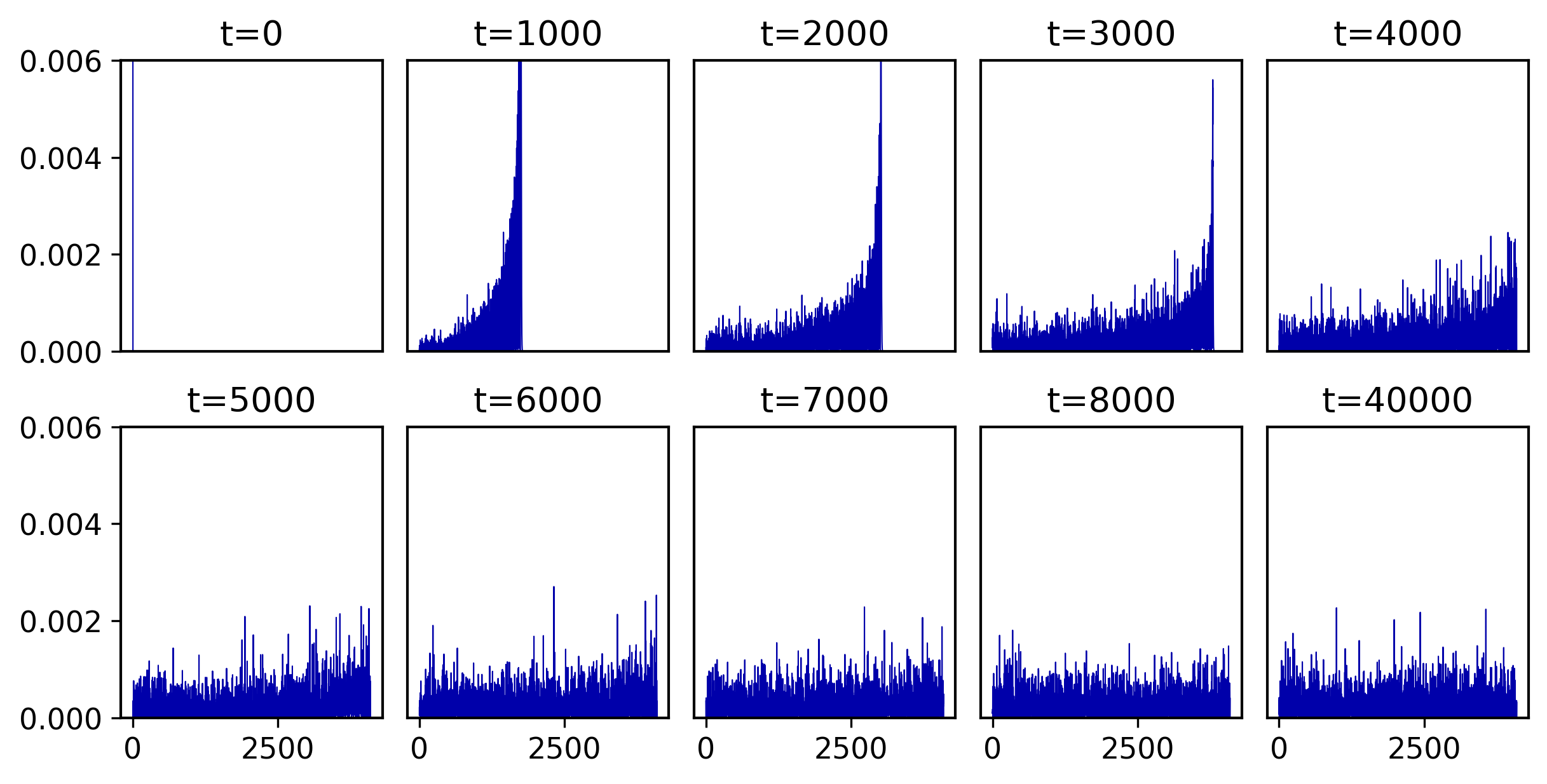} 
 \end{tabular}
\caption{Snapshots of the probability distribution in the Krylov basis of the time evolved TFD for a range of times as specified above each panel. This plot corresponds to $\beta=0$, $N=4096$ and the GUE ensemble. The horizontal axis shows the index of the Krylov basis elements from $1$ to $4096$ and the y-axis shows the probability that the initial state has evolved so that it is found in the given basis state.  At $t=0$ the y-axis runs from 0 to 1 and all the probability weight is on the initial state.  At $t=40000$ the mean probability is $1/4096$. Thus, we arranged the scale of each panel to better show the spread of the wavefunction over the Krylov basis.  
}
\label{snap0}
\end{figure}

\begin{figure}
\centering
\begin{tabular}{c}
  \includegraphics[width=0.98\linewidth]{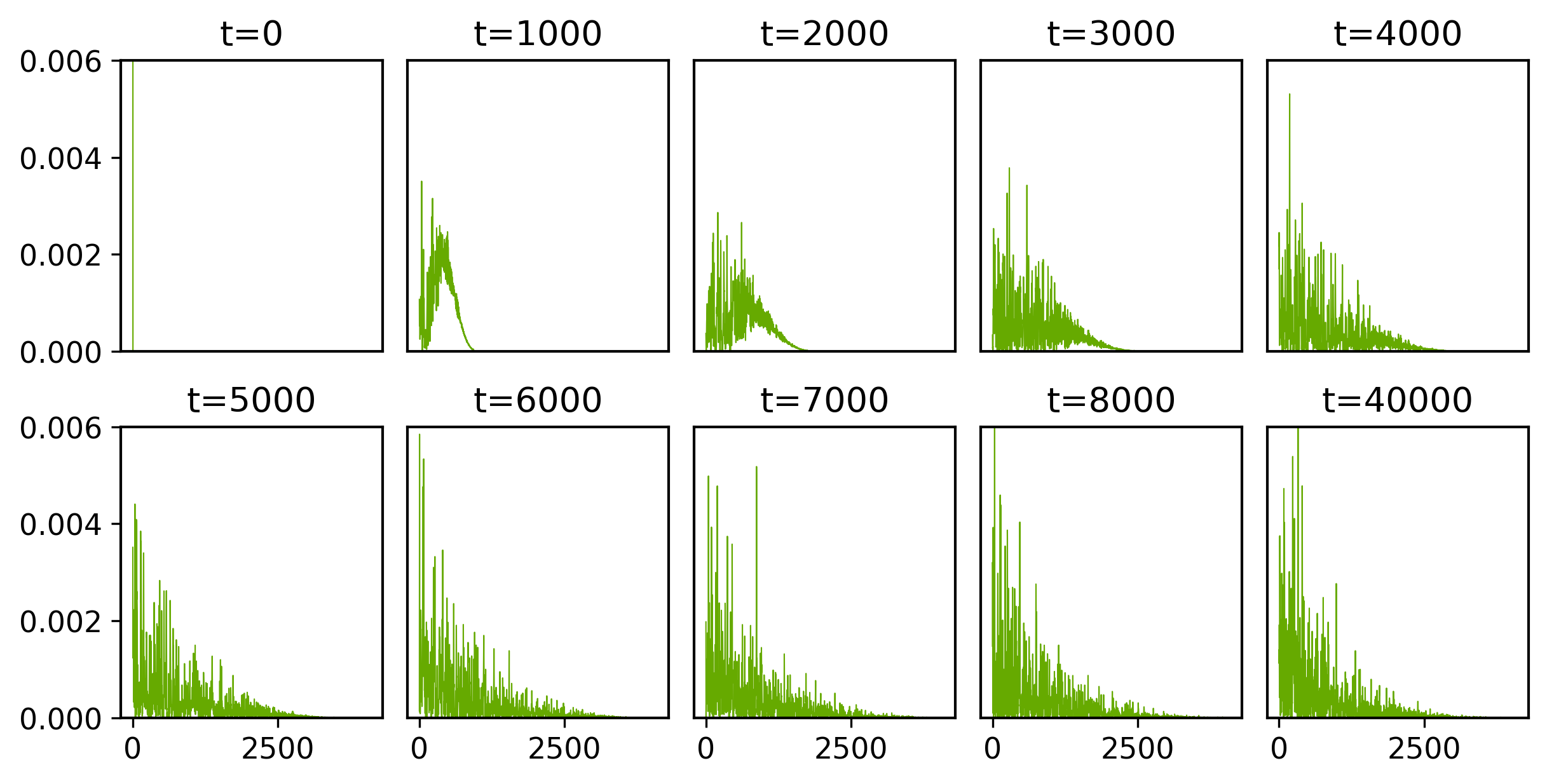}  
\end{tabular}
\caption{Snapshots of the probability distribution in the Krylov basis of the time evolved TFD for a range of times as specified above each panel. This plot corresponds to $\beta=5$, $N=4096$ and the GUE ensemble.  The horizontal axis shows the index of the Krylov basis elements and the y-axis shows the probability that the initial state has evolved so that it is found in the given basis state. The y-axis scales differ in each panel (see caption of Fig.~\ref{snap0} for an explanation of this choice). 
}
\label{snap5}
\end{figure}

We can also characterize the spread of the wavefunction across the Krylov basis in terms of the  entropic definition of complexity~(\ref{ecom2}) or the variance of the distribution of probabilities of the basis states. These quantities are displayed in Figs.~\ref{KEnt_GUE_plateau} and~\ref{KVar_GUE_plateau} and also show a ramp, a peak,  slope, and plateau.

It is also illuminating to examine the explicit form of the wavefunction in the Krlov basis at different moments of time. Figs.~\ref{snap0} and~(\ref{snap5}) show the spread of wavefunction over the Krylov basis for $\beta=0,5$ for a range of times from $t=0$ until late times when the complexity has plateaued. At $t=0$ the wavefunction is localized on the initial TFD state which is also the first Krylov basis element.  The dynamics then looks like a probability shockwave that starts on the initial state and propagates outward to higher basis elements, leaving a tail of probability behind. For high temperatures ($\beta \to 0$), the probability is initially concentrated at the shockwave front, while for intermediate and low temperatures, the probability distribution over the Krylov basis is more concentrated in the middle of the distribution.  But in both cases, when the shockwave reaches the last Krylov basis vector, it is far from being stationary. The wave bounces back and this gives rise to the downward slope after the peak in state complexity.  In the entropic definition of complexity, there is also a downward slope after the bounce of the shockwave (Fig.~\ref{KEnt_GUE_plateau}) for most temperatures.  However, at infinite temperature the probability is so concentrated at the shockwave front that the distribution actually continues to spread after bouncing from the edge of the Krylov chain so that the entropic complexity does not show a peak and download slope in this limit (dark blue line in Fig.~\ref{KEnt_GUE_plateau}).

\begin{figure}
\centering
\begin{tabular}{c}
  \includegraphics[width=0.98\linewidth]{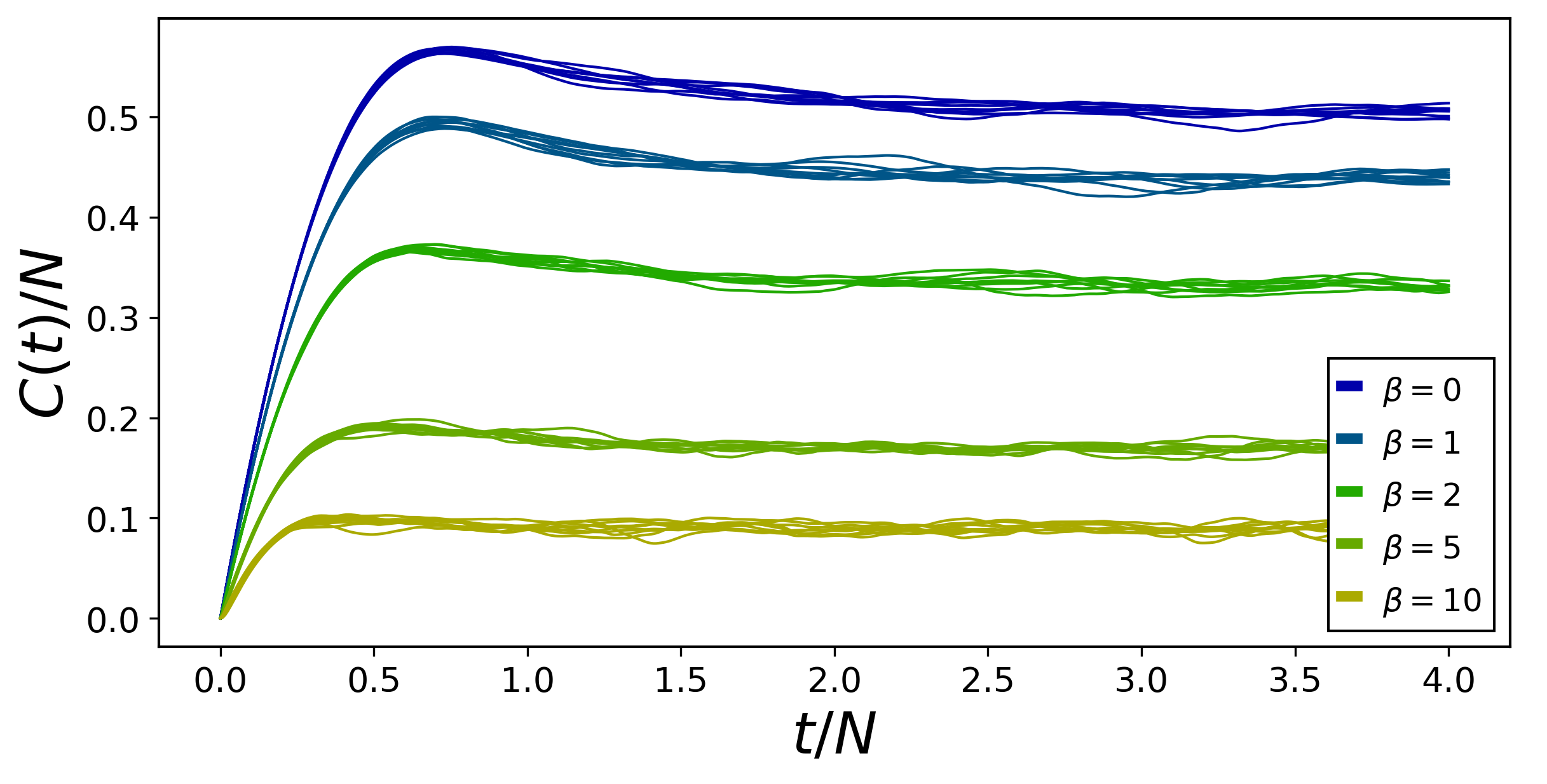}
\end{tabular}
\caption{Spread complexity of the time evolved TFD state in the GOE ensemble over exponentially large time, for different values of $\beta=\{0, 1, 2,5,10\}$.  In each case we have plotted ensembles with $N=\{1024, 1280, 1536, 1792, 2048, 2560, 3072, 3584, 4096\}$. Notice that after rising to a peak, the complexity decays smoothly to the plateau value.
}
\label{GOENc}
\end{figure}

\begin{figure}
\centering
\begin{tabular}{c}
  \includegraphics[width=0.98\linewidth]{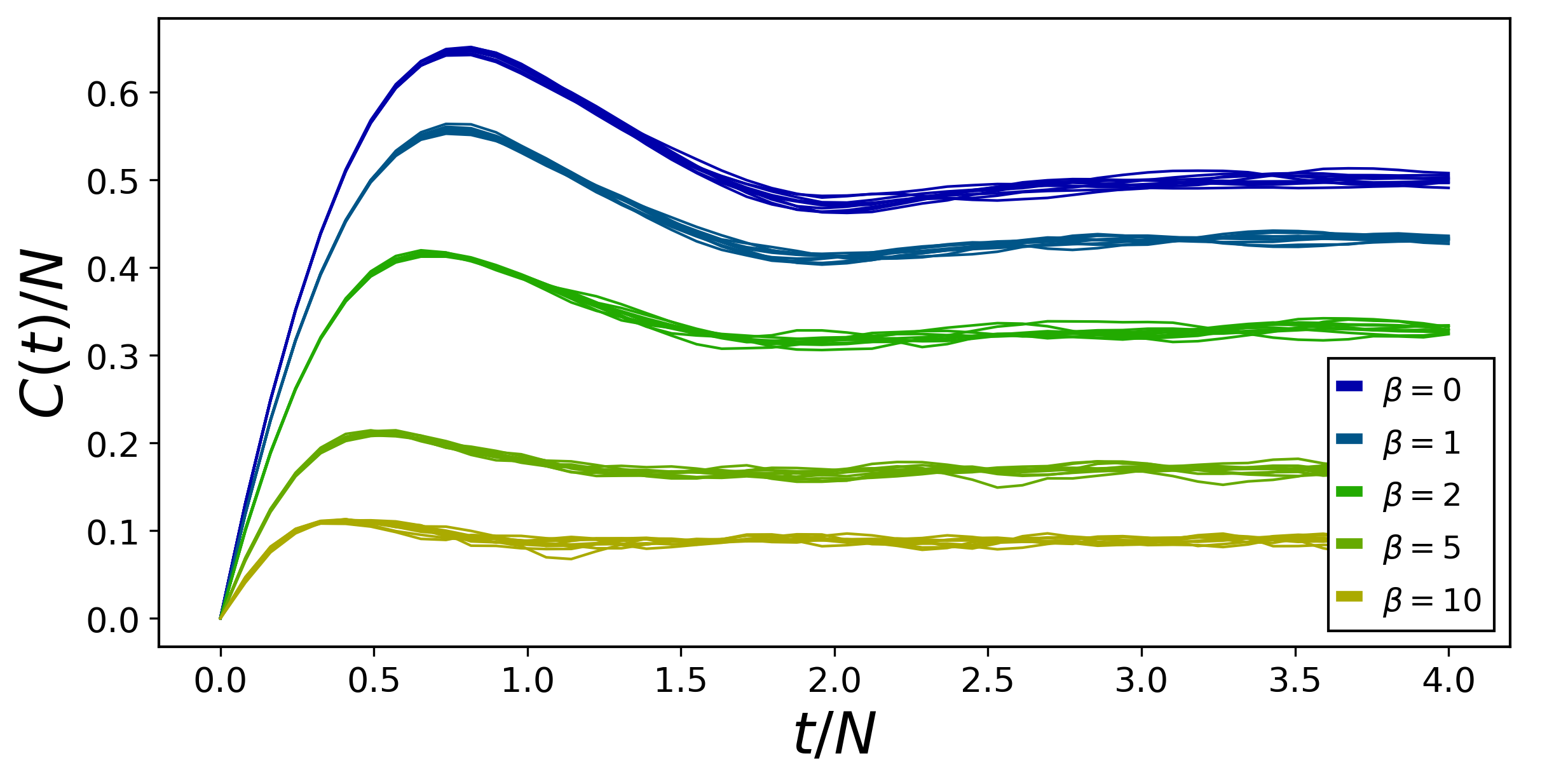}\end{tabular}
\caption{Spread complexity of the time evolved TFD state in the GSE ensemble over exponentially large time, for different values of $\beta=\{0, 1, 2,5,10\}$.  In each case we have plotted ensembles with $N=\{1024, 1280, 1536, 1792, 2048, 2560, 3072, 3584, 4096\}$. Notice that after rising to a peak, the complexity decays smoothly to the plateau value.
}
\label{GSENc}
\end{figure}

We can repeat our computations for the GOE ensemble, defined as an ensemble of real symmetric $N\times N$ matrices $H$ with Gaussian measure
\be 
\frac{1}{Z_{\textrm{GOE}(N)}}e^{-\frac{N}{4} \textrm{Tr}(H^2)}
\;,\ee
and the GSE ensemble, defined as an ensemble of $N\times N$ Hermitian quaternionic matrices with Gaussian measure
\be 
\frac{1}{Z_{\textrm{GSE}(N)}}e^{-N \textrm{Tr}(H^2)}\;.
\ee
The details of the computation are the same as for the GUE ensemble. As reviewed above, these ensembles mainly differ in the specific universal correlation functions between nearby and far away energy eigenvalues. In fact, as described in \cite{GUHR1998189}, spectral rigidity of the matrix ensembles, related to the correlations of far away energy eigenvalues, controls the ramp in the spectral form factor. The shape of this ramp and particularly the way it transitions to the plateau strongly depend on the particular matrix ensemble (see Fig.~$10$ in \cite{GUHR1998189}). In particular the transition to the plateau is sharp for GUE, smooth for GOE, and displays a kink for the GSE. This was also verified for SYK models recently \cite{SFF1}. 

We thus might expect that the universality classes of matrix models will also differ in the dynamics of complexity.  Figs.~\ref{GOENc} and~\ref{GSENc}  plot the quantum state complexity of the time evolved TFD over an exponentially large period of time for the GOE and GSE ensembles, and the same values of $N$ and $\beta$ as those used for the GUE ensemble.  The three ensembles clearly differ, in parallel to the differences that appear in the ramp structure of the spectral form factor in these three cases: the transition to the plateau is sharper for GUE, smoother for GOE, and displays a kink for the GSE.  In particular, following the behavior of the spectral form factor, and  controlled by spectral rigidity, the wavefunction of the TFD state in the Gaussian Symplectic Ensemble, and hence the complexity, bounces twice before reaching approximate stationarity.  This suggests that the complexity peak and slope originate in spectral rigidity in the random matrix model.

\subsection{SYK Model}

\begin{figure}
\centering
\begin{tabular}{c}
  \includegraphics[width=0.98\linewidth]{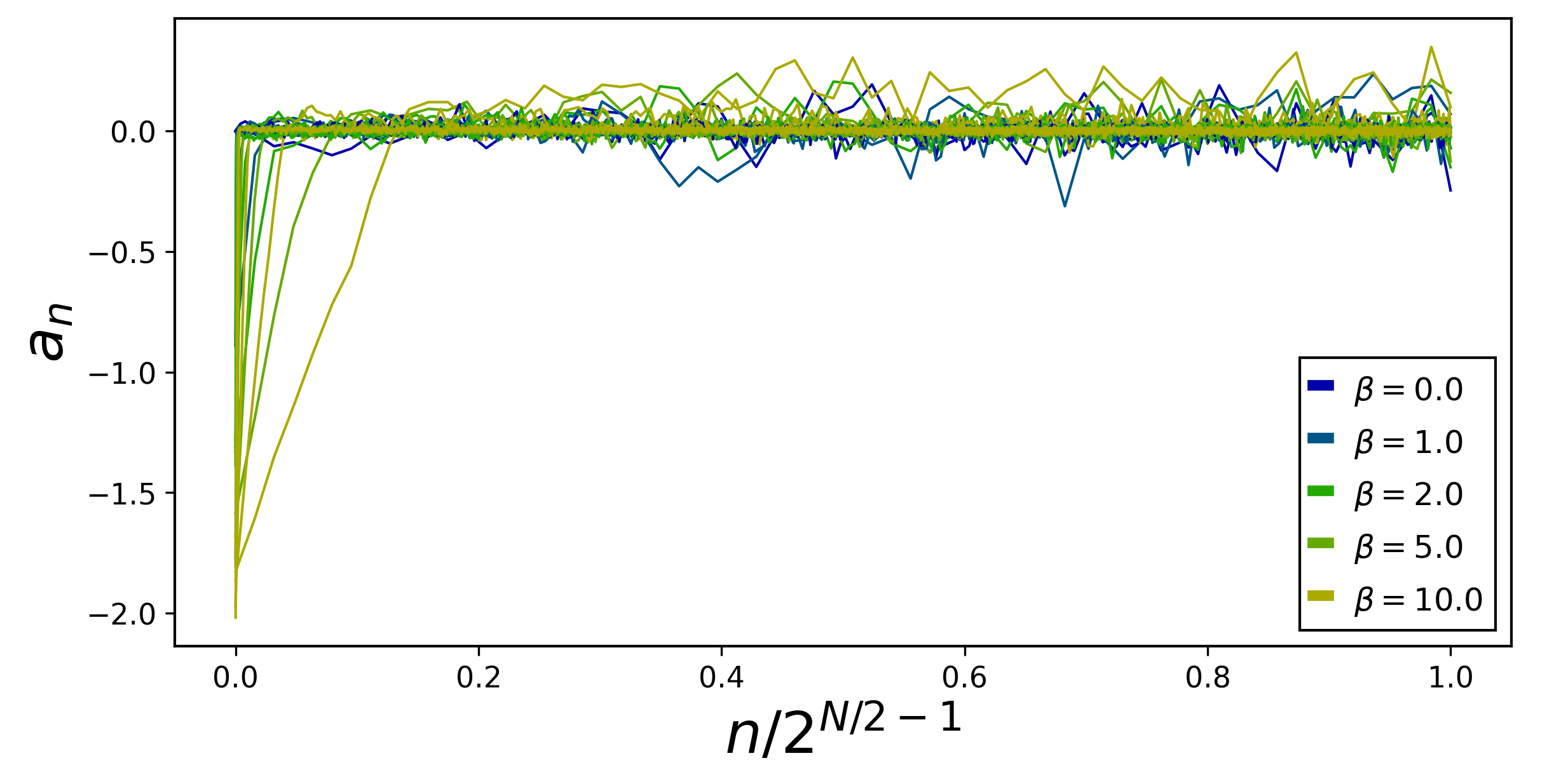} 
 \end{tabular}
\caption{Full set of Lanczos coefficients $a_n$ as a function of $n/N$ for time evolution of the TFD state in the SYK model, and for different values of $N=\{14,18,22,26\}$ and temperature $\beta=\{0, 1, 2,5,10\}$. Values of $N$ within the GUE universality class. Transition between regimes analyzed in Fig.~(\ref{fig_LZan_SYK_smalln}).
}
\label{fig_LZan_SYK_full}
\end{figure}

\begin{figure}
\centering
\begin{tabular}{c}
  \includegraphics[width=0.9\linewidth]{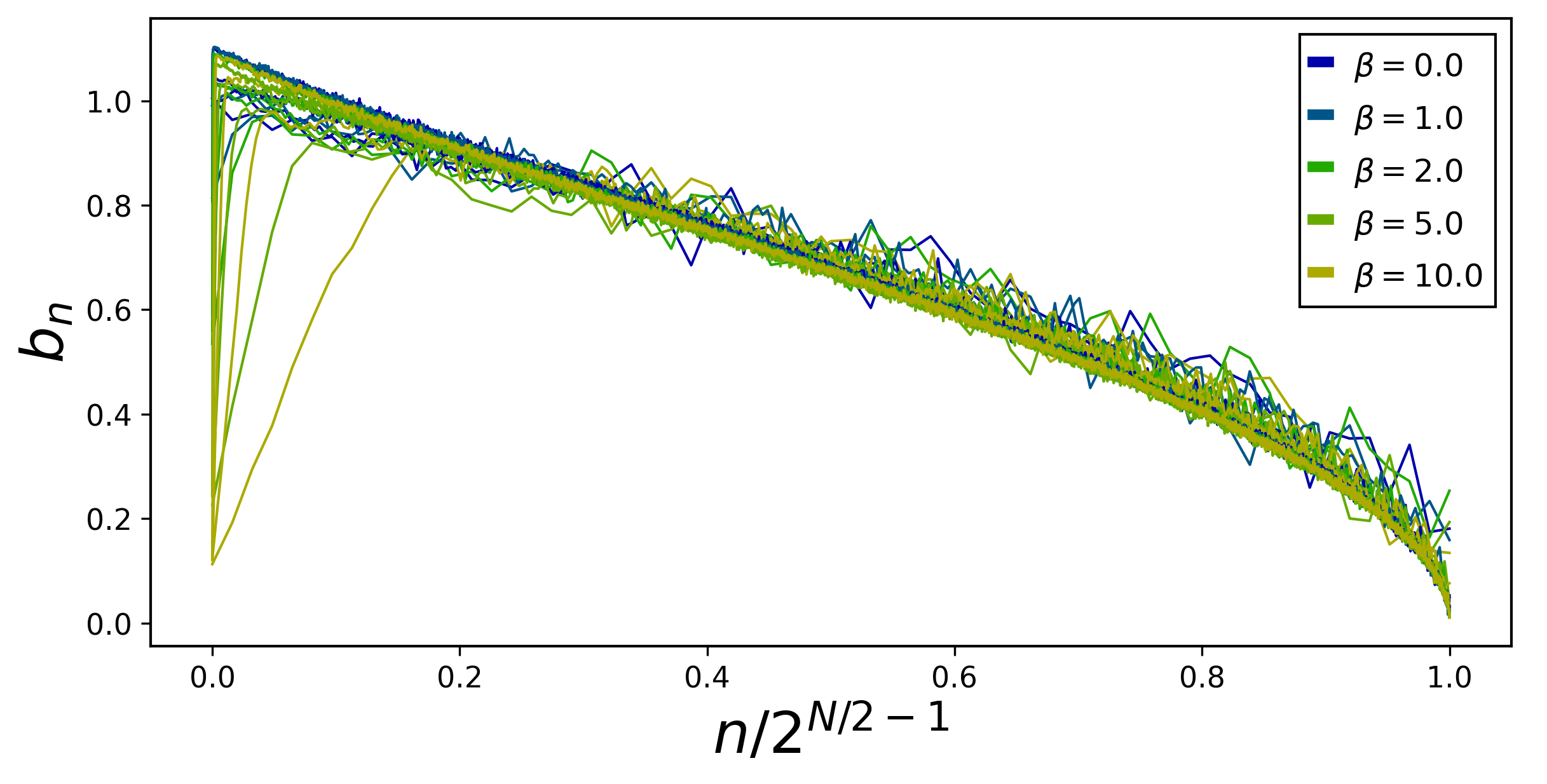}
\end{tabular}
\caption{Full set of Lanczos coefficients $b_n$ as a function of $n/N$ for time evolution of the TFD state in the SYK model, and for different values of $N=\{14,18,22,26\}$ and temperature $\beta=\{0, 1, 2,5,10\}$. Values of $N$ within the GUE universality class. Transition between regimes analyzed in Fig.~(\ref{fig_LZbn_SYK_smalln}).
}
\label{fig_LZbn_SYK_full}
\end{figure}

The conjectures of \cite{Dyson,Bohigas,GUHR1998189,RM1,RM2,RM3} tell us that random matrices are good approximations for the structure of the fine grained spectrum of chaotic systems. As argued above, the energy-time uncertainty relation then tells us that these approximations will work well for describing dynamical processes at large time scales.  But the detailed density of states in specific models can differ strongly from the universal statistics of random matrices, and the differences manifest themselves in  different  thermalization/relaxation processes at small times. We might then worry  the quantum state complexity that we have defined will differ dramatically for specific chaotic Hamiltonians as compared to the random matrix models.

To examine this possibility we considered the SYK model, which is simultaneously an interesting many-body quantum system and an accurate model of the dynamics of 2d quantum black holes \cite{kitaev,sachdev,sachdev2,Polchinski:2016xgd,Maldacena:2016hyu,Kitaev:2017awl,GABORREVIEW}. The SYK model is a model of $N$ Majorana fermions $\psi_i$, $i=1,\cdots,N$. We normalize them by
\be 
\left\lbrace \psi_i,\psi_j  \right\rbrace = \psi_i\psi_j+\psi_j\psi_i=2\delta_{ij}\;.
\ee
The SYK model is defined by an ensemble of Hamiltonians 
\be 
H=-\frac{1}{\sqrt{\binom{N}{4}}}\sum\limits_{1\leq i_1<i_2<i_3<i_4\leq N}\,J_{i_1i_2i_3i_4}\,\psi_{i_{1}}\psi_{i_{2}}\psi_{i_{3}}\psi_{i_{4}}\;,
\ee
where the couplings $J$ are real,  independently distributed, Gaussian random variables with zero mean and unit variance. These conventions were used in \cite{feng2018spectrum}. Notice that the dimension of the Hilbert space here is $2^{N/2}$. Below we will normalize physical quantities by this dimension.

We  now follow the same steps as before. We draw instances of the Hamiltonian from the ensemble and construct the TFD state. Each instance gives a particular $2^{N/2}\times 2^{N/2}$ matrix, and we apply the algorithm for computing its Hessenberg form \cite{python-SciPy}\cite{LAPACK-Hessenberg}, from which we read off the Lanczos coefficients. We then exponentiate the Hessenberg form, apply it to the initial state, and compute the wavefunction in the Krylov basis. From the wavefunction, we then compute the probabilities of the various Krylov basis elements, and hence the  complexity.

\begin{figure}
\centering
\begin{tabular}{c}
  \includegraphics[width=0.98\linewidth]{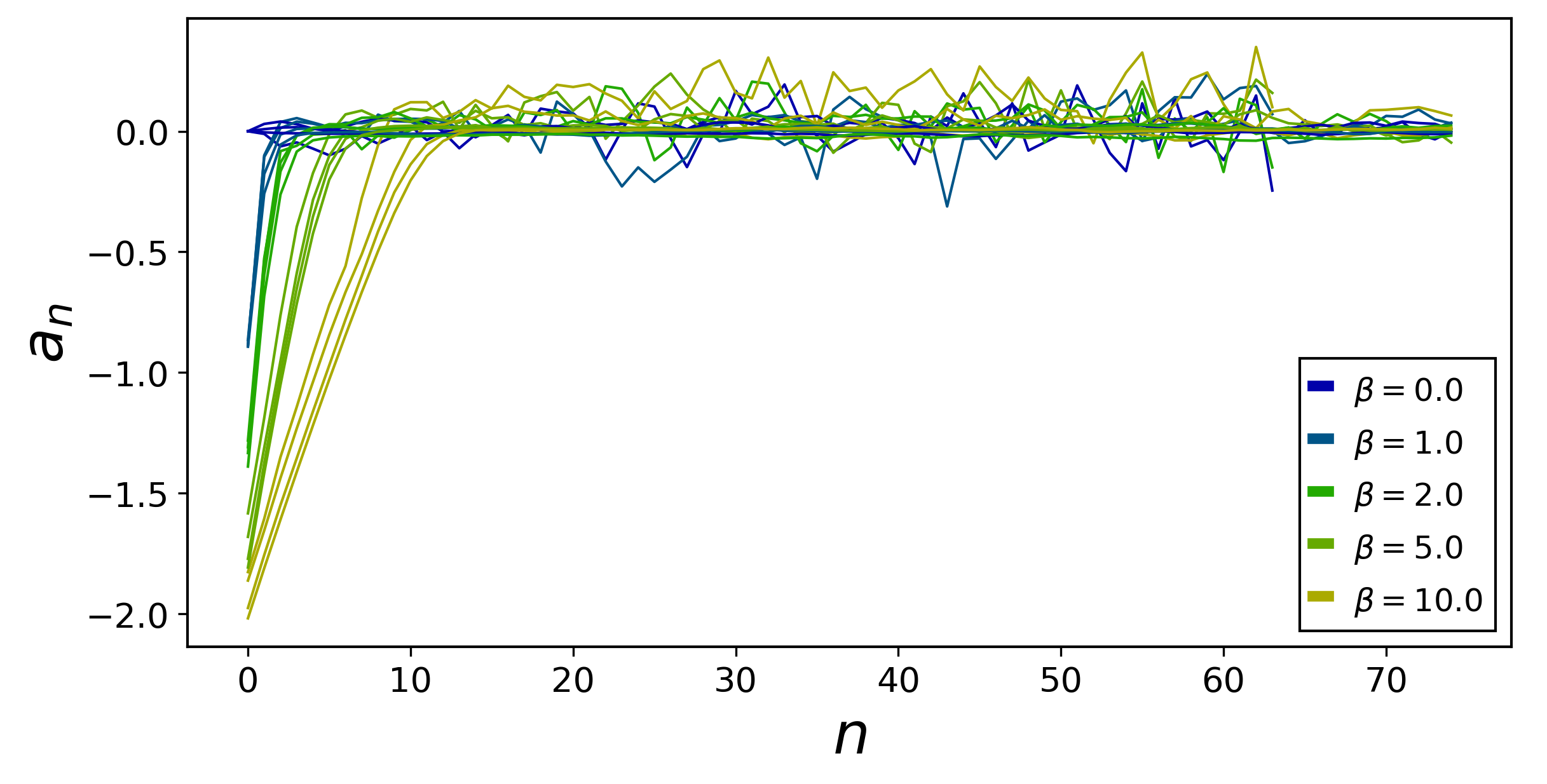}
\end{tabular}
\caption{Lanczos coefficients $a_n$ for time evolution of the TFD state in the SYK model, shown as a function of small values of $n$,  and for different values of $N=\{14,18,22,26\}$ and temperature $\beta=\{0, 1, 2,5,10\}$. Values of $N$ within the GUE universality class. The transition to the plateau occurs at $n\sim\mathcal{O}(1)$.
}
\label{fig_LZan_SYK_smalln}
\end{figure}

\begin{figure}
\centering
\begin{tabular}{cc}
   \includegraphics[width=0.98\linewidth]{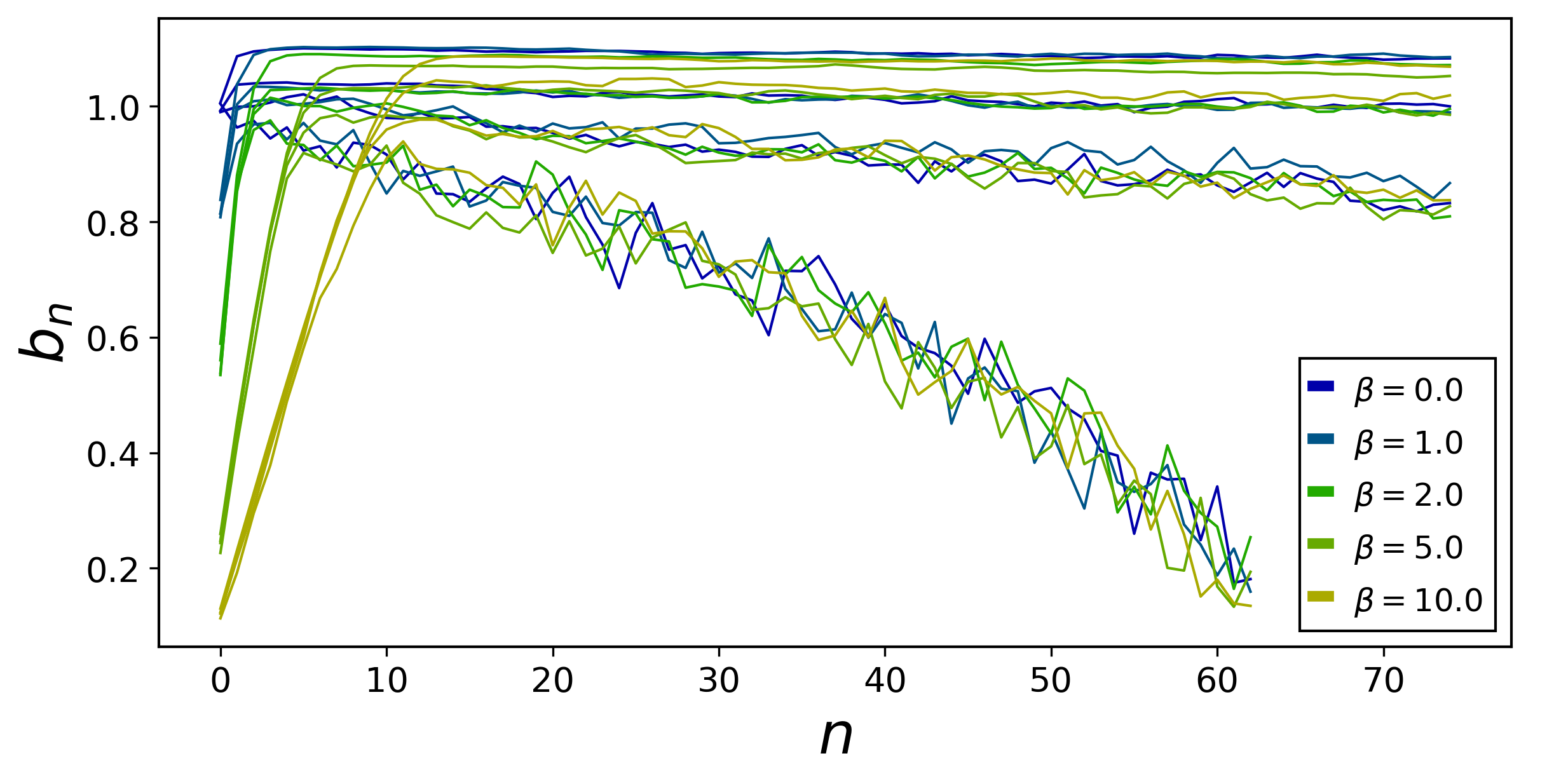}\end{tabular}
\caption{Lanczos coefficients $b_n$ for time evolution of the TFD state in the SYK model, shown as a function of small values of $n$, and for different values of $N=\{14,18,22,26\}$ and temperature $\beta=\{0, 1, 2,5,10\}$. Values of $N$ within the GUE universality class. The transition to the plateau occurs at $n\sim\mathcal{O}(1)$.
}
\label{fig_LZbn_SYK_smalln}
\end{figure}

\begin{figure}
\centering
\begin{tabular}{cc}
    \includegraphics[width=0.9\linewidth]{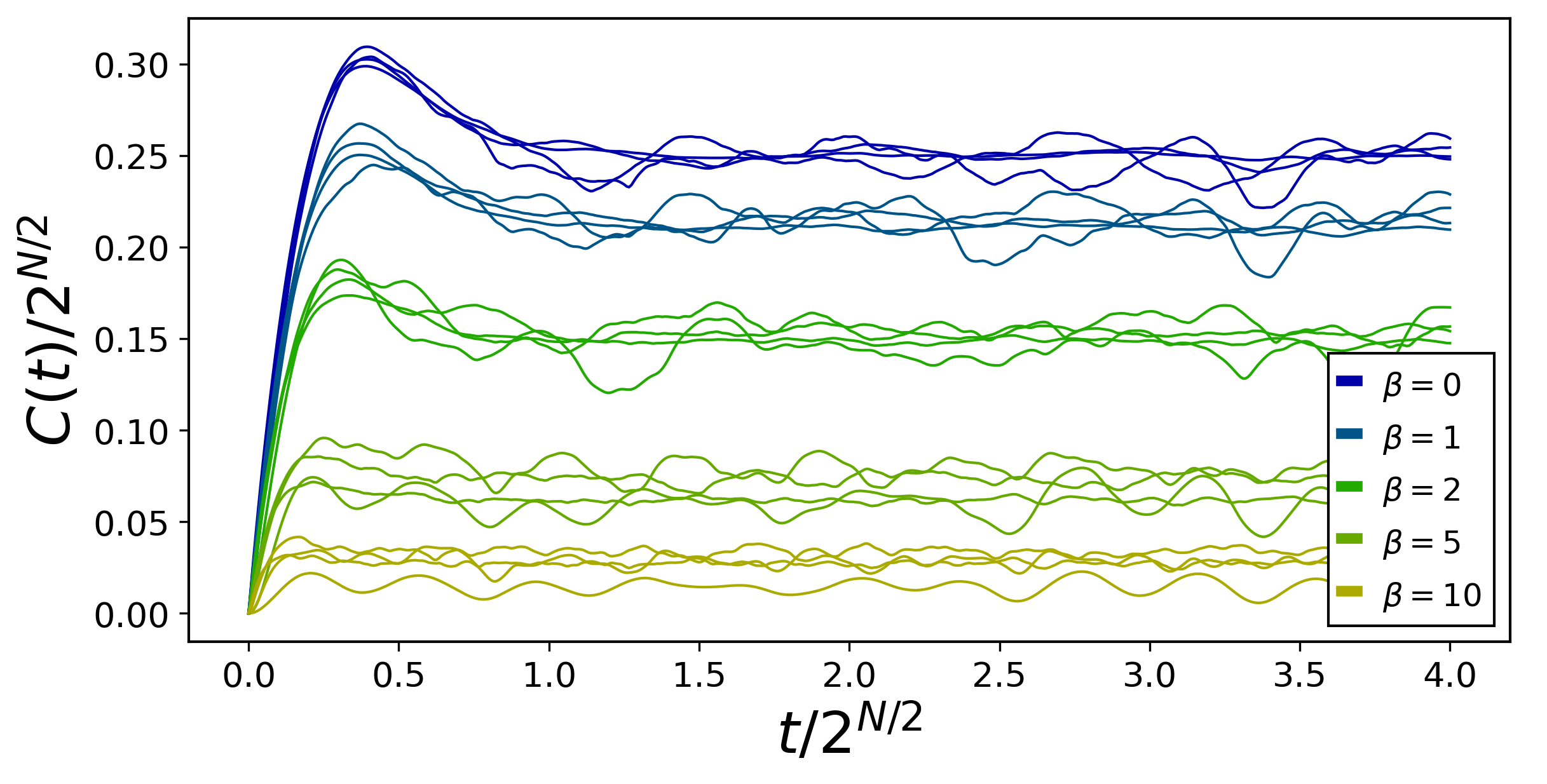}\end{tabular}
\caption{Spread complexity of the time evolved TFD in the SYK model over exponentially large period of time and for different values of $N=\{14, 18, 22, 26\}$ and temperature $\beta=\{0, 1, 2,5,10\}$. Values of $N$ within the GUE universality class.
}
\label{fig_KC_SYK_plateau}
\end{figure}

There is an important subtlety relating the SYK to matrix models \cite{PhysRevB.95.115150}:  the SYK model displays the different GUE, the GOE or the GSE ensembles, depending on the number of Majorana fermions and hence the nature of the particle-hole symmetry in the model. Concretely, for $N$ mod $8$ equal to $2$ or $6$ the GUE appears, for $N$ mod $8$ equal to zero the GOE appears, and finally for $N$ mod $8$ equal to $4$ the GSE appears. As we described above, the dynamics of complexity depends on the specific ensemble.  Thus, for clarity,  we choose SYK models in the GUE universality class and plot the results for $N=14,18,22,26$ in Figs.~\ref{fig_LZan_SYK_full} --
\ref{fig_KC_SYK_plateau}.  The  other cases that are related to GOE and GSE ensembles proceed similarly and are not shown here.  We have checked that the results in those cases are consistent with  observations in the previous section about complexity and universality, and the results of \cite{PhysRevB.95.115150}.

Figs.~\ref{fig_LZan_SYK_full} and~\ref{fig_LZbn_SYK_full}  show the global structure of Lanczos coefficients. As before, there are two regimes, one showing linear growth of Lanczos coefficients and one in which the $a_n$ and the $b_n$ are approximately zero and constant, respectively. In Fig.~\ref{fig_LZan_SYK_smalln} and Fig.~\ref{fig_LZbn_SYK_smalln} we focus on the small $n$  regime, to verify that transition between these behaviors occurs at  $n\sim\mathcal{O}(1)$, as before. One can again verify that in the linear regime $a_n+d=2b_n\sim n$ and the dynamics resembles that of the free oscillator discussed in earlier sections.

Since the behavior of the Lanczos coefficients is parallel to the GUE matrix model, we also expect that the dynamics of complexity will behave similarly.  We verify this in Fig.~\ref{fig_KC_SYK_plateau}, where we analyze temperatures $\beta=\{0, 1, 2,5,10\}$ with $N=\{14,18,22,26\}$ as above. We clearly see the same regimes as for the matrix model ensembles. We have an initial period of quadratic growth in time, followed by a linear ramp which ends at a peak. Next we have an approximately linear downward slope which ends as  complexity equilibrates at a plateau. Given the results of the previous section, the presence of the slope is signalling that the SYK model shares the universal Hamiltonian eigenvalue statistics of random matrix models, as shown previously in \cite{PhysRevB.95.115150,SFF1} using other methods. We also see from the shape of the slope and the transition to the plateau that we are analyzing SYK models in the GUE universality class.

It is interesting to compare these results with the analysis in \cite{perm}. There, a similar notion of complexity based on the spread of the wavefunction, but in the computational basis, was studied for SYK. The resulting complexity saturates at a exponential value at a time of the order of the entropy of the system, instead of a time exponentially large in the entropy. This shows the importance of the minimization over all bases put forward in this article. Also, this gives hope that in actual physical situations, the minimization performed in sec~\ref{II.II} is actually good for sufficiently long times.

\subsection{Black holes from collapse} 

Above, we analyzed complexity growth in the TFD states of several theories that are related to two dimensional gravity.  The TFD states of these theories are related to the physics of eternal black holes.  We showed above that Schwarzian theory has a TFD state complexity that grows quadratically in time for all time, while the matrix model and SYK theories display a transition to linear behavior followed eventually by a slope and a plateau. 

It is interesting to instead consider states that model black holes made from collapse.  In this case, the energy band of the system is fixed, although we do not know the precise microstate -- thus, we are in the microcanonical ensemble, rather that the canonical ensemble. The latter ensemble is the one modeled by the conventional TFD after a partial trace. To model a black hole formed by collapse we can consider a ``microcanonical TFD'' by entangling only states in a  small energy window, all of them weighted with roughly the same amplitude, namely one  over the microcanonical degeneracy.  This limits the range of energies in which to compute energy moments.

Alternatively, we can consider a modified Gibbs ensemble, in which we introduce a maximum energy $E_{max}$ in the definition of the TFD, so that
\be
\vert\psi_{\beta_{E_{max}}}\rangle \equiv\frac{1}{\sqrt{Z_{\beta_{E_{max}}}}}\sum_n^{n_{E_{max}}} e^{-\frac{\beta E_n}{2}}\vert n,n\rangle\;,
\ee
where
\be
Z_{\beta_{E_{max}}}=
\sum_n^{n_{E_{max}}}
e^{-\beta \,E_n}\;.
\ee
Following the previous examples, the moments
\be
\mu_n^S=i^n\frac{\textrm{Tr}\rho_{\beta_{E_{max}}}\,H^n}{Z_{\beta_{E_{max}}}}\;,
\ee
are expected to have characteristic behaviors in different ranges.
Suppose $E_{max}\geq E$ and $n\ll E\beta$, where $E$ is the thermal energy
\be
E=\frac{\textrm{Tr}\rho_{\beta}\,H^n}{Z_\beta}\simeq\frac{\textrm{Tr}\rho_{\beta_{E_{max}}}\,H^n}{Z_{\beta_{E_{max}}}}\;.
\ee
Then,  in a theory with a semiclassical limit, we expect the the moments will factorize 
and can be written in terms of the variance of the energy in the thermal state.

Equivalently, this corresponds to small times $t\ll \beta$, where we can approximate the survival probability by
\be 
\vert \psi_t\rangle =\,e^{-iHt}\vert \psi\rangle\Rightarrow \vert C_t\vert^2\simeq e^{-\sigma_E^2\,t^2}\Longrightarrow C_t\simeq e^{-\frac{\sigma_E^2\,t^2}{2}}e^{i\,\phi_t}\;,
\ee
with a time-dependent phase $e^{i\,\phi_t}$. This implies that for sufficiently short times the system effectively behaves as if it is  a particle moving in the  Weyl-Heisenberg group (see \cite{Caputa:2021sib} and above). As we discussed in Sec.~\ref{sec:SL2Rsection}, this regime is equivalently found within the $SL(2,R)$ paradigm. Then the $b_n$'s satisfy~(\ref{leadingshorttime}) and Krylov complexity grows quadratically in time~(\ref{eq:shortime}).

The next regime appears  if there is a hierarchical separation $E_{max}\gg E$. In such cases we can consider moments for which  $E_{max}\beta\gg n\gg E\beta\sim S(E)$. We can understand the behavior of the moments from the moment integral 
\be 
\int\limits_{0}^{E_{max}}dE \,e^{S(E)-\beta E}\,E^n\simeq \int\limits_{0}^{E_{max}}dE \,e^{-\beta E +n\log E}\;.
\ee
To evaluate this integral by the method of saddlepoints
we can neglect the contribution from the entropy $S(E)$ as in the second expression above since $S(E) \ll n$ in this regime. The saddlepoint equation is then $E = n/\beta$. This gives
\be 
\mu_n\propto \left( \frac{n}{\beta}\right) ^n\;,
\ee
implying that the $b_n$'s and/or the $a_n$'s grow linearly for sufficiently large $n$ as discussed earlier. In this regime we cannot say for sure what is the behavior of complexity, since to do so we need to find the ratio between both sets of Lanczos coefficients. In the models studied above we always found $a_n+d=2b_n$, with $d$ a constant, which implies quadratic growth of complexity, at least for an intermediate range of times.  We may conjecture that this quadratic growth is universal, but we do not have an argument establishing this.

Finally, whenever there is an upper bound to the energy $E_{\textrm{max}}$, there is a further regime in which $n \gg E_{ max} \, \beta$. The behavior of the moments in this regime is very simple since the moment integral $\int\limits dE \, e^{S(E)-\beta E}E^n $ concentrates on the upper limit.  Using this fact, the moments scale as 
\begin{equation}
    \mu_n \sim E_{\rm max}^n\;.
\end{equation}
Following our reasoning above, it follows that both $a_n$ and $b_n$ must saturate to a constant value, implying that complexity grows linearly in time, with a rate of $\mathcal{O}(E_{ max})$. A similar phenomenon was noted for operator growth complexity in  \cite{Barbon:2019wsy,Rabinovici:2020ryf,Kar:2021nbm}.  

For systems describing black holes made from collapse we would expect $E_{\textrm{max}}$ to be of order the thermal energy $E$, which in turn should be the mass of the black hole.  In this case, the intermediate window of time with model dependent complexity growth that depends on the precise relation between the $a_n$ and $b_n$ coefficients is absent. We are left with two universal regimes -- the initial quadratic growth of complexity controlled by the variance of the energy in the initial state, and a late regime in which complexity grows linearly at a rate controlled by the mass of the black hole. This linear growth lasts for exponentially larges times in the entropy. As argued before, at those times we expect that the dynamics is well approximated by that of a random matrix, and displays a peak, followed by a downward slope terminating at the plateau.

\section{Discussion} 

We have provided a definition of quantum state complexity that avoids the basis ambiguity present in previous work by minimizing over all choices of basis. Usually, quantities defined through minimization are very difficult to compute.  Surprisingly, we showed the minimization required by our definition can be carried out explicitly via the Lanczos method for representing quantum dynamics, and the associated Krylov basis. We show that the quantum state complexity we defined is intimately related to basic physical quantities such as the survival amplitude and the energy spectrum. We exploited these relations to explore how quantum complexity of unitary evolution behaves for chaotic systems, and, in particular, how it is related to spectral statistics \cite{Dyson,Bohigas}.  In chaotic systems, we found a  linear growth of complexity for exponential time, followed by a plateau, as expected from the conjecture in \cite{SusskindQC}, and further extended the conjecture by demonstrating finer grained dynamics at exponentially long time scales. Specifically, the exponential growth overshoots the final plateau, reaches a peak, and then decays back.  Using matrix models as examples, we show that the form of the decaying slope and the smoothness with which it merges with the plateau are controlled by universal statistics, namely spectral rigidity.  Indeed, the form of this complexity decay to the plateau distinguishes between the three universality classes of matrix models: the Gaussian Unitary, Gaussian Orthogonal, and Gaussian Symplectic ensembles.

We have demonstrated an explicit connection between the dynamics of quantum state complexity and quantum chaos. It would be interesting to similarly analyze integrable systems, especially ones with finite Hilbert spaces where our methods can be applied most easily.  This would help to construct a  bridge from the classification of phases of quantum matter via partition functions, to a novel characterization of these phases in terms of the dynamics of quantum complexity.

Another important avenue is to explore the structural complexity of Hamiltonian eigenstates.  This complexity lies at the heart of the black hole information paradox \cite{PhysRevD.14.2460}, as explored some time ago in \cite{2005Vijay}.  It is also central to the Eigenstate Thermalization Hypothesis \cite{PhysRevA.43.2046,PhysRevE.50.888,2008rigol} that seeks to explain why the energy eigenstates of some many-body systems may be well-described by thermal ensembles. Indeed, the Eigenstate Complexity Hypothesis of \cite{Balasubramanian:2019wgd} shows that a quantum system in which energy eigenstates cannot be easily transported into each other through simple operations will have linear complexity growth for an exponentially long time, in Nielsen's geometric sense of geodesic lengths \cite{Nielsen}. In this vein it is worthwhile to analyze the relations between eigenstate complexity and the Eigenstate Thermalization Hypothesis in SYK, expanding on \cite{2016Magan,Sonner:2017hxc,Hunter-Jones:2017raw,Haque:2017bts}. 

Finally, it would be interesting to use these results, in particular the time evolved TFD wavefunction, to better understand the physics of the black hole interior, expanding on \cite{Hartman:2013qma}, and perhaps the conjectures relating complexity in quantum field theory to spatial volumes and action in dual theory of gravity \cite{SStanford,Brown1,Brown2,belin2021complexity}.  Concretely, recent work has suggested that it might be possible to understand the  black hole interior volume  as a notion of phase space volume in a dual mechanical theory \cite{iliesiu2021volume}.  Quantum complexity, as we have defined it, is by construction the minimal notion of phase space volume explored by the system, since Hilbert space equipped with the canonical metric and symplectic structure
can be understood as a phase space, and quantum dynamics become classical dynamics in this phase space \cite{ashtekar1997geometrical}, a feature exploited for quantum complexity in \cite{Bueno:2019ajd}. In this context, it is natural to explore the connections between our approach and the ideas in \cite{iliesiu2021volume}. A related promising path is  to use the geometric approach to Krylov complexity developed recently in \cite{Caputa:2021sib}, applied  to the TFD quantum state complexity, making the connections between symmetries, geometry and complexity manifest. It would also be interesting to study how the black hole interior may be sensitive to the quantum chaotic universality class of the underlying theory, since we have shown that the exponential time dynamics of complexity differs between these classes. Finally, another avenue is to analyze from the present light other types of interior processes, such as the recently considered collisions in the black hole interior \cite{2021zhao,2021hael,haehl2022collisions}. 

\medskip
{\bf Acknowledgements.} We are grateful to Alex Streicher for conversations about numerical evaluation of the Lanczos algorithm. We also thank Dongsheng Ge, M\'{a}rk Mezei, Dimitrios Patramanis, G\'{a}bor S\'{a}rosi and Joan Simon for interesting discussions. The work of P.C. is supported by NAWA “Polish Returns 2019” PPN/PPO/2019/1/00010/U/0001 and NCN Sonata Bis 9 2019/34/E/ST2/00123 grants.  The work of Q.W., V.B. and J.M is supported by a DOE QuantISED grantDE-SC0020360 and the Simons Foundation It From Qubit collaboration (385592).

\appendix


\begin{thebibliography}{99}
\bibitem{Kolmogorov}
A.~N.~Kolmogorov. ``Three approaches to the quantitative definition of information.''
Problems of Information and Transmission, 1(1):1--7, 1965

\bibitem{Rissanen}
J.~Rissanen, 1986. ``Stochastic complexity and modeling.''
The annals of statistics, pp.1080-1100.


 \bibitem{Nielsen}
  M.~A.~Nielsen,
``A geometric approach to quantum lower bounds,''
 [arXiv:0502070 [quant-ph]].\\
    M.~A.~Nielsen, M.~R.~Dowling, M.~Gu and A.~C.~Doherty
`` Quantum computation as geometry,''
Science {\bf 311}, 1133 (2006),
 [arXiv:0603161 [quant-ph]].\\
 M.~R.~Dowling and M.~A.~Nielsen,
`` The geometry of quantum computation,''
 [arXiv:0701004 [quant-ph]].
 
 
 \bibitem{Brown:2016wib}
A.~R.~Brown, L.~Susskind and Y.~Zhao,
``Quantum Complexity and Negative Curvature,''
Phys. Rev. D \textbf{95} (2017) no.4, 045010
[arXiv:1608.02612 [hep-th]].
 
\bibitem{Magan2018}
J.~Magan,
  ``Black holes, Complexity and quantum chaos'',
 JHEP \textbf{09} (2018), 043
arXiv:1805.05839 [hep-th].


\bibitem{Balasubramanian:2019wgd}
V.~Balasubramanian, M.~Decross, A.~Kar and O.~Parrikar,
``Quantum Complexity of Time Evolution with Chaotic Hamiltonians,''
JHEP \textbf{01}, 134 (2020)
[arXiv:1905.05765 [hep-th]].



\bibitem{Balasubramanian:2021mxo}
V.~Balasubramanian, M.~DeCross, A.~Kar, Y.~Li and O.~Parrikar,
``Complexity growth in integrable and chaotic models,''
JHEP \textbf{07}, 011 (2021)
[arXiv:2101.02209 [hep-th]].

 \bibitem{Bueno:2019ajd}
P.~Bueno, J.~M.~Magan and C.~S.~Shahbazi,
``Complexity measures in QFT and constrained geometric actions,''
JHEP \textbf{09} (2021), 200
[arXiv:1908.03577 [hep-th]].


\bibitem{Brandao:2019sgy}
F.~G.~S.~L.~Brand\~ao, W.~Chemissany, N.~Hunter-Jones, R.~Kueng and J.~Preskill,
``Models of Quantum Complexity Growth,''
PRX Quantum \textbf{2} (2021) no.3, 030316


\bibitem{Bulchandani2021}
Bulchandani, Vir B. and Sondhi, S. L.,
``How smooth is quantum complexity?,''
JHEP \textbf{10} (2021), 230
[arXiv:2106.08324 [quant-ph]].



\bibitem{brown2021quantum}
Adam R. Brown,
``A Quantum Complexity Lowerbound from Differential Geometry,''
[arXiv:2112.05724 [hep-th]].




  \bibitem{SusskindQC}
L.~Susskind,
`` Computational Complexity and Black Hole Horizons,''
  Fortsch. Phys. {\bf 64}, 44-48 (2016),
  [arXiv:1402.5674 [hep-th]].
  
  
  \bibitem{Lanczosbook}
The Recursion Method: Application to Many-Body Dynamics, vol. 23 Science  Business Media
V. Viswanath, G. Müller\\
C.~Lanczos,
``An iteration method for the solution of the eigenvalue problem of linear differential and integral operators,''
J. Res. Natl. Bur. Stand. B \textbf{45} (1950), 255-282






\bibitem{GUHR1998189}
Thomas Guhr and Axel Müller–Groeling and Hans A. Weidenmüller,
``Random-matrix theories in quantum physics: common concepts,''
Physics Reports, \textbf{299} (1998) 189-425
[arXiv:9707301 [hep-th]].


\bibitem{SFF1}
J.~S.~Cotler, G.~Gur-Ari, M.~Hanada, J.~Polchinski, P.~Saad, S.~H.~Shenker, D.~Stanford, A.~Streicher and M.~Tezuka,
``Black Holes and Random Matrices,''
JHEP \textbf{05} (2017), 118
[erratum: JHEP \textbf{09} (2018), 002]
[arXiv:1611.04650 [hep-th]].



\bibitem{RM1}
M.~L.~Mehta,
``On the statistical properties of the level-spacings in nuclear spectra,''
Nuclear Physics \textbf{18} (1960) 395-419

\bibitem{RM2}
M.~Gaudin,
``Sur la loi limite de l’espacement des valeurs propres d’une matrice al\'e
atoire,''
Nuclear Physics \textbf{25} (1961) 447-458

\bibitem{RM3}
F.~J.~Dyson,
``Statistical theory of the energy levels of complex systems. III,''
Journal of Mathematical Physics 3 \textbf{1} (1962) 166-175


\bibitem{REVIEW1995}
Francesco, P.Di and Ginsparg, P. and Zinn-Justin, J,
``2D Gravity and Random Matrices,''
Physics Reports, \textbf{254} (1995) 1–133
[arXiv:9306153 [hep-th]].







\bibitem{Jefferson:2017sdb}
R.~Jefferson and R.~C.~Myers,
``Circuit complexity in quantum field theory,''
JHEP \textbf{10} (2017), 107
[arXiv:1707.08570 [hep-th]].

\bibitem{Chapman:2017rqy}
S.~Chapman, M.~P.~Heller, H.~Marrochio and F.~Pastawski,
``Toward a Definition of Complexity for Quantum Field Theory States,''
Phys. Rev. Lett. \textbf{120} (2018) no.12, 121602
[arXiv:1707.08582 [hep-th]].

  \bibitem{Caputa:2018kdj}
P.~Caputa and J.~M.~Magan,
``Quantum Computation as Gravity,''
Phys. Rev. Lett. \textbf{122} (2019) no.23, 231302
[arXiv:1807.04422 [hep-th]].



\bibitem{Balasubramanian:2018hsu}
V.~Balasubramanian, M.~DeCross, A.~Kar and O.~Parrikar,
``Binding Complexity and Multiparty Entanglement,''
JHEP \textbf{02}, 069 (2019)
[arXiv:1811.04085 [hep-th]].

\bibitem{Erdmenger:2020sup}
J.~Erdmenger, M.~Gerbershagen and A.~L.~Weigel,
``Complexity measures from geometric actions on Virasoro and Kac-Moody orbits,''
JHEP \textbf{11} (2020), 003
[arXiv:2004.03619 [hep-th]].

\bibitem{Chagnet:2021uvi}
N.~Chagnet, S.~Chapman, J.~de Boer and C.~Zukowski,
``Complexity for Conformal Field Theories in General Dimensions,''
Phys. Rev. Lett. \textbf{128} (2022), 5;
[arXiv:2103.06920 [hep-th]].


\bibitem{Erdmenger:2021wzc}
Erdmenger, Johanna and Flory, Mario and Gerbershagen, Marius and Heller, Michal P. and Weigel, Anna-Lena,
``Exact Gravity Duals for Simple Quantum Circuits,''
[arXiv:2112.12158 [hep-th]].





\bibitem{Caputa:2017yrh}
P.~Caputa, N.~Kundu, M.~Miyaji, T.~Takayanagi and K.~Watanabe,
``Liouville Action as Path-Integral Complexity: From Continuous Tensor Networks to AdS/CFT,''
JHEP \textbf{11} (2017), 097; P.~Caputa, N.~Kundu, M.~Miyaji, T.~Takayanagi and K.~Watanabe,
``Anti-de Sitter Space from Optimization of Path Integrals in Conformal Field Theories,''
Phys. Rev. Lett. \textbf{119} (2017) no.7, 071602


\bibitem{2018Bartek}
Czech, Bartłomiej,
``Einstein Equations from Varying Complexity,''
Phys. Rev. Lett. \textbf{120} (2018) no.3



 \bibitem{perm}
  J.~M.~Magan,
`` Decoherence and microscopic diffusion at the Sachdev-Ye-Kitaev model,'' 
  Phys. Rev. D. \textbf{98} (2018) no.2, 071602
  [arXiv:1612.06765 [hep-th]].
  
  
  
   \bibitem{czech2022holographic}
 Bartlomiej Czech,
`` Holographic State Complexity from Group Cohomology,'' 
  [arXiv:2201.01303 [hep-th]].
  
  
\bibitem{haferkamp2021linear}
Jonas Haferkamp and Philippe Faist and Naga B. T. Kothakonda and Jens Eisert and Nicole Yunger Halpern,
``Linear growth of quantum circuit complexity,''
[arXiv:2106.05305 [hep-th]].

  
  


\bibitem{python-SciPy}
P.~Virtanen et al,
``SciPy 1.0: Fundamental Algorithms for Scientific Computing in Python''
Nature Methods, 17(3), 261-272.

\bibitem{LAPACK-Hessenberg}
Quintana-Ortí, Gregorio, and Robert van de Geijn. 
``Improving the performance of reduction to Hessenberg form.'' 
ACM Transactions on Mathematical Software (TOMS) 32.2 (2006): 180-194.

\bibitem{python-mpmath}
F.~Johansson et al,
``mpmath: a Python library for arbitrary-precision floating-point arithmetic (version 0.18)''
, December 2013.


\bibitem{ohya2004quantum}
Ohya, Masanori and Petz, D{\'e}nes,
``Quantum entropy and its use'',
Springer Science \& Business Media.




\bibitem{Maldacena:2001kr}
J.~M.~Maldacena,
``Eternal black holes in anti-de Sitter,''
JHEP \textbf{04} (2003), 021
[arXiv:hep-th/0106112 [hep-th]].



\bibitem{delCampo:2017bzr}
 A.~del Campo, and J.~Molina-Vilaplana, and J.~Sonner,
``Scrambling the spectral form factor: unitarity constraints and exact results,''
Phys. Rev. D \textbf{95} (2017), 12
[arXiv:hep-th/1702.04350 [hep-th]].





\bibitem{Saad:2019lba}
Saad, Phil and Shenker, Stephen H. and Stanford, Douglas,
``JT gravity as a matrix integral,''
[arXiv:hep-th/1903.11115 [hep-th]].


\bibitem{Stanford:2019vob}
Stanford, Douglas and Witten, Edward,
``JT gravity and the ensembles of random matrix theory,''
Adv. Theor. Math. Phys.\textbf{24} (2020), 6, 1475-1680
[arXiv:hep-th/1907.03363 [hep-th]].



\bibitem{Johnson:2019eik}
Johnson, Clifford V.
``Nonperturbative Jackiw-Teitelboim gravity,''
Phys. Rev. D \textbf{101} (2020), 10
[arXiv:hep-th/1912.03637 [hep-th]].


\bibitem{Maxfield:2020ale}
Maxfield, Henry and Turiaci, Gustavo J.
``The path integral of 3D gravity near extremality; or, JT gravity with defects as a matrix integral,''
JHEP \textbf{01} (2021), 118
[arXiv:hep-th/2006.11317 [hep-th]].

\bibitem{Mertens:2020hbs}
Mertens, Thomas G. and Turiaci, Gustavo J.
``Liouville quantum gravity -- holography, JT and matrices,''
JHEP \textbf{01} (2021), 073
[arXiv:hep-th/2006.07072 [hep-th]].


\bibitem{Witten:2020wvy}
Witten, Edward,
``Matrix Models and Deformations of JT Gravity,''
Proc. Roy. Soc. Lond. A\textbf{476} (2020), 2244,
[arXiv:hep-th/2006.13414 [hep-th]].

 
  \bibitem{SStanford}
D.~Stanford and L.~Susskind,
`` Complexity and Shock Wave Geometries,''
  Phys.~Rev.~D {\bf 90}, 12 (2014),
  [arXiv:1406.2678 [hep-th]].
  
   \bibitem{Brown1}
A.~Brown, D.~Roberts, L.~Susskind, B~Swingle, Y.~Zhao,
`` Holographic Complexity Equals Bulk Action?,''
 Phys.~Rev.~Lett {\bf 116}, 19 (2016),
[arXiv:1509.07876 [hep-th]].

\bibitem{Brown2}
A.~Brown, D.~Roberts, L.~Susskind, B~Swingle, Y.~Zhao,
`` Complexity, action, and black holes,''
 Phys.~Rev.~D {\bf 93}, 19 (2016),
[arXiv:1512.04993 [hep-th]].



\bibitem{Parker:2018yvk}
D.~E.~Parker, X.~Cao, A.~Avdoshkin, T.~Scaffidi and E.~Altman,
``A Universal Operator Growth Hypothesis,''
Phys. Rev. X \textbf{9} (2019) no.4, 041017






\bibitem{Barbon:2019wsy}
J.~L.~F.~Barb\'on, E.~Rabinovici, R.~Shir and R.~Sinha,
``On The Evolution Of Operator Complexity Beyond Scrambling,''
JHEP \textbf{10} (2019), 264


\bibitem{Avdoshkin:2019trj}
Avdoshkin, Alexander and Dymarsky, Anatoly,
``Euclidean operator growth and quantum chaos,''
PhysRevRes \textbf{2} (2020), 4
[arXiv:1911.09672 [hep-th]].


\bibitem{Dymarsky:2019elm}
Dymarsky, Anatoly and Gorsky, Alexander,
``Quantum chaos as delocalization in Krylov space,''
PhysRevB \textbf{8} (2020), 102
[arXiv:1912.12227 [hep-th]].


\bibitem{Magan:2020iac}
J.~M.~Mag\'an and J.~Simon,
``On operator growth and emergent Poincar\'e symmetries,''
JHEP \textbf{05} (2020), 071
[arXiv:2002.03865 [hep-th]].

\bibitem{Jian:2020qpp}
S.~K.~Jian, B.~Swingle and Z.~Y.~Xian,
``Complexity growth of operators in the SYK model and in JT gravity,''
JHEP \textbf{03} (2021), 014


\bibitem{Rabinovici:2020ryf}
E.~Rabinovici, A.~S\'anchez-Garrido, R.~Shir and J.~Sonner,
``Operator complexity: a journey to the edge of Krylov space,''
JHEP \textbf{06} (2021), 062
[arXiv:2009.01862 [hep-th]].


\bibitem{Dymarsky:2021bjq}
A.~Dymarsky and M.~Smolkin,
``Krylov complexity in conformal field theory,''
PhysRevD \textbf{8} (2021), 104
[arXiv:2104.09514 [hep-th]].


\bibitem{Kar:2021nbm}
A.~Kar, L.~Lamprou, M.~Rozali and J.~Sully,
``Random Matrix Theory for Complexity Growth and Black Hole Interiors,''
JHEP \textbf{01} (2022), 016
[arXiv:2106.02046 [hep-th]].


\bibitem{Caputa:2021sib}
P.~Caputa, J.~M.~Magan and D.~Patramanis,
``Geometry of Krylov Complexity,''
PhysRevResearch \textbf{04} (2022), 1
[arXiv:2109.03824 [hep-th]].


\bibitem{Kim:2021okd}
Kim, Joonho and Murugan, Jeff and Olle, Jan and Rosa, Dario,
``Operator delocalization in quantum networks,''
PhysRevA \textbf{01} (2022), 105
[arXiv:2109.03824 [hep-th]].


\bibitem{Caputa:2021ori}
P.~Caputa and S.~Datta,
JHEP \textbf{12} (2021), 188
[arXiv:2110.10519 [hep-th]].


\bibitem{Patramanis:2021lkx}
Patramanis, Dimitrios,
``Probing the entanglement of operator growth,''
[arXiv:2111.03424 [hep-th]].


\bibitem{Trigueros:2021rwj}
Trigueros, Fabian Ballar and Lin, Cheng-Ju,
``Krylov complexity of many-body localization: Operator localization in Krylov basis,''
[arXiv:2112.04722 [hep-th]].



\bibitem{Rabinovici:2021qqt}
Rabinovici, E. and S\'anchez-Garrido, A. and Shir, R. and Sonner, J.,
``Krylov Localization and suppression of complexity,''
[arXiv:2112.12128 [hep-th]].

\bibitem{hornedal2022ultimate}
Niklas Hörnedal, Nicoletta Carabba, Apollonas S. Matsoukas-Roubeas and Adolfo del Campo,
``Ultimate Physical Limits to the Growth of Operator Complexity,''
[arXiv:2202.05006 [hep-th]].

\bibitem{Adhikari:2022whf}
Adhikari, Kiran and Choudhury, Sayantan and Roy, Abhishek,
``${\cal K}$rylov ${\cal C}$omplexity in ${\cal Q}$uantum ${\cal F}$ield ${\cal T}$heory,''
[arXiv:2204.02250 [hep-th]].






\bibitem{Maldacena:2015waa}
J.~Maldacena, S.~H.~Shenker and D.~Stanford,
``A bound on chaos,''
JHEP \textbf{08} (2016), 106
[arXiv:1503.01409 [hep-th]].



\bibitem{RevModPhys.62.867}
Zhang, Wei-Min and Feng, Da Hsuan and Gilmore, Robert
Coherent states: Theory and some applications,
  Rev. Mod. Phys. 62, 4, 867--927, 1990

  
     \bibitem{coherent1}
 J.~ R.~Klauder and E.~ C.~ G.~ Sudarshan,
 Fundamentals of quantum optics, (Benjamin, New York) 1968. 
  
 \bibitem{coherent2}
A.~M.~Perelomov,
 Commun.\ Math.\ Phys.\ {\bf 26} (1972) 222.  
  
\bibitem{kitaev}
  A.~Kitaev,
  ``A simple model of quantum holography,''
  Talks at KITP, April 7, 2015 and May 27, 2015.

\bibitem{sachdev}
   S.~Sachdev and J.~Ye,
   `` Gapless spin fluid ground state in a random, quantum Heisenberg ferromagnet,''
   Phys. \ Rev.\ Lett.\ {\bf 70} (1993) 3339,
   arXiv:cond-mat/9212030 [cond-mat]. 
   
\bibitem{sachdev2}
S.~Sachdev,
  {\em Bekenstein-Hawking Entropy and Strange Metals},
  Phys.\ Rev.\ X {\bf 5} (2015) 041025,
  [arXiv:1506.05111 [hep-th]].



\bibitem{Polchinski:2016xgd}
Polchinski, Joseph and Rosenhaus, Vladimir,
  ``The Spectrum in the Sachdev-Ye-Kitaev Model,''
  JHEP {\bf 04} (2016) 001,
  [arXiv:1601.06768 [hep-th]].


\bibitem{Maldacena:2016hyu}
J.~Maldacena and D.~Stanford,
``Remarks on the Sachdev-Ye-Kitaev model,''
Phys. Rev. D \textbf{94} (2016) no.10, 106002
[arXiv:1604.07818 [hep-th]].



\bibitem{Kristan2016}
Jensen, Kristan,
``Chaos in AdS2 Holography,''
Phys. Rev. Lett \textbf{11} (2016) 117
[arXiv:1605.06098 [hep-th]].

\bibitem{Maldacena:2016upp}
Maldacena, Juan and Stanford, Douglas and Yang, Zhenbin,
``Conformal symmetry and its breaking in two dimensional Nearly Anti-de-Sitter space,''
PTEP \textbf{12} (2016) 12C104
[arXiv:1606.01857 [hep-th]].



\bibitem{SFF2}
A.~M.~Garc\'\i{}a-Garc\'\i{}a and J.~J.~M.~Verbaarschot,
``Spectral and thermodynamic properties of the Sachdev-Ye-Kitaev model,''
Phys. Rev. D \textbf{94} (2016) no.12, 126010
[arXiv:1610.03816 [hep-th]].


\bibitem{Stanford:2017thb}
Stanford, Douglas and Witten, Edward,
``Fermionic Localization of the Schwarzian Theory,''
JHEP \textbf{10} (2017) 008
[arXiv:1703.04612 [hep-th]].


\bibitem{Mertens:2017mtv}
Mertens, Thomas G. and Turiaci, Gustavo J. and Verlinde, Herman L.,
``Solving the Schwarzian via the Conformal Bootstrap,''
JHEP \textbf{08} (2017) 136
[arXiv:1705.08408 [hep-th]].


\bibitem{Kitaev:2017awl}
Kitaev, Alexei and Suh, S. Josephine,
``The soft mode in the Sachdev-Ye-Kitaev model and its gravity dual,''
JHEP \textbf{05} (2018) 183
[arXiv:1711.08467 [hep-th]].

\bibitem{GABORREVIEW}
Sarosi, Gabor,
``AdS$_{2}$ holography and the SYK model,''
PoS(Modave2017), Sissa Medialab, (2018)
[arXiv:1711.08482 [hep-th]].


\bibitem{Dyson}
Dyson,Freeman J.,
``Statistical Theory of the Energy Levels of Complex Systems. I,''
Journal of Mathematical Physics \textbf{3} (1962) 1.



\bibitem{Bohigas}
O Bohigas, M. J. Giannoni, and C. Schmit,
``Characterization of chaotic quantum spectra
and universality of level fluctuation laws,''
Phys. Rev. Lett. \textbf{52} (1984) 1-4.




\bibitem{VijayGabor}
V.~Balasubramanian, B.~Craps, B.~Czech and G.~S\'arosi, ``Echoes of chaos from string theory black holes,'' JHEP \textbf{03}, 154 (2017)
[arXiv:1612.04334 [hep-th]].

\bibitem{astreicher}
A.~Streicher, Personal Communication.

\bibitem{feng2018spectrum}
Renjie Feng, Gang Tian, and Dongyi Wei,
``Spectrum of SYK model,''
[arXiv:1801.10073 [hep-th]].


\bibitem{PhysRevB.95.115150}
Yi-Zhuang You, Andreas Ludwig, and Cenke Xu,
``Sachdev-Ye-Kitaev model and thermalization on the boundary of many-body localized fermionic symmetry-protected topological states,''
Phys. Rev. B \textbf{95} (2017) 11.
[arXiv:1602.06964 [cond-mat]].



\bibitem{PhysRevD.14.2460}
Hawking, S. W.,
``Breakdown of predictability in gravitational collapse,''
Phys. Rev. D \textbf{14} (1976) 10.





\bibitem{2005Vijay}
Balasubramanian, Vijay, Boer, Jan de, Jejjala, Vishnu, and Simón, Joan,
``The library of Babel: on the origin of gravitational thermodynamics,''
JHEP \textbf{12} (2005) 1029-8479
[arXiv:0508023 [hep-th]].


\bibitem{PhysRevA.43.2046}
Deutsch, J. M.,
``Quantum statistical mechanics in a closed system,''
Phys. Rev. A \textbf{43} (1991) 4.



\bibitem{PhysRevE.50.888}
Srednicki, Mark,
``Chaos and quantum thermalization,''
Phys. Rev. E \textbf{50} (1994) 2.
[arXiv:9403051 [cond-mat]].

\bibitem{2008rigol}
Marcos Rigol, Vanja Dunjko, and Maxim Olshanii,
``Thermalization and its mechanism for generic isolated quantum systems,''
Nature \textbf{452} (2008) 7189.
[arXiv:0708.13241 [cond-mat]].

\bibitem{2016Magan}
Magan, Javier M.,
``Random Free Fermions: An Analytical Example of Eigenstate Thermalization,''
Phys. Rev. Lett \textbf{3} (2016) 116.
[arXiv:1508.05339 [quant-ph]].




\bibitem{Sonner:2017hxc}
Sonner, Julian and Vielma, Manuel,
``Eigenstate thermalization in the Sachdev-Ye-Kitaev model,''
JHEP \textbf{11} (2017) 149.
[arXiv:1707.08013 [hep-th]].



\bibitem{Hunter-Jones:2017raw}
Hunter-Jones, Nicholas and Liu, Junyu and Zhou, Yehao,
``On thermalization in the SYK and supersymmetric SYK models,''
JHEP \textbf{02} (2018) 142.
[arXiv:1710.03012 [hep-th]].


\bibitem{Haque:2017bts}
Haque, Masudul and McClarty, Paul,
``Eigenstate thermalization scaling in Majorana clusters: From chaotic to integrable Sachdev-Ye-Kitaev models,''
Phys. Rev. B \textbf{11} (2019) 100.
[arXiv:1711.02360 [cond-mat.stat-mech]].




\bibitem{Hartman:2013qma}
T.~Hartman and J.~Maldacena,
``Time Evolution of Entanglement Entropy from Black Hole Interiors,''
JHEP \textbf{05} (2013), 014
[arXiv:1303.1080 [hep-th]].




\bibitem{belin2021complexity}
Alexandre Belin, Robert C. Myers, Shan-Ming Ruan,  Gábor Sárosi, and Antony J. Speranza,
``Complexity Equals Anything?,''
[arXiv:2111.02429 [hep-th]].


\bibitem{iliesiu2021volume}
Luca V. Iliesiu, Márk Mezei and Gábor Sárosi,
``The volume of the black hole interior at late times,''
[arXiv:2107.06286 [hep-th]].


\bibitem{ashtekar1997geometrical}
Abhay Ashtekar and Troy A. Schilling,
``Geometrical Formulation of Quantum Mechanics,''
[arXiv:9706069 [hep-th]].

\bibitem{2021zhao}
Zhao, Ying,
``Collision in the interior of wormhole,''
JHEP \textbf{03} (2021), 144
[arXiv:2011.06016 [hep-th]].



\bibitem{2021hael}
Haehl, Felix M. and Streicher, Alexandre and Zhao, Ying,
``Six-point functions and collisions in the black hole interior,''
JHEP \textbf{08} (2021), 134
[arXiv:2105.12755 [hep-th]].


\bibitem{haehl2022collisions}
Felix M. Haehl and Ying Zhao,
``Collisions of localized shocks and quantum circuits,''
[arXiv:2202.04661 [hep-th]].
















































\end{thebibliography}
\end{document}